# How much perceived-stress signal is in the maternal ECG? An exclusion bound across deep-learning and heart-rate-variability representations in pregnancy

Martin G. Frasch[1,2], Marlene J.E. Mayer[3], Clara Becker[3], Peter Zimmermann[3], Camilla Zelgert[3], Nicolas B. Garnier[4], Marta C. Antonelli[5,6], Silvia M. Lobmaier[3,*]

[1]Institute on Human Development and Disability, University of Washington, Seattle, WA, USA
[2]JoyBeat Medical, Seattle, WA, USA
[3]Department of Obstetrics and Gynaecology, TUM University Hospital, Technical University of Munich, Munich, Germany
[4]CNRS, ENS de Lyon, LPENSL, UMR5672, 69342, Lyon CEDEX 07, France
[5]Technical University of Munich; Institute for Advanced Study, Garching, Germany
[6]Instituto de Biología Celular y Neurociencia "Prof. Eduardo De Robertis," Facultad de Medicina, Universidad de Buenos Aires, Argentina

[*]**Corresponding author:** Prof. Dr. med. Silvia M. Lobmaier, Department of Obstetrics and Gynaecology, TUM University Hospital, Technical University of Munich, Ismaninger Straße 22, 81675 Munich, Germany. Email: silvia.lobmaier@tum.de

# Abstract

Deep learning can read physiological states from the electrocardiogram (ECG), prompting attempts to detect psychological stress. We asked how much information about perceived stress the maternal ECG carries. In 144 pregnant women, with every cross-validation split taken over participants, across 94 arms spanning 21 self-supervised, foundation-model, and heart-rate-variability architectures, the best discrimination of moderate-to-high stress (PSS-10 ≥ 19) was AUC 0.604 (95% CI 0.511–0.696), and AUC ≥ 0.70 was rejected in all eleven arms tested. Three pre-registered tests in a second cohort from the same center, recorded on different hardware, were null, at 30% power. Which of the three simultaneously recorded ECG channels was analyzed accounted for 96% of the predicted score's variance, against 4% between women; across 20 combinations of channel rule, label variant, and visit, the external correlation with perceived stress ranged from $r = -0.095$ to $r = +0.307$, a spread no larger than such a search produces by chance (raw one-sided $p$ of the maximum 0.011; calibrated max-statistic $p = 0.583$). Hair cortisol was not predicted either. In the second cohort, PSS-10 scores were not stable across administrations spanning pregnancy and the postpartum period (ICC 0.06–0.16), so one administration is a weak measure of a woman's stress level at another visit. A single-recording ECG screener for perceived stress is excluded in this population, and multi-channel studies must pre-register their channel rule.

**Keywords:** ECG; self-supervised learning; digital marker; cross-device validation; autonomic nervous system; pregnancy.

# Introduction

Deep learning applied to the surface electrocardiogram (ECG) can identify disease states and physiological patterns that may not be apparent on conventional expert ECG interpretation. Some of these findings lie genuinely beyond routine human reading of the ECG — for example asymptomatic left-ventricular dysfunction,(Attia, Kapa, et al. 2019) silent atrial fibrillation captured during sinus rhythm,(Attia, Noseworthy, et al. 2019) anemia(Kwon et al. 2020) and one-year all-cause mortality.(Raghunath et al. 2020) Other tasks at which deep models perform well, such as rhythm classification(Hannun et al. 2019) and full 12-lead diagnostic labeling,(Ribeiro et al. 2020) fall within the scope of experienced clinical interpretation and are better understood as automation of existing expert tasks. Taken together, this work indicates that the ECG carries physiological information beyond what conventional interpretation extracts. A natural question is whether *non-cardiac* states with autonomic correlates — most prominently psychological stress and affect — leave a recoverable imprint on the same signal, and whether any such marker behaves robustly under the conditions in which it would be deployed: different recording hardware, different sampling rates, and different operators.

Two limitations recur in the ECG-AI literature relevant to this question. First, cross-device external validation is uncommon: most models are trained and tested within a single device family or institutional dataset, leaving open whether reported performance reflects a genuine waveform-level signal or hardware-correlated confounds.(Strodthoff et al. 2021; Sangha et al. 2024; Siontis et al. 2015) Second, ECG-AI models of *psychological* states have largely been pursued in laboratory or wearable-stress paradigms with short-duration ground truth and within-session evaluation,(Schmidt et al. 2018; Gedam and Paul 2021) and rarely tested for sensitivity to a controlled intervention. The field has therefore not yet shown that an ECG-AI correlate of a non-cardiac autonomic state both generalizes across hardware and tracks change.

Pregnancy provides a clinically relevant setting in which to evaluate ECG-derived correlates of psychological stress. Perceived stress is common in pregnancy — elevated levels are reported in roughly 15–25% of pregnancies(Dunkel Schetter 2011; Howland et al. 2020) — and has well-characterized autonomic correlates in maternal and fetal heart rate.(Koenig et al. 2016; DiPietro, Costigan, and Gurewitsch 2003; Dunkel Schetter and Tanner 2012) It is also a target for an emerging class of behavioral interventions whose efficacy is difficult to quantify with self-report alone.(Guardino and Dunkel Schetter 2014) The reference instrument used here, the 10-item Perceived Stress Scale (PSS-10),(Cohen, Kamarck, and Mermelstein 1983) is validated but, as a self-report measure, is vulnerable to recall and reporting bias and lacks the temporal resolution to track response to intervention; it indexes *perceived* stress rather than an objectively defined autonomic state. Pregnancy is additionally a setting in which both research-grade abdominal ECG arrays and lower-channel ambulatory recorders are occasionally deployed, providing a built-in test of transfer across recording hardware.

Self-supervised learning offers a way to learn ECG representations from unlabeled recordings, which suits pregnancy cohorts in which labeled outcomes are few but hours of signal are available per woman. Contrastive and predictive methods adapted from computer vision and speech, among them SimCLR, BYOL, and contrastive predictive coding, learn 12-lead ECG representations that approach supervised performance on diagnostic labels with little labeled data,(Mehari and Strodthoff 2022) and variants that define positive pairs by physiological augmentation or by patient identity extend this.(Gopal et al. 2021; Diamant et al. 2022) Self-supervised ECG representations have been applied in pregnancy once before, by our group on an earlier analysis of this cohort (107 women): a transformation-recognition network pretrained on four public affect datasets (AMIGOS, DREAMER, WESAD, and SWELL) and transferred to maternal ECG reported an AUROC of 0.982 for stressed versus control pregnancies,(Sarkar et al. 2021) and the same network had reported 0.969 accuracy for laboratory affect states in WESAD.(Sarkar and Etemad 2022) Both evaluations drew cross-validation folds within each person's recording: shuffled windows in WESAD, and each woman's data divided into five parts in the pregnancy study. We chose SimCLR with a one-dimensional ResNet-34 encoder as a well-characterized baseline of this family rather than for a task-specific reason and evaluated it alongside a transformer pretrained on 1.4 million ECGs, a diagnostic foundation model, and heart-rate-variability panels, so that the question is whether any such representation carries participant-level information about perceived stress.

We set out to address three questions of direct relevance to ECG-AI marker development: which ECG signal source carries the strongest stress-related information; whether a model developed on one device generalizes to a cohort recorded on different hardware; and whether the resulting score is sensitive to a non-pharmacological autonomic perturbation. All three depend on how much perceived-stress information the maternal ECG carries, so we first establish that quantity and bound it. We report a systematic evaluation of 94 analysis arms spanning 21 architectures — self-supervised contrastive encoders, a transformer pretrained on 1.4 million ECGs, a diagnostic foundation model, and conventional and extended heart-rate-variability panels — on a single shared participant-level fold definition, together with formal non-superiority bounds, three pre-registered tests in a second cohort from the same center recorded on different hardware, a pre-registered test against an orthogonal biological construct, and an estimate of the reference standard's stability across pregnancy. We use two pregnancy cohorts from a single tertiary obstetric center (FELICITy 1, *n*=151 enrolled and 144 analyzed, for development; the FELICITy 2 randomized controlled trial, *n*=28, for a within-center, cross-device evaluation). Throughout, we describe the model output as an *ECG-derived correlate* rather than a validated biomarker, reserving the latter term for markers with objective biological validation.(FDA-NIH Biomarker Working Group 2016)

# Results

## Participant flow

Of 151 pregnant women enrolled in FELICITy 1 between June 2016 and July 2019, one was excluded because of an incomplete PSS-10 response, leaving 150 participants with paired PSS-10 and ECG data. Three pairs of these women (six participants) had byte-identical ECG files in all three signals despite different PSS-10 scores, indicating a copied recording; all six were excluded, leaving 144 participants for every FELICITy 1 analysis. Maternal hair cortisol was available for 125 women, of whom 109 also had usable, non-duplicated maternal ECG. In FELICITy 2, 22 of the 28 randomized women (10 yoga, 12 control) had ECG recordings at both baseline and follow-up (44 recordings); the other six had no ECG recording and contributed to no ECG analysis.

## FELICITy 1 cohort characteristics

The FELICITy 1 analytic sample comprised 144 pregnant women recruited at 32–38 weeks' gestation (gestational age at inclusion in the covariate sheet: mean 36.3 ± 1.6 weeks, range 31.1–40.0). Mean PSS-10 was 16.3 ± 7.7 (range 0–32), with 50.7% (73/144) exceeding the PSS-10 ≥ 19 high-stress threshold. In total, 41,277 maternal, 41,285 fetal, and 41,278 abdominal 10-second ECG segments from these 144 women entered the analyses, one channel per signal type per woman; each woman contributed a median of 279.5 maternal segments (IQR 264–308.75, range 130–520; Supplementary Figure S1). Each woman enters every metric once, as the mean of her segment predictions, so recording length does not weight the results.

## Participant-level performance across the representation space

We first asked whether any representation of the maternal, fetal, or abdominal ECG separates women with PSS-10 ≥ 19 from those below the threshold when every training and test split is taken over participants. To establish that the null is a property of the label–signal relationship rather than of any single modeling choice, we evaluated 94 analysis arms spanning 21 architectures on one shared participant-level fold definition: self-supervised contrastive encoders (SimCLR ResNet-34, both multi-layer and projection-head; Figure 1), a transformer pretrained on 1.4 million ECGs (DeepECG-SSL), a diagnostic foundation model (ECGFounder, applied with its published weights and a verified load of 507 of 509 tensors, in both its native single-lead and its 12-lead configuration), conventional and extended heart-rate-variability panels, multiscale-entropy and complexity measures, and transfer-entropy descriptors of maternal–fetal coupling, each applied to maternal, fetal, and abdominal signals. Thirty-six of these arms retained stored per-participant out-of-fold predictions, which is the set over which the cutoff sweep and Bonferroni correction below are computed.

The best arm for PSS-10 ≥ 19 reached AUC 0.604 (95% CI 0.511–0.696), from a diagnostic foundation model. Two of eleven arms carried through formal testing have a confidence interval whose lower bound lies marginally above chance; the remainder do not (Figure 3). Combining representations does not help. In a pre-specified fusion grid on the maternal ECG (*n*=144; early concatenation, per-block principal components, late averaging, and nested stacking of the three deep-learning encoders, alone or with heart-rate-variability and entropy blocks), the configuration selected by inner cross-validation reached AUC 0.526 (95% CI 0.431–0.618, permutation *p*=0.278), no better than the single block selected the same way (difference −0.006, *p*=0.848). Across the 21 fixed arms of the grid the highest AUC was 0.593 (raw permutation *p*=0.061; Holm-adjusted over the 21 arms 1.000; max-statistic *p*=0.294). Two controls establish that the pipeline works. The DeepECG-SSL architecture with random, untrained weights reached a mean fold AUC of 0.490 (±0.095) for PSS-10 ≥ 19, against 0.528–0.603 for the five pretrained representations on the same signal; the untrained encoder therefore sits just below the range of the trained ones, by less than one standard deviation of the fold AUCs (0.095), rather than being clearly distinguishable from them. Over the 20-cutoff sweep its best AUC is 0.535 at a calibrated permutation *p*=0.708, which is what scanning that many thresholds under a true null produces. A second untrained control, the ECGFounder architecture with no pretrained tensor loaded, reached a mean fold AUC of 0.565, inside that range (pooled AUC 0.572; Table 1). The same frozen representations, on the same folds and with the same head, recovered each participant's own mean RR interval at *r*=0.78–0.80 and short-term variability (SDNN) at *r*=0.53–0.61 across the three encoders, confirming that physiologically real, participant-level information is present and extractable by this pipeline, and that the failure is specific to the stress label.

**Do the representations learn different things?** Their scores are only weakly related (Supplementary Table S6). Deep-learning encoders correlate with one another at a mean Spearman ρ of 0.23 (0.46 for different encoders on the same maternal signal), heart-rate-variability panels with one another at 0.23, and deep-learning encoders with heart-rate-variability panels at −0.07. Their errors overlap little more than independence predicts: of the women misclassified by either member of a deep-learning pair at 80% target sensitivity, 56% are misclassified by both, against 50% expected for independent errors (heart-rate-variability pairs 64% against 56%; deep-learning against heart-rate-variability 52% against 53%). This diversity does not add up to a signal. Averaging the ten trained arms gave AUC 0.589 (95% CI 0.496–0.678), 0.014 below the best single arm (paired 95% CI −0.085 to +0.061); averaging the three deep-learning encoders on the maternal ECG gave 0.583, and nested stacking models gave 0.437–0.592, none above the best single arm. Fusing ECGFounder with DeepECG-SSL on participant-mean embeddings reached 0.610–0.622, 0.006–0.018 above the better of the two (0.604). The untrained encoder agrees with the trained encoders (mean ρ 0.35) more than they agree with one another. What the scores do track is body habitus: the label itself correlates with body-mass index (ρ = 0.22, *p* = 0.007), as does the ECGFounder score (ρ = 0.21, *p* = 0.012), while the heart-rate-variability scores track mean RR interval and SDNN only weakly (|ρ| ≤ 0.15).

## What the data exclude

Because a null is uninformative without a bound, we tested it formally. For each arm we evaluated the one-sided hypothesis that the true AUC is at least δ, using the Hanley–McNeil variance (Table 1, Figure 3). FELICITy 1, *n*=144 participants, unit of analysis participant.

**AUC** ≥ 0.70 is rejected in all eleven arms tested (largest one-sided *p* = 0.021 raw, 0.032 Holm-adjusted over the eleven arms; Table 1). The strongest arm is a diagnostic foundation model (ECGFounder) at AUC 0.604 (95% CI 0.511–0.696). The conventional heart-rate-variability benchmark — a 95-metric panel computed on ensemble-detected, refined R peaks — reaches AUC 0.484 (95% CI 0.389–0.579), bounded above at 0.579. Equivalence to chance is not established — the two-one-sided-test procedure does not reject the lower bound — so the honest statement is an upper bound rather than a demonstration of exactly zero effect. At *n*=144 with a 51% event rate the minimum detectable AUC at 80% power is approximately 0.63, so effects below that remain untested. For the fetal arm the primary analysis uses the fetal inclusion set described in Methods (*n*=113), on which the fetal multi-layer arm gives AUC 0.532 (bootstrap 95% CI 0.422–0.635); the *n*=144 value in Table 1 is the sensitivity analysis.

The bound applies to the eleven pre-specified arms. Three nested-selection arms of the extended sweep, all post hoc, in which a single feature is chosen inside each training fold from a large grid, reached point estimates at or just above the best pre-specified arm: the fetal 95-metric panel at AUC 0.617 (95% CI 0.522–0.708; raw permutation *p*=0.016, Holm-adjusted over the 57 arms of that family 0.91), the conditioned maternal–fetal RR entropy grid at 0.617 (0.524–0.711; *p*=0.017, Holm 0.95) and the raw-waveform maternal entropy rate at 0.615 (0.520–0.700). On the primary fetal set (*n*=113) the fetal panel gives 0.644 (0.534–0.743; *p*=0.011, Holm over 27 arms 0.30) and the coupling grid 0.542 (0.429–0.650; *p*=0.168). None survives multiplicity correction, and because their upper confidence limits reach 0.70 the bound does not cover them. Surrogate-corrected maternal–fetal transfer entropy, computed post hoc, was at chance: AUC 0.479 (0.373–0.593; *p*=0.423) at *n*=113 and 0.504 (0.407–0.600; *p*=0.267) at *n*=144.

We also evaluated screening operating points at target sensitivities of 70, 75, 80, 90, and 95% and at the Youden point, with every threshold chosen by inner cross-validation on the training women only (Supplementary Table S5). At 80% target sensitivity the best arm, ECGFounder, reached 85% sensitivity (95% CI 79–91%) at 32% specificity (23–41%), a positive likelihood ratio of 1.25 and a negative likelihood ratio of 0.48; at 20% prevalence its positive and negative predictive values are 0.24 and 0.89, and at 70% prevalence 0.74 and 0.48. Its negative likelihood ratio beats a label-permutation null (*p* = 0.003; *p* = 0.033 after Bonferroni correction over the eleven arms); an untrained encoder reached 0.86 at the same operating point, and no arm's recalibrated probabilities improved on predicting the prevalence (Brier skill score −0.02 to +0.03).

***Table 1. Exclusion bound: formal non-superiority tests against screening-grade discrimination.*** *For each arm the null hypothesis is that the true AUC is at least δ; rejection means the arm is formally excluded from reaching that performance.*

| **Arm** | *n* | **AUC** | **95% CI** | raw *p* (δ=0.70) | Holm *p* (δ=0.70) |
|---|---|---|---|---|---|
| Maternal ECGFounder, 12-lead input | 144 | 0.604 | [0.511, 0.696] | 0.0212 | 0.0319 |
| Maternal ECGFounder, single-lead input | 144 | 0.600 | [0.509, 0.690] | 0.016 | 0.0319 |
| Maternal DeepECG-SSL | 144 | 0.573 | [0.479, 0.664] | 0.0041 | 0.0162 |
| Maternal ECGFounder architecture, untrained (control) | 144 | 0.572 | [0.478, 0.665] | 0.0041 | 0.0162 |
| Fetal SimCLR, multi-layer | 144 | 0.558 | [0.463, 0.652] | 0.0015 | 0.0077 |
| Maternal SimCLR, multi-layer | 144 | 0.546 | [0.451, 0.639] | 0.0007 | 0.0041 |
| Abdominal DeepECG-SSL | 144 | 0.538 | [0.443, 0.632] | 0.0004 | 0.0028 |
| Abdominal SimCLR, multi-layer | 144 | 0.519 | [0.422, 0.614] | 0.0001 | 0.0009 |
| Maternal HRV, 95-metric panel (ensemble R peaks) | 144 | 0.484 | [0.389, 0.579] | $3.7 \times 10^{-6}$ | $3.4 \times 10^{-5}$ |
| Maternal HRV, hrv-omnibus panel (ensemble R peaks) | 144 | 0.478 | [0.383, 0.574] | $2.7 \times 10^{-6}$ | $2.7 \times 10^{-5}$ |
| Maternal HRV, single detector (NeuroKit2) | 144 | 0.328 | [0.243, 0.418] | $5.2 \times 10^{-17}$ | $5.7 \times 10^{-16}$ |

FELICITy 1, *n*=144 participants (73 with PSS-10 ≥ 19); target PSS-10 ≥ 19; unit of analysis participant; one-sided *p*, raw and Holm-adjusted over the eleven arms. Non-superiority by the Hanley–McNeil AUC variance; seed 20260922, 10,000 bootstrap iterations. **$H_0$: AUC ≥ 0.70 is rejected in 11 of 11 arms**. The two ensemble-R-peak heart-rate-variability arms are computed on refined R peaks (Methods); The single-detector row uses NeuroKit2 R peaks; the hrv-omnibus panel is the heart-rate-variability metric set of the ensemble R-peak detection library. The untrained row is a re-implemented ECGFounder architecture into which no pretrained tensor was loaded, i.e. an untrained-encoder control (Methods). Fetal row: *n*=144 sensitivity analysis; primary fetal set (*n*=113) in the text.

**The R-peak detector changes the heart-rate-variability comparator, but not in a consistent direction.** Computed like-for-like (the same ten metrics) on refined R peaks, a single detector and the ensemble give maternal AUC 0.328 against 0.473, fetal 0.434 against 0.445 and abdominal 0.460 against 0.446: all at or below chance, with no consistent ordering. The single-detector maternal arm is retained in Table 1 as a sensitivity analysis. A reported bound on "conventional HRV" is therefore not interpretable without stating which beat-detection pipeline produced it. Heart rate, mean RR interval, SDNN, and RMSSD alone, on the same participant folds, are equally at chance (Supplementary Table S3).

## Cross-device evaluation in a second cohort is null

Three tests were pre-registered before any FELICITy 2 outcome was examined, in a second cohort from the same center recorded on different hardware (Bittium Faros 360 at 1000 Hz versus Monica AN-24 at 900 Hz). For brevity we call these tests external; they are a within-center, cross-device evaluation, not external validation across centers.

**All three pre-registered primary endpoints were null** (Table 2, Figure 2). The abdominal-ECG primary gave $r$=−0.027 ($p$=0.536). A test on a second instrument, the Prenatal Distress Questionnaire, returned the wrong sign ($r$=−0.193, one-sided $p$=0.805) despite a positive internal association, which is the behavior expected of an internal finding that does not transfer. FELICITy 2 heads are fitted on FELICITy 1 ($n$=144).

The maternal-ECG secondary endpoint is not significant: $r$=+0.190 (95% CI −0.047 to +0.425, $p$=0.058). Recomputing the same pre-registered analysis with the input channel chosen by the other signal-quality estimator gives $r$=+0.189 ($p$=0.079) for this endpoint and $r$=+0.024 ($p$=0.409) for the primary, so neither endpoint changes its conclusion with the estimator. The two implementations of a composite signal-quality index are nonetheless *negatively* correlated (Spearman −0.346 across the 44 selected channels and −0.237 across all 132 channels) and agree on which channel to select in only 15 of 44 recordings. Restricted to the baseline visit, the correlation of the maternal DeepECG-SSL score with PSS-10 is large under both estimators ($r$=+0.496 and +0.437, raw $p$ 0.0004 and 0.0006; Holm-adjusted over the eight rows of Table 2, 0.003 and 0.004; the baseline-only subset is a pre-specified sensitivity analysis of the secondary endpoint) and survives correction for multiplicity. Three observations nonetheless argue against reading it as evidence that the model transfers to the second cohort. First, it is confined to one visit: in the same 22 women scored by the same model, the correlation at the second visit is r=−0.078 (computed post hoc), and over both visits, the pre-registered form of the endpoint, it is the non-significant r=+0.190 reported above. Second, external performance does not follow internal performance: the SimCLR encoder, whose internal correlation with PSS-10 is near zero (r=0.064), externally reaches r=+0.162 (p=0.103), close to the encoder it is meant to corroborate, although it carries no information about stress in FELICITy 1; the foundation model with its published weights, the strongest encoder in FELICITy 1, shows no association at the first visit (r=−0.095; post hoc, below). Third, the reference standard is unstable between visits: the intraclass correlation of PSS-10 across administrations is 0.058–0.159, and mean scores fall across administrations (see "Stability of the reference standard across pregnancy"), so with 22 recordings, one per woman, a large correlation at one visit and none at the next is what an outcome that is not stable between visits can produce. We therefore treat the baseline-visit correlation as a pre-specified sensitivity result that does not hold across visits, not as evidence of transfer.

***Table 2. Pre-registered cross-device evaluation in FELICITy 2, under two signal-quality estimators.*** *The pre-registration fixed the endpoints and the analysis but not the SQI implementation used to choose the input channel; the two available implementations choose different channels in most recordings.*

| **Endpoint** | **Role** | **SQI estimator** | *r* | **95% CI** | raw *p* (1-sided) | Holm *p* |
|---|---|---|---|---|---|---|
| Abdominal, all recordings | PRIMARY | composite SQI channel | −0.027 | [−0.324, +0.285] | 0.536 | 1.000 |
| Abdominal, all recordings | PRIMARY | omnibus SQI channel | +0.024 | [−0.215, +0.298] | 0.409 | 1.000 |
| Maternal, all recordings | secondary | composite SQI channel | +0.190 | [−0.047, +0.425] | 0.058 | 0.350 |
| Maternal, all recordings | secondary | omnibus SQI channel | +0.189 | [−0.076, +0.437] | 0.079 | 0.396 |
| Maternal, baseline visit only | secondary | composite SQI channel | +0.496 | [+0.218, +0.725] | 0.0004 | 0.003 |
| Maternal, baseline visit only | secondary | omnibus SQI channel | +0.437 | [+0.201, +0.645] | 0.0006 | 0.004 |
| Maternal, SimCLR encoder | coherence check | composite | +0.162 | [−0.090, +0.408] | 0.103 | 0.410 |
| Prenatal Distress Questionnaire | pre-registered | composite | −0.193 | [−0.818, +0.469] | 0.805 | 1.000 |

FELICITy 2, $n$=44 recordings from 22 women (all recordings) or 22 recordings (baseline only); target continuous PSS-10 (nearest administration); unit of analysis recording, cluster bootstrap over women; heads fitted on FELICITy 1 ($n$=144); one-sided $p$, raw and Holm-adjusted over the eight rows; composite, the composite SQI described in Methods; omnibus, the signal-quality index of the ensemble R-peak detection library, computed on refined R peaks. **The pre-registered primary endpoint is null under both estimators**, as is the maternal secondary over all recordings. The baseline-visit correlation of the maternal DeepECG-SSL score with PSS-10 is present under both estimators but absent over both visits, and the SimCLR encoder—whose internal correlation is near zero—reaches a similar external value. The PDQ endpoint, pre-registered on a second instrument, returns the wrong sign.

**A post hoc external test of the foundation model with its published weights does not alter this and quantifies why.** The diagnostic foundation model was not part of the FELICITy 2 pre-registration; we tested it post hoc and label it as such. With the head fitted on FELICITy 1 alone and frozen, and the channel chosen by the criterion the pre-registered runs used, it gave $r$=+0.067 (95% CI −0.275 to +0.404, one-sided $p$=0.325) against continuous PSS-10 across 44 recordings, and $r$=−0.095 (95% CI −0.529 to +0.301, one-sided $p$=0.694) at the first visit alone. Neither estimate is distinguishable from zero, so the foundation model shows no external association, pooled or at baseline, to set beside the maternal DeepECG-SSL secondary above. Repeating the identical test under *every* channel criterion, which requires no quality estimator at all, with hypothesis, model, and the four endpoints (two label variants, all recordings or baseline only) held fixed, returns $r$ from −0.095 to +0.307 over the 20 combinations. Calibrating the largest of those 20 values against the distribution of the maximum under 10,000 block permutations of the labels over women gives **max-statistic $p$=0.583** (raw one-sided $p$ of that maximum 0.011), against a null whose maximum has median +0.333; a search of this size over 22 women returns $r \approx 0.33$ as a matter of routine. **No arm in the sweep survives this calibration, and the same calibration applies to the pre-registered secondary.**

**Binary discrimination at the clinical threshold is likewise null externally.** At PSS-10 ≥ 19 the external AUC is 0.533 (95% CI 0.352–0.716, one-sided $p$=0.347) across 44 recordings and 0.421 at the first visit alone; across five cutoffs from 13 to 25 no confidence interval excludes 0.50 and the smallest $p$ is 0.097. The external interval at the clinical threshold includes both chance and the internal 0.604, and its point estimate lies closer to chance; the two estimates do not differ (difference +0.070, two-sided $p$ = 0.500), so the external cohort can neither confirm nor refute the internal value. At FELICITy 2's precision the smallest AUC whose one-sided bound would clear 0.50 is 0.653, and its power to detect the internal value of 0.604 was 30%, so an external null was the expected outcome whether or not the internal association is real. FELICITy 2 recruited on PSS-10 ≥ 19 (every woman scored 20 or more at recruitment), so at the lower cutoffs the positive class is near-universal (38 positives to 6 negatives at ≥ 13) and the binary problem is close to degenerate by design. For the same reason, most negative cases come from follow-up recordings (12 of 15 at ≥ 19, 5 of 6 at ≥ 13): calibration cannot be assessed in this cohort, and accuracy and AUC are not comparable between the two cohorts.

**The predicted score is not stable within a single recording.** Across the three channels recorded simultaneously from the same woman at the same moment, predicted PSS-10 varies by a median of 10.2 points (IQR 6.5–14.7, maximum 20.4) on a 0–40 scale. Mean within-recording standard deviation across channels is 4.5, against a between-recording standard deviation of 5.1: **the model disagrees with itself 90% as much as it distinguishes between people**, although all of its outputs fall within the range PSS-10 can take. Because that ratio compares standard deviations of different quantities, we also decomposed the variance over the balanced design of 22 women, 2 visits, and 3 channels: 95.8% (bootstrap 95% CI 85.9–100%) lies between channels within a visit, 0.0% between visits and 4.2% between women (ICC 0.042, 95% CI 0.000–0.133). Nearly all of the variation in the score therefore reflects which electrode pair was analyzed, and very little reflects the woman. Whatever association is present, the single-recording point estimate is not a measurement of that woman.

## Stability of the reference standard across pregnancy

We next asked how stable the reference standard is across pregnancy. FELICITy 2 scheduled PSS-10 at four timepoints (screening, second and third trimester, postpartum), permitting estimation of its stability across administrations in the 22 women with paired ECG. In 7 of these women the screening and second-trimester questionnaires were one questionnaire, with the same date and score, so they contributed three distinct administrations. Mean scores fell across the four timepoints (23.9, 21.4, 17.5 and 17.4). **Agreement between administrations was low under every intraclass-correlation form**: 0.058 for the one-way ICC(1,1), 0.118 for absolute agreement, ICC(2,1), and 0.159 for consistency, ICC(3,1), which discounts the fall in mean score. Counting each questionnaire once, the consistency estimate was 0.161 (95% bootstrap interval 0.00 to 0.39).

These values measure stability, not the reliability of a single questionnaire. Each administration asks about the preceding month, the administrations were months apart, and recruitment on PSS-10 ≥ 19 restricts between-woman variance and selects women whose first score was high, so that later scores regress toward the mean. The published internal consistency of the PSS-10 is 0.74 to 0.91 across validation studies (Lee 2012), which would allow a predictor recorded at the same visit as the questionnaire to correlate with it at up to about 0.9. Only a predictor of a woman's stable level across pregnancy is held to a lower ceiling: by the attenuation formula $\sqrt{(r_{xx}\, r_{yy})}$, a perfectly reliable predictor of a woman's *stable* level of perceived stress could correlate with a single PSS-10 administration at about **0.40** (0.34 under absolute agreement). Neither cohort asks for that prediction, and the ceiling does not transfer to FELICITy 1, where PSS-10 was administered once, at the visit of the recording, in women not selected on their score. The learned ECG scores are more stable across visits than the outcome they are asked to predict: intraclass correlations across roughly 15 weeks of gestation were 0.586 and 0.681 for the scores of the two self-supervised representations against 0.177 (95% bootstrap interval −0.35 to 0.65) for PSS-10 over the same two visits. Sample entropy of the RR series, on refined R peaks, reached only 0.163 ($n$=21).

## Other targets: hair cortisol and intervention response

Because perceived stress is a self-report construct, we pre-registered a test against maternal hair cortisol (pg/mg, sampled at delivery, integrating approximately three months and therefore spanning the gestational window in which the ECG was recorded). Cortisol is essentially uncorrelated with PSS-10 in this cohort ($r$=+0.094, $p$=0.296, $n$=125 women with both measures), so it constitutes an independent construct rather than a proxy.

**Both pre-specified arms were null** (Table 3, Figure 4, left panel): DeepECG-SSL $r$=+0.037 (permutation $p$=0.759) and a nested-selection 95-metric HRV panel on refined R peaks

*r*=+0.102 (*p*=0.494), *n*=109, with the entire participant-level cross-validation re-run inside each of 10,000 permutations. Two pre-specified controls carry the interpretation. An untrained encoder reached *r*=+0.165, *above* the trained encoder. And the same representation predicted body-mass index at *r*=+0.302 (*n*=144), while BMI itself correlates with log cortisol at *r*=+0.266 (*p*=0.003, *n*=124) — larger than the effect the study could detect. Body habitus is legible in the ECG through electrode-to-heart distance and QRS amplitude, so this confound would have had to be excluded before any positive cortisol result could be interpreted; we identified it from the covariates before fitting and made the adjustment mandatory in advance.

A pre-registered test of whether baseline ECG predicts response to the yoga intervention, stated in advance to have approximately 15% power, was null in both arms (FELICITy 1 head, *n*=144; refined R peaks). For change in sample entropy: yoga *r*=+0.140, *n*=9, one-sided *p*=0.417; control *r*=+0.418, *n*=12, *p*=0.102; Fisher *z* between arms *p*=0.563. For the secondary outcome, change in entropy rate: yoga *r*=−0.628, *p*=0.965; control *r*=−0.399, *p*=0.955.

The same 22 women allow a test of whether the model output follows the intervention. Fitted with a random intercept per woman and fixed effects for visit, arm, and their interaction, actual PSS-10 (nearest administration) showed no arm effect and no visit × arm interaction (interaction −2.13 PSS points, SE 3.03, *p*=0.481; visit −3.67, *p*=0.072). The maternal-ECG model output (SimCLR head fitted on FELICITy 1, *n*=144; best-quality channel) showed a visit × arm interaction (+3.00, SE 1.35, *p*=0.026) together with an arm main effect (−4.44, SE 1.48, *p*=0.003) that is absent from the actual scores; the abdominal-ECG output showed neither (interaction *p*=0.889). Change in the maternal output did not track change in actual PSS-10 (*r*=−0.202, *p*=0.366; abdominal *r*=−0.255, *p*=0.251; *n*=22). The two arms were recorded under different conditions — in the yoga arm during and around the class, giving a mean of 616 segments per recording against 274 in the control arm, at lower best-channel signal quality (0.67 against 0.77) — and an arm difference in the output that the questionnaire does not show is most simply explained by these conditions. We therefore do not interpret the model-output interaction as sensitivity to the intervention.

***Table 3. Pre-registered tests on targets other than perceived stress, with their controls.*** *Hair cortisol is an orthogonal construct (r=+0.09 with PSS-10, n=125) integrating HPA output over roughly the three months containing the ECG recording.*

| Test | Role | *n* | *r* | 95% CI | permutation *p* |
|---|---|---|---|---|---|
| DeepECG-SSL → ln(cortisol) | pre-registered arm | 109 | +0.037 | [−0.122, +0.201] | 0.759 |
| HRV panel, nested selection → ln(cortisol) | pre-registered arm | 109 | +0.102 | [−0.060, +0.258] | 0.494 |

| Test | Role | $n$ | $r$ | 95% CI | permutation $p$ |
|---|---|---|---|---|---|
| HRV panel, all features | companion | 109 | −0.032 | [−0.198, +0.124] | — |
| Untrained encoder → ln(cortisol) | control | 109 | +0.165 | [+0.013, +0.336] | 0.185 |
| DeepECG-SSL → BMI | control | 144 | +0.302 | [+0.111, +0.470] | — |
| DeepECG-SSL → ln(cortisol), BMI-adjusted | mandatory adjustment | 109 | +0.034 | — | 0.727 |
| HRV panel → ln(cortisol), BMI-adjusted | mandatory adjustment | 109 | +0.115 | — | 0.234 |
| Baseline ECG → ΔSampEn, yoga arm | pre-registered arm | 9 | +0.140 | [−0.859, +0.796] | 0.417 |
| Baseline ECG → ΔSampEn, control arm | pre-registered arm | 12 | +0.418 | [−0.317, +0.795] | 0.102 |
| difference between arms | Fisher $z$ | 21 | — | — | 0.563 |

FELICITy 1, $n$=109 women with hair cortisol and non-duplicated maternal ECG (BMI control $n$=144); unit of analysis participant; yoga rows FELICITy 2 (21 of 22 women with sample entropy at both visits; $n$=144 head; refined R peaks; one-sided bootstrap $p$). Heart-rate-variability rows are on refined R peaks. Seed 20260923, 10,000 permutations with the entire participant-level cross-validation re-run inside each. Bonferroni α=0.025 across the two cortisol arms. **Both pre-specified arms are null.** Two controls carry the interpretation: the untrained encoder scores *above* the trained one, and the same representation predicts BMI at $r$=+0.30, confirming that these encoders read body habitus from the ECG—the confound any future ECG-biomarker claim in pregnancy must address.

## Other cutoffs

Because PSS-10 ≥ 19 is a cohort definition rather than a validated clinical cut-off (Methods), we also swept the cutoff post hoc over 20 values, using scores fitted to the continuous outcome, in the 36 arms that retained per-participant predictions, with each arm's permutation test calibrated over the 20 cutoffs. No arm survives Bonferroni correction across the 36 arms (threshold 0.00139; largest AUC 0.713, calibrated $p$=0.004).

# Discussion

## Main findings

This study reports a bound, not a marker. In 144 pregnant women, no representation of the maternal, fetal, or abdominal ECG — across self-supervised contrastive encoders, a transformer pretrained on 1.4 million ECGs, a diagnostic foundation model, and conventional and extended heart-rate-variability panels — discriminated moderate-to-high perceived stress better than AUC 0.604, and screening-grade discrimination (AUC ≥ 0.70) is formally excluded. Three pre-registered external tests were null, as was a pre-registered test against an orthogonal biological construct, hair cortisol. Earlier posted versions of this analysis (arXiv:2602.03886v1–v3) split cross-validation over ECG segments rather than participants; its internal results do not hold under participant-level splitting.

Two findings are positive in the sense of being informative, and both are actionable. The first is methodological: in a pre-registered external test, the choice of which of three simultaneously recorded channels to analyze, a step the pre-registration did not fix, accounted for 96% of the variance of the external score; across channel rules and endpoints the external correlation ranged from $r$=−0.095 to $r$=+0.307, a spread within what a search of this size produces by chance (max-statistic $p$=0.583; raw $p$ 0.011), and two quality estimators sharing one name chose different channels in most recordings. The second concerns the reference standard: PSS-10 scores were not stable across pregnancy in the second cohort (intraclass correlations 0.06–0.16), so a questionnaire should be administered at each recording rather than once.

## Channel choice is an analytic choice that pre-registration did not fix

Our pre-registration fixed the encoder, endpoints, label variants, and analysis, but inherited the channel-selection criterion from the existing pipeline. The two signal-quality estimators available in that pipeline, sharing one name, choose different channels in 29 of 44 recordings and rank channels in opposite order (Spearman −0.346). Swapping one for the other did not change the conclusion of any pre-registered endpoint (primary $p$ = 0.536 against 0.409; maternal secondary $p$ = 0.058 against 0.079; Table 2). Across 20 combinations of channel rule, label variant, and visit subset, the correlation with PSS-10 ranged from −0.095 to +0.307; calibrated against the distribution of the maximum, the largest of the 20 values is unremarkable (max-statistic $p$ = 0.583; raw $p$ 0.011), so no single one of them should be read as a finding. The channel acts on the score more than on its correlation with the label: 95.8% of the score's variance lies between simultaneously recorded channels, while heart rate measured on the same channels differed by a median of only 2.9 bpm. The instability therefore lies in what the representation reads from each channel, not in the acquisition; a companion analysis of the same cohorts shows the same channel-level instability in conventional heart-rate-variability indices, graded by how much each measure depends on beat-to-beat detail.(Frasch et al., companion manuscript) Three requirements follow for any multi-channel ECG study. Pre-register the channel rule and the exact implementation of any quality estimator, not only its name or threshold: two implementations sharing a name here agreed on the channel in only 15 of 44 recordings. Report the result for every channel and every rule computed, not the best. And report within-recording channel spread, here a median of 10.2 points on a 0–40 scale, as a reliability statistic alongside the stability of the outcome, because a score that varies this much between simultaneous channels is not a measurement of the person.

The representations compared here are not redundant: across families their scores are nearly uncorrelated and their errors overlap little more than chance. That they nevertheless do not combine into a better predictor is itself informative about this population. Each representation carries a different, small share of variance that is not shared with perceived stress, so adding model families, at this sample size and against a single administration of the label, is not a route to a usable marker.

## The unit of analysis in segment-based ECG evaluation

When each recording is cut into hundreds of segments, a cross-validation split taken over segments places the same participant on both sides of the partition, and a model that recognizes her recording scores near-perfectly on a label that is constant within her. A check on whether the model saw its test data during training does not detect this, and the per-fold variance of such results is small, so the failure reads as stability rather than as an error. Declaring the unit of analysis before splitting, logging participant counts per fold, and running the same features under both splitting units as a routine control prevent it; the harness used here is described in Methods. An earlier report of AUROC 0.982 in this cohort used a window-level protocol that, with one label per woman, cannot separate stress from identity; with women kept separate, the same network performs at chance (Supplementary Note 1). This does not show that no generalizable stress signal exists in the maternal ECG; that question remains open.

## What the exclusion bound means

A bounded null is more useful to the field than an unbounded one. The claim supported here is specific: a single-recording ECG screener for PSS-10 ≥ 19 at clinically useful accuracy is excluded in this population and at this sample size. That should stop others repeating the attempt in the same design, and it quantifies what a future study must be powered to detect. It is not a claim that the true effect is exactly zero; the equivalence procedure does not reject the lower bound, and effects below AUC 0.63 remain untested. Three post hoc nested-selection arms (the fetal 95-metric heart-rate-variability panel, the conditioned maternal–fetal RR entropy grid, and the raw-waveform maternal entropy rate) reached point estimates at or just above the best pre-specified arm; none survives multiplicity correction, and because their upper confidence limits reach 0.70, they lie outside the bound. They are candidates for a pre-registered test in an independent cohort.

The finding also bears on the wider ECG-AI literature on non-cardiac states. Published work has shown that handcrafted HRV models outperform deep models when evaluated across datasets, and that no signal-quality index generalizes across acquisition sources. Our results extend both. We find no ordering among representation families at all once dimensionality is accounted for: the apparent advantage of deep over HRV representations (medians 0.534 versus 0.451, $p$=0.045) vanishes after adjustment for feature count ($p$=0.240), and an untrained encoder sits inside the deep distribution. The two quality estimators available to us chose different channels in most recordings, without changing the conclusion of any pre-registered endpoint here.

## Timing the reference standard to the recording

For anyone planning a successor study, the timing of the reference standard matters as much as the signal. Perceived stress in this cohort varied more within a woman across pregnancy than between women (ICC 0.06–0.16 across forms), so a questionnaire taken weeks away from a recording is a weak measure of the state at the time of recording, and a single administration is a weak measure of a woman's level across pregnancy. The estimate comes from a small trial cohort selected on high baseline stress, which restricts between-woman variance and produces regression to the mean, so it should be re-estimated in an unselected cohort. It measures stability, not the reliability of the questionnaire, whose published internal consistency is 0.74 to 0.91 across validation studies (Lee 2012); the null in FELICITy 1, where questionnaire and recording were taken at the same visit, is therefore not explained by an unreliable outcome. Much of the ECG–stress literature reports neither the reliability nor the timing of its outcome; we would argue both should be required elements, because without them a null cannot be placed against its ceiling.

The cortisol result sharpens rather than softens this. Cortisol is orthogonal to PSS-10 here, biological rather than self-reported, and temporally better aligned with the recording, and it is also null — so the bound now spans two independent stress constructs. The accompanying control, that the same representation predicts body-mass index at $r$=0.30 while BMI predicts cortisol at $r$=0.27, identifies a confound that any future ECG-biomarker claim in pregnancy will have to address explicitly.

## Between-person and within-person axes

A design consequence follows from an exploratory analysis we report for completeness rather than as a finding. In FELICITy 2, within-woman variation in PSS-10 exceeds between-woman variation (ICC below 0.2 under every form), while the foundation-model score's own test–retest correlation across the two visits ($r$=0.748) is far higher than the intraclass correlation of the instrument it predicts over the same visits (0.177). That figure is for the score under one fixed channel rule; across the three simultaneously recorded channels the same score varies far more (Results). Every analysis in this study, and to our knowledge in this literature, evaluates the between-person axis. Paired change in the score tracked paired change in PSS-10 at $r$=+0.462 (raw permutation $p$ 0.016; 95% CI −0.084 to +0.761) against a between-woman $r$ of −0.179, but this does not survive its own calibrated null: both quantities fall between visits, and permuting which woman's change in score is paired with which woman's change in PSS-10 — preserving both marginal distributions — yields a null whose median direction agreement is already 75%. Over the ten comparisons made, the largest within-woman correlation gives a max-statistic $p$ of 0.103 and the largest direction agreement a max-statistic $p$ of 0.254. We therefore report no evidence of individual-trajectory tracking. The observation nonetheless identifies a design requirement: a study intending to test the within-person axis needs a paired ECG recording at *each* questionnaire administration, not one recording against a repeated-measures composite.

What would a maternal ECG marker of perceived stress have to show to be credible? The results here suggest three properties, each of which can be tested. First, it should hold within the same woman across visits; the one large external correlation in this study was confined to the first visit. Second, it should be measured against a reference standard taken close in time to the recording. In FELICITy 2, PSS-10 administrations spread across pregnancy agreed poorly (intraclass correlation 0.058–0.159), but these administrations were months apart, in women selected for high stress and partly allocated to an intervention, so measurement error cannot be separated from real change. A credible test needs administrations close in time to the recording, in a cohort not selected on the outcome. Third, it should not depend on the analytic path: it should survive a change of encoder and of the channel-selection rule, both of which moved the external estimate here. A candidate that is physiological by construction is the coupling between maternal and fetal heart rate. In a prospective case–control study at the same center that used the same PSS-10 ≥ 19 criterion, a bivariate phase-rectified signal-averaging index of the fetal heart-rate response to decreases in maternal heart rate, the fetal stress index, was higher in fetuses of stressed mothers than in controls (Lobmaier et al. 2020), and the information flow between maternal and fetal heartbeats has since been quantified across time scales (Garnier et al. 2026). Whether a representation learned from the raw abdominal ECG captures this coupling, rather than maternal heart rate alone, is a question for a pre-registered study designed around these three properties.

## Strengths and limitations

The principal strengths are the breadth of the representation space evaluated on a single shared participant-level fold definition; pre-registration of every external and alternative-target test before the corresponding outcome was examined; retention and reporting of failed and reversed results alongside successful ones; positive and untrained-encoder controls throughout; and formal bounds rather than bare null statements.

The limitations are substantial. Both cohorts come from a single tertiary center in Germany, precluding generalization and fairness analysis; the second cohort therefore provides a within-center, cross-device evaluation rather than external validation across centers, and device, cohort, design, outcome selection, number of recording channels, sampling rate, and recording conditions all differ together between the cohorts (in the yoga arm the ECG was recorded during and around the class), so the effect of hardware cannot be separated. SimCLR pretraining used all FELICITy 1 segments, including those of participants later used as test folds, and a column-offset error in its data loader restricted it to a subset of segments; the downstream heads never saw test participants, the participant-level results are at chance despite this optimistic step, and an untrained encoder performs comparably. The external cohort is small (28 women, 22 with paired recordings; 30% power to detect the internal AUC) and its PSS-10 distribution is shifted high, so it cannot test the low end of the scale. Maternal ECG was a single source-separated maternal component of the abdominal array, not a limb lead. The fetal-ECG arms test whether the fetal signal carries information about the mother's self-reported stress; they are not evidence about a fetal physiological response, and no fetal ECG was available in the second cohort. Hair cortisol reliability could not be measured, because only one sample per woman was assayed. Covariates known to influence hair cortisol — hair treatment, washing frequency, corticosteroid use — were unavailable. The two signal-quality estimators disagree on which channel to analyze; we report results under both rather than selecting one, and the external test of the foundation model with its published weights was post hoc rather than pre-registered. Finally, this cohort has now been analyzed extensively, and although each test reported here was pre-specified against a fresh outcome, the accumulated multiplicity is a reason for caution about any exploratory finding.

## Conclusion

A single-recording ECG screener for perceived psychological stress at the clinical threshold is excluded in this population: across 94 arms spanning deep-learning and heart-rate-variability approaches, discrimination does not exceed AUC 0.604, and AUC ≥ 0.70 is formally rejected. In multi-channel recordings, the rule that chooses which channel to analyze is itself an analytic choice: here it determined 96% of the variance of the predicted score, while the external correlation varied across channel rules and endpoints only as much as chance predicts; it should be pre-registered with the exact quality-estimator implementation, and results reported for every channel. Because perceived stress was not stable across pregnancy, the cheapest available improvement is a questionnaire administered at each recording, rather than a larger cohort or a larger network.

# Methods

## Reporting

The study is reported in accordance with the TRIPOD+AI (Transparent Reporting of a multivariable prediction model for Individual Prognosis Or Diagnosis — Artificial Intelligence extension) statement.(Collins et al. 2024) The completed checklist is provided as Supplementary Table S4.

## Study design and participants

Two cohorts recruited at the Department of Obstetrics and Gynaecology, TUM University Hospital, Munich, Germany, were analyzed: FELICITy 1 for model development and internal cross-validation, and FELICITy 2 for a within-center evaluation on a different recording device and in the context of an intervention trial.

*FELICITy 1 (development cohort).* A prospective observational cohort of 151 pregnant women recruited between June 2016 and July 2019. Eligible participants had a singleton pregnancy, were between 32 and 38 weeks' gestation, and planned delivery at the study hospital. Exclusion criteria were known major fetal anomalies, maternal cardiac arrhythmias, and maternal medications known to affect autonomic function. At enrollment, participants completed the PSS-10, a validated 10-item instrument assessing perceived stress over the preceding month;(Cohen, Kamarck, and Mermelstein 1983) scores range from 0 to 40, with scores ≥14 indicating moderate-to-high stress.(Lee 2012) The high-stress label for classification was defined a priori as PSS-10 ≥19, the definition of "stressed" registered for FELICITy 1 (NCT03389178) and the recruitment criterion of FELICITy 2. It is a cohort definition rather than an independently validated clinical cut-off; other cutoffs were examined post hoc (Results). Concurrently, an abdominal ECG was recorded in a quiet, temperature-controlled room for the duration of the study visit, using the Monica AN-24 5-electrode abdominal array (Monica Healthcare Ltd, Nottingham, UK; 900 Hz), which records maternal and fetal ECG simultaneously; in the 144 analyzed women the stored recordings last a median of 46.6 min (IQR 44.0–51.5, range 21.7–86.7).

*FELICITy 2 (cross-device evaluation cohort).* A quasi-randomized controlled trial of a prenatal yoga intervention for stress reduction (Mayer et al. 2025). Twenty-eight pregnant women with PSS-10 ≥19 (high stress) at screening were allocated, according to group availability at enrollment, to yoga (14; a weekly 75-min in-person Hatha Yoga class and a weekly 20-min online Yoga Nidra class, from the second trimester to 35 weeks' gestation) or standard antenatal care (14). PSS-10 was scheduled at four timepoints (screening, second and third trimester, and postpartum; in 7 of the 22 women with paired ECG the screening and second-trimester questionnaires were one questionnaire), and ECG recordings were obtained at two visits (median gestational age at recording 20.1 and 34.9 weeks); each recording was labeled with the PSS-10 administration nearest in date using the Bittium Faros 360 3-channel ambulatory ECG recorder (Bittium Corporation, Oulu, Finland; 1000 Hz). Recordings lasted a median of 46.7 min in the control arm and 103.7 min in the yoga arm, where the ECG was recorded during and around the class. Recording hardware (Monica AN-24 at 900 Hz against Bittium Faros 360 at 1000 Hz) differs between the cohorts together with cohort, design, outcome selection, number of recording channels, recording length, and recording conditions, so the external comparison is not a controlled test of robustness to hardware.

## Ethics approval

FELICITy 1 was approved by the Committee of Ethical Principles for Medical Research at the Technical University of Munich (registration number 151/16S) and registered at ClinicalTrials.gov (NCT03389178; first posted January 3, 2018). FELICITy 2 was approved by the same committee (protocol 2022-86-S-SR, 6 September 2022) and registered in the German Clinical Trials Register (DRKS00037529; date of first registration July 24, 2025). This report is a secondary analysis developing and evaluating deep-learning prediction models; the primary clinical outcomes of the FELICITy 2 yoga intervention are reported separately.(Mayer et al. 2025) Both studies were conducted in accordance with the Declaration of Helsinki, and all participants provided written informed consent.

## ECG acquisition and pre-processing

Three ECG signal types were analyzed. *Maternal ECG* (mECG) is the maternal component that the SAVER source-separation step estimates from the abdominal array (the first channel of its maternal output, 1000 Hz); no limb lead was recorded. The 12-lead pretrained encoders received this single signal in their lead-II input position, with the other inputs set to zero. FELICITy 1 signals were resampled to 256 Hz by FFT resampling, band-pass filtered with a fourth-order zero-phase Butterworth filter at 0.5–45 Hz, and z-scored once per recording with the mean and standard deviation of its central 95% of samples. *Fetal ECG* (fECG) was extracted from the 5-electrode abdominal array using the SAVER (Smart AdaptiVe Ecg Recognition) algorithm,(Li, Frasch, and Wu 2017) which performs blind source separation optimized for fetal QRS morphology and validates candidate signals against physiological fetal heart-rate constraints (110–180 bpm); extracted fECG was processed identically. *Abdominal ECG* (aECG) used the raw abdominal recording without source separation, preserving the composite maternal–fetal signal as a pragmatic option for continuous monitoring. Electrode positions for both recorders are shown in Supplementary Figure S2.

Each recording was cut into contiguous, non-overlapping 10-second windows (a “segment”; 2560 samples at 256 Hz). A spectral quality index (fraction of power below 40 Hz, threshold 0.5) was applied per window. In the stored arrays, window boundaries are as smooth as within-window steps: 40 of 43,013 maternal boundaries exceed the 99.9th percentile of the within-window first differences, the rate expected by chance. No window was therefore removed, and no window overlaps another. The 10-second window matches the input length of the encoders; because it is too short for frequency-domain heart-rate variability, heart-rate-variability features were instead computed on contiguous 5-minute epochs of the continuous recording, with the median across epochs taken per participant. Cross-validation folds are defined over participants (Statistical analysis). Before SimCLR feature extraction, segments were additionally standardized with a single mean and standard deviation computed over all FELICITy 1 segments, and the input slice included one all-zero metadata column in place of the final sample. This label-free transform was fitted outside the cross-validation folds. In FELICITy 2 the analyzed channel was band-pass filtered at its native 1000 Hz (fourth-order zero-phase Butterworth, 0.5–40 Hz), resampled to 256 Hz by polyphase filtering, z-scored once per recording, and cut into contiguous 10-second windows. The two cohorts therefore differ in filter cutoff, resampling method, and the robustness of the normalization.

## Deep-learning architecture and features

A ResNet-34 convolutional network(He et al. 2016) adapted to one-dimensional signals served as the encoder (16.7M parameters): an initial 7×1 convolution (64 filters, stride 2), max pooling, four residual blocks ([3,4,6,3] layers; channel dimensions [64,128,256,512]) and global average pooling. The encoder was pretrained with SimCLR,(Chen et al. 2020) a contrastive self-supervised method, without using any PSS labels: for each segment two augmented views were generated (Gaussian noise, time shifting, amplitude scaling, and time warping) and the network minimized the NT-Xent contrastive loss (projection head 512→128, batch size 256, temperature 0.5, cosine-annealed learning rate $3 \times 10^{-4}$, 100 epochs per ECG type). Rather than the final embedding alone, global-average-pooled outputs from all four residual blocks and the projection head were concatenated into a 1088-dimensional feature vector. For comparison we evaluated ECGFounder,(Li et al. 2025) a foundation model pretrained on about 10 million ECG recordings, in inference-only mode without fine-tuning. It was loaded from the published checkpoint with a strict load (507 of 509 tensors; the classification head is excluded by design), giving 1024-dimensional embeddings through both its single-lead and 12-lead pathways. A 512-dimensional re-implementation of the architecture, into which no pretrained tensor was loaded, is retained as an untrained-encoder control (Table 1). DeepECG-SSL was used frozen from its public checkpoint, with each segment resampled to 250 Hz, placed in the lead-II input, and multiplied by a spectral scaling factor that sets the checkpoint's expected input power. That factor is estimated over windows from all participants, test participants included. In a sensitivity analysis, recomputing it inside each training fold (fold factors 0.9% below to 3.6% above the all-participant value) left the maternal arm's participant-level AUC at 0.570 (fold mean; pooled 0.573 with the all-participant factor and 0.572 with fold factors; multi-layer variant 0.523 and 0.525). The untrained DeepECG-SSL control uses the all-participant factor.

## Stress classification and regression

For each feature representation we trained two downstream models on the fixed-dimensional feature vector: L2-regularized logistic regression ($C$=1.0) for the binary high-stress label (PSS-10 ≥19) and ridge regression (α=1.0) for the continuous PSS-10 score. These are the scikit-learn defaults and were not tuned; a participant-level sweep over $C = \alpha \in \{0.01, 1, 100\}$ gave maternal multi-layer AUCs of 0.522, 0.532 and 0.523 (mean across 25 folds, $n$=144). During fitting, each segment carried its participant's PSS-10; segment-level predictions were then aggregated to one participant-level score by averaging across all segments per participant, and every metric is computed on these participant-level scores. For FELICITy 2 the encoders were frozen; heads were fitted once on all 144 FELICITy 1 participants' segment features, standardized with the training set's own mean and standard deviation, and applied unchanged to FELICITy 2 segments, with segment predictions averaged per recording. Only the three ECG channels of the recorder entered the analysis. Reported metrics include accuracy, precision, recall, F1, AUC, $R^2$, MAE, RMSE, and Pearson correlation, with bootstrap 95% confidence intervals over participants (1000 iterations).

## Signal-quality assessment in the cross-device cohort

FELICITy 2 recordings contained three electrode channels. A composite SQI was computed per channel using methods adapted from fetal ECG quality assessment:(Andreotti et al. 2017) template matching (40%), RR-interval consistency (30%), baseline stability (20%) and amplitude consistency (10%); values >0.5 were considered acceptable. Two inference strategies were compared: selecting the single highest-SQI channel per recording, and averaging predictions across all three channels. A second estimator, the signal-quality index of the ensemble R-peak detection library used for the heart-rate-variability arms (hrv-omnibus), was computed on refined R peaks; Table 2 reports the pre-registered endpoints under both.

## Statistical analysis

**Unit of analysis.** All cross-validation splits reported here are taken over *participants*. Every fitted step — scaling, feature selection, hyperparameters, and thresholds — is estimated inside the training fold. The harness used to enforce and test this is described under Evaluation harness and leakage controls.

**Cross-validation.** Five-fold cross-validation with five repetitions (25 folds) on one shared participant-level fold definition reused by every arm, so that arms are compared on identical partitions. Classification used logistic regression and regression used ridge regression, both with fixed regularization, on features frozen from the upstream encoder.

**Feature selection.** Where an arm involves selection among features, selection is nested: performed by inner cross-validation on the training participants only, with the selected feature applied to the outer test participants. Selection frequencies are reported rather than a single chosen feature. Entropy, complexity, and transfer-entropy features were computed with the k-nearest-neighbor estimators of Garnier, as in the parent trial (Mayer et al. 2025) and in the published information-flow analysis of this cohort (Garnier et al. 2026), using the public implementation at github.com/martinfrasch/entropy (commit 44fbf35). Maternal–fetal transfer entropy and entropy rate were computed with that implementation's C library, which handles the embedding lag explicitly, on refined R peaks, with k-nearest-neighbor ($k$=5) and Gaussian estimators and surrogate correction; these coupling arms are post hoc. R peaks for all heart-rate-variability and coupling arms are the ensemble detections merged so that each beat is counted once ("refined").

**Fetal inclusion.** Fetal series with a fetal heart rate below 100 bpm indicate maternal contamination after source separation. For all fetal-ECG and maternal–fetal coupling arms, the primary analysis uses the inclusion set of the published analysis of this cohort (Garnier et al. 2026) without the duplicated recordings, $n$=113 (57 with PSS-10 ≥ 19); $n$=144 is reported as a sensitivity analysis. Maternal and abdominal arms use all 144 participants.

**Missing data.** One FELICITy 1 participant without a PSS-10 score and six whose recordings were byte-identical duplicates of another participant's (three pairs) were excluded before any modeling. In FELICITy 2, six randomized women without an ECG recording were excluded from all ECG analyses. No outcome value was imputed, and no window or recording was excluded on signal quality; in the nested-selection grids, rare undefined feature values were replaced by the feature's median over participants, a label-free step taken before the split. Where a feature could not be computed for a participant (for example, no valid coupling window), that participant is absent from that arm only, and the arm's $n$ is stated.

**Rounding.** Every reported value is the stored run value rounded half-up to three decimals; $p$ values below 0.001 are given to one significant figure; raw $p$ is shown beside every multiplicity-adjusted value, and every result with raw $p < 0.05$ is listed with its adjusted value in Supplementary Table S7. **Inference.** Confidence intervals for correlations and AUCs are bootstrap intervals over participants. Null distributions are obtained by permuting the participant-level outcome and re-running the entire cross-validation inside each permutation, so that any selection performed within the pipeline is included in the null. Where a search over thresholds or arms was performed, the permutation is calibrated over that same search space, and Bonferroni correction across arms is reported alongside.

**Bounds.** Because a null result is uninterpretable without a bound, one-sided non-superiority tests were computed against δ=0.70 using the Hanley–McNeil AUC variance, complemented by two-one-sided tests and a Bayesian bootstrap with a region of practical equivalence of 0.45–0.55 (seed 20260922, 10 000 iterations).

**Controls.** Every reported association is accompanied by a label-permutation null and, for deep representations, an untrained-encoder control using the same architecture with random weights. A positive control — recovery of each participant's own heart rate from the same frozen features — establishes that the pipeline extracts real physiological information.

**Mixed models.** In FELICITy 2, actual and model-predicted PSS-10 were each fitted with a linear mixed model with a random intercept per woman and fixed effects for visit, group, and their interaction (REML; `statsmodels mixedlm`); change scores were correlated between paired visits.

**Pre-registration.** All external-cohort tests and all tests against targets other than PSS-10 were pre-registered in dated documents before the corresponding outcome was examined; the pre-registrations, including those whose results were null or reversed, are released with the analysis code.

Stability. Intraclass correlations for the reference instrument were estimated from the four PSS-10 timepoints in FELICITy 2 in the 22 women with paired ECG, under the one-way

ICC(1,1), absolute-agreement ICC(2,1) and consistency ICC(3,1) forms. Because the first two timepoints were one questionnaire in 7 women, the consistency estimate was repeated on distinct administrations only, in a linear mixed model with a random intercept for each woman and the administration as a fixed effect (restricted maximum likelihood); intervals are 95% percentiles of 2,000 bootstrap resamples of women. Test–retest stability of ECG-derived quantities used the paired recordings approximately 15 weeks apart. The attenuation ceiling on an observable correlation is reported as $\sqrt{(r_{xx}\, r_{yy})}$.

**Sample size.** Sample sizes were determined by full enrollment of the parent FELICITy 1 and FELICITy 2 studies rather than by a prediction-model-specific calculation. Minimum detectable effects are therefore reported for each analysis in place of a prospective power statement.

## Evaluation harness and leakage controls

All analyses reported here were run through a single evaluation harness. It enforces four boundaries at which leakage can enter even when no model is fitted on its evaluation data.

**Unit of analysis.** Folds are defined once at the participant level and stored (five-fold participant partitions, stratified on the binary label when drawn and repeated over five seeds, of the 144 analyzed participants; the fetal inclusion set uses the same partitions restricted to its 113 participants). Every subsequent analysis loads this file and asserts that each participant in its feature matrix appears in it; no code path falls back to row-level splitting. Because all feature families use the same stored fold definition, arms are directly comparable and fold assignment cannot differ between them. Each fold logs the number of *participants* as well as the number of segments on both sides of the split, so a violation is visible in the run log before any metric is read.

**Both splits computed and reported.** For every segment-level feature matrix the harness additionally computes the row-level result on identical features and writes both, so the size of any splitting-unit leak is measured within each analysis rather than asserted.

**Selection inside the training fold.** Where a grid of features was searched, selection was nested: the single best feature was chosen by inner five-fold cross-validation on the training participants only, then evaluated on the held-out participants. The complete per-feature out-of-fold table is written alongside each result so that selection instability is visible; across the entropy and coupling grids the most frequently selected feature typically changed between folds.

**Controls that can falsify the result.** Three were run routinely. A *positive control*: the same frozen embeddings, folds, and head used to predict each participant's own mean RR interval and SDNN, recovering these at $r$=0.78–0.80 and $r$=0.53–0.61 respectively across the three encoders. An *untrained-encoder control*: the DeepECG-SSL architecture with random weights (mean fold AUC 0.490 ±0.095 for PSS-10 ≥ 19, just below the 0.528–0.603 range of the pretrained representations on the same signal; best-over-sweep AUC 0.535 with a calibrated permutation $p$=0.708). And *permutation calibrated over the search* *actually performed*: where a threshold or grid was scanned, the reference distribution is the maximum over that scan under permuted labels, not a single test.

**Aggregation fixed across arms.** How segment-level predictions are aggregated to the participant is itself an analytic choice. Every comparison reported here uses a single aggregation, the mean over a participant's segments, applied identically to every arm.

### Computational environment and reproducibility

Model training used PyTorch 2.1 on a single NVIDIA A100 GPU (40 GB); downstream analyses used Python 3.10 with scikit-learn 1.3, statsmodels 0.14 and scipy 1.11. Random seeds for data splitting, weight initialization, and augmentation were fixed and are stored with the analysis code, allowing deterministic re-execution of the full pipeline.

## Data availability

Individual-level data from the FELICITy 1 and FELICITy 2 cohorts are available from the Department of Obstetrics and Gynaecology, TUM University Hospital, on reasonable request, subject to institutional data-sharing agreements and participant consent.

## Code availability

The analysis code, including the evaluation harness, is not publicly released at this time: parts of it are the subject of pending patent applications (see Competing interests). The code, the fold definition file, per-participant out-of-fold predictions for every arm and the scripts that generate every table and figure are available for verification to editors, reviewers and, on reasonable request, other researchers under a confidentiality agreement. The pre-registration documents, including those whose results were null or reversed, are available from the corresponding author on request. Every procedure is specified in Methods in enough detail to be re-implemented. The weights of the SimCLR encoders trained here are withheld owing to third-party intellectual property; ECGFounder and DeepECG-SSL were used with their authors' public checkpoints. The TRIPOD+AI checklist is provided as Supplementary Table S4.

## Author contributions

M.G.F. and S.M.L. conceived and designed the study. S.M.L., M.J.E.M., C.B., P.Z. and C.Z. collected the FELICITy 1 and FELICITy 2 data. M.G.F. developed the deep-learning framework and performed the analyses. M.C.A. contributed to interpretation of the fetal autonomic findings. N.B.G. contributed the entropy, complexity, and transfer-entropy estimators and advised on their application. M.G.F. drafted the manuscript with input from S.M.L. All authors critically revised the manuscript, approved the final version, and agree to be accountable for all aspects of the work.

## Competing interests

M.G.F. holds patents on fetal monitoring, is named on a pending US provisional patent application relating to ECG-derived stress assessment and holds equity in pregnancy-health start-ups. That provisional application was filed before the analyses reported here were completed; its claims rest in part on internal performance figures that do not hold under participant-level splitting, and it is disclosed here for completeness. M.G.F. is also named on a second US provisional patent application (filed October 2026) on signal-quality assessment of multi-channel ECG that draws on the channel-disagreement observations reported here. All other authors declare no competing interests.

## Funding

This work was supported by a Hans Fischer Senior Fellowship from the Institute for Advanced Study, Technical University of Munich (IAS-TUM), awarded to M.C.A.; by funds from the Department of Obstetrics and Gynaecology, TUM University Hospital, Technical University of Munich, to S.M.L.; and by the Canadian Institutes of Health Research (CIHR; grant number 123489) to M.G.F. The funders had no role in study design, data collection and analysis, decision to publish, or preparation of the manuscript.

## Acknowledgments

The authors thank the participants of the FELICITy study for their contribution to this research, and the clinical staff at the Department of Obstetrics and Gynaecology, TUM University Hospital, Technical University of Munich, for their assistance with data collection.

## Figure legends

**Figure 1. Model architecture and multi-layer feature extraction.** (A) ResNet-34 encoder adapted for 1D ECG, showing initial convolution, four residual blocks with channel dimensions [64,128,256,512], and global average pooling. SimCLR pretraining optimizes a contrastive loss between augmented views. (B) Multi-layer feature extraction: outputs from all four residual blocks plus the projection head are concatenated into a 1088-dimensional vector. (C) Downstream heads: logistic regression for binary stress classification and ridge regression for continuous PSS prediction.

**Figure 2. Pre-registered endpoints in FELICITy 2, under both signal-quality estimators.** Points are correlations with PSS-10 and bars are 95% confidence intervals; the pre-registered primary endpoint is shown in gold. The primary is null under both estimators, as is the maternal secondary over all recordings; the baseline-visit correlation of the maternal DeepECG-SSL score with PSS-10 is present under both estimators but absent over both visits. Heads fitted on FELICITy 1 (*n*=144); estimators as defined in Table 2. FELICITy 2, *n*=44 recordings from 22 women; unit of analysis recording, cluster bootstrap over women.

**Figure 3. The exclusion bound.** FELICITy 1, n=144 participants, target PSS-10 ≥ 19, unit of analysis participant. Participant-level AUC with 95% confidence intervals for the arms carried through formal testing. The shaded region marks screening-grade discrimination (AUC ≥ 0.70), which is rejected by one-sided non-superiority tests in every arm shown. Dotted line, chance. The lowest arm is the single-detector heart-rate-variability comparator; its large negative deviation is the expected behavior of cross-validated performance under a true null, and it sits below the ensemble-derived panel on the same signal.

**Figure 4. Other targets, their controls, and the full representation sweep.** FELICITy 1 (cortisol *n*=109; BMI control and sweep *n*=144), unit of analysis participant. Left: the pre-registered hair-cortisol arms (blue) with the two mandatory controls (gray) — an untrained encoder, which scores above the trained one, and the same representation predicting body-mass index, which confirms that these encoders read body habitus from the ECG. Bars are 95% confidence intervals. Right: cumulative distribution of the best AUC achieved by each of the 36 arms retaining stored per-participant predictions, against chance (dotted) and the screening-grade threshold (red).

# Figures

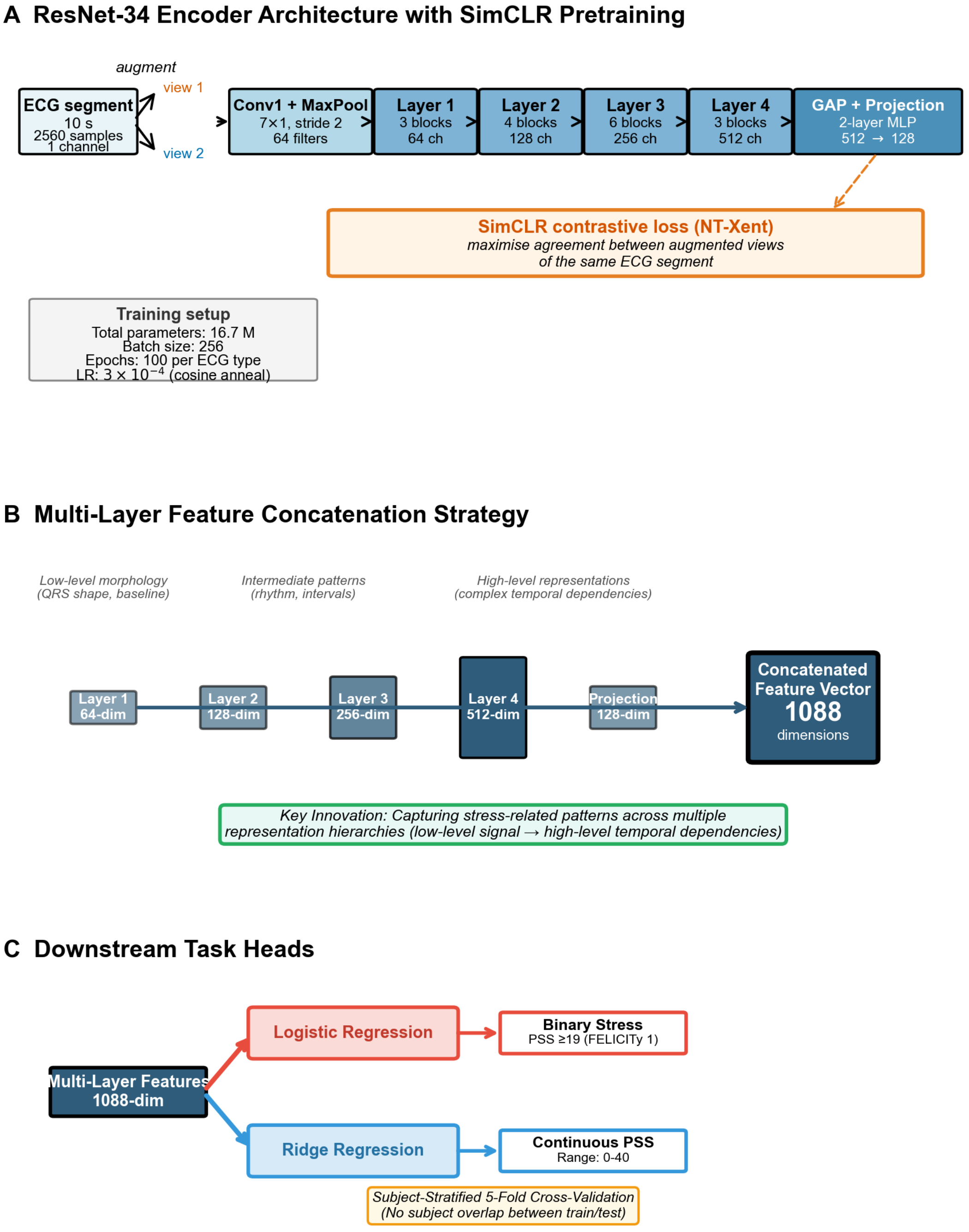

***Model architecture and multi-layer feature extraction.***

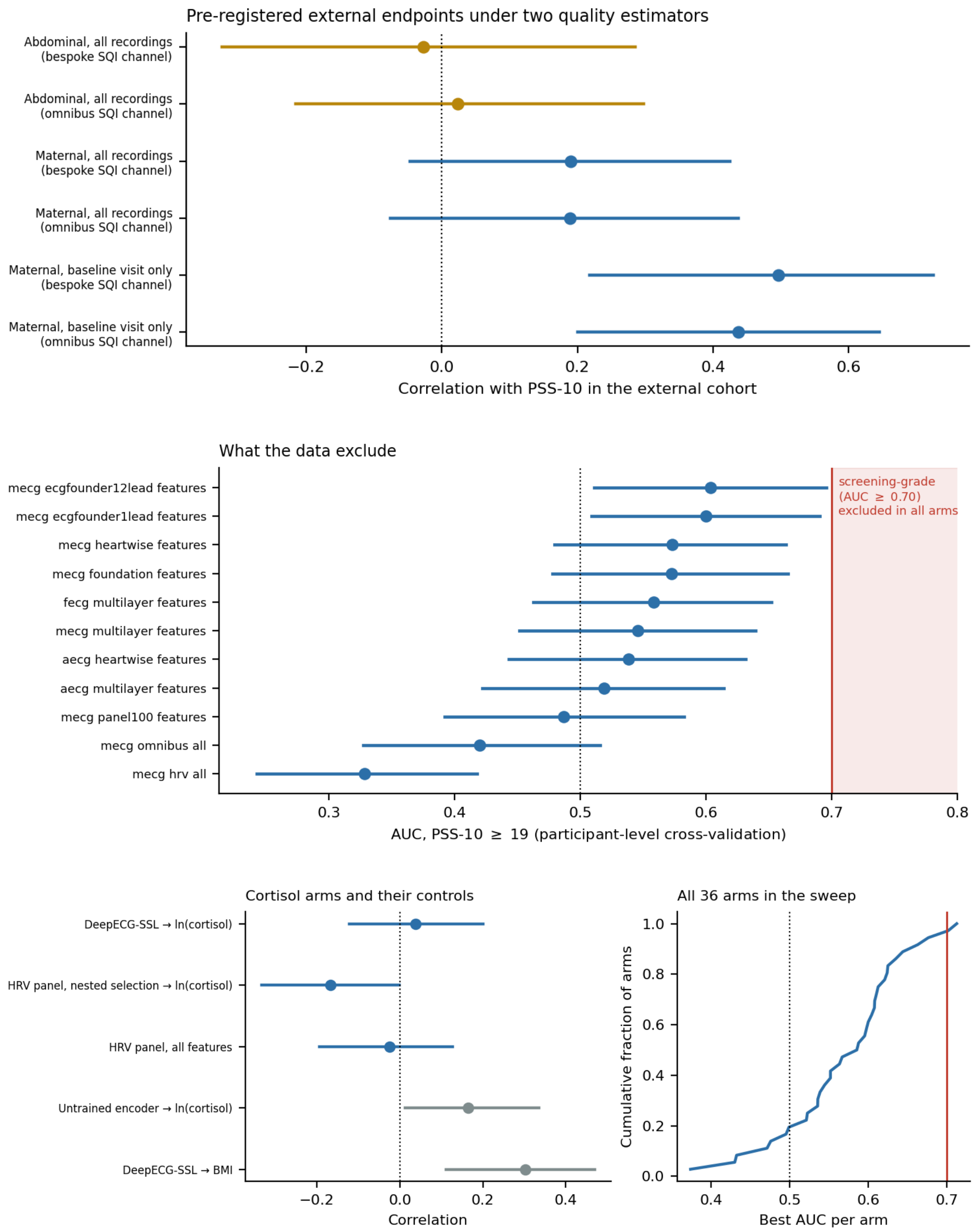

***Other targets, their controls, and the full arm sweep.*** *Left: the pre-registered hair-cortisol arms with the two mandatory controls — an untrained encoder, which scores above the trained one, and the same representation predicting BMI, which confirms that these encoders read body habitus from the ECG. Right: cumulative distribution of the best AUC achieved by each arm of the representation sweep, against chance (dotted) and the screening-grade threshold (red). FELICITy 1, n=144.*

# Supplementary Information

*How much perceived-stress signal is in the maternal ECG? An exclusion bound across deep-learning and heart-rate-variability representations in pregnancy*.

## Supplementary Figure S1. Segments per participant

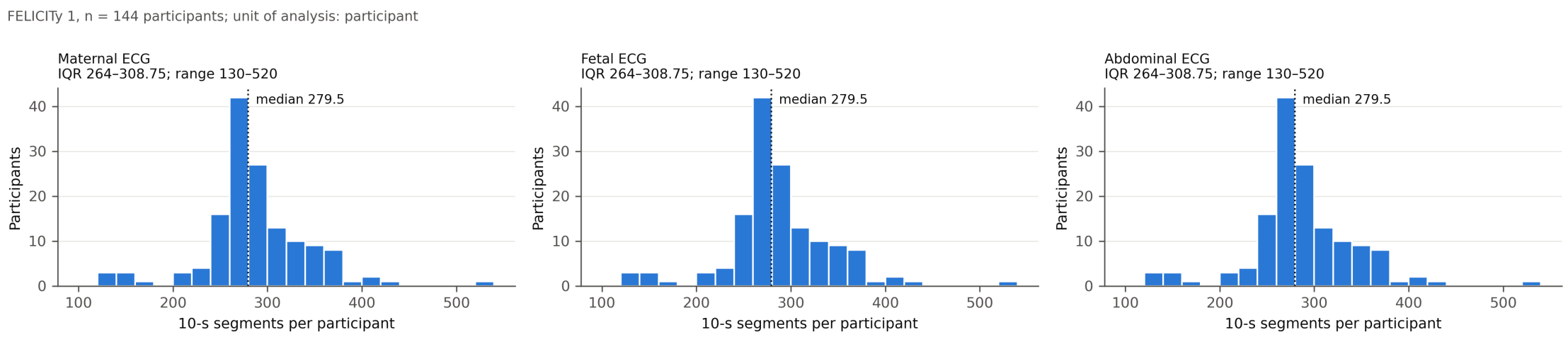

## Supplementary Figure S2. Electrode positions

The same transabdominal electrode placement was used in both cohorts: FELICITy 1 recorded with the Monica AN-24 and FELICITy 2 with the Bittium Faros 360 (Mayer et al., Bioengineering 2025;12:1141).

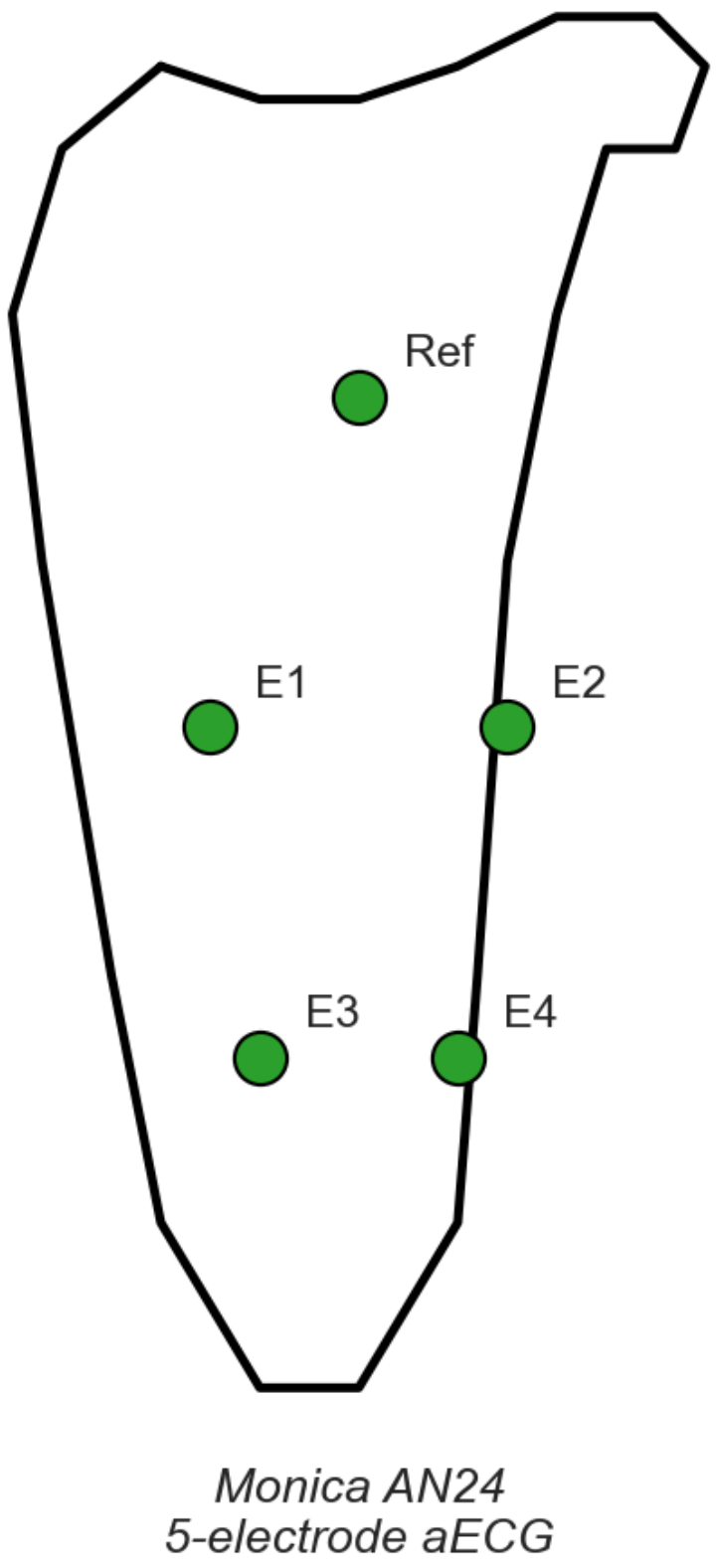

*FELICITy 1. Schematic of the Monica AN-24 five-electrode abdominal array: a reference electrode (Ref) and four recording electrodes (E1–E4) on the maternal abdomen. The*

*maternal and fetal ECG are the source-separated components that SAVER estimates from the four abdominal channels; the abdominal ECG is one raw abdominal channel. No limb lead was recorded.*

## Supplementary Table S3. Heart rate and heart-rate variability on the participant folds used for every arm

*FELICITy 1, PSS-10 ≥ 19, unit = participant; pooled out-of-fold AUC. Maternal and abdominal n = 144; fetal n = 113 (primary), n = 144 in brackets; — = not run at n = 113.*

| Arm | Maternal | Fetal | Abdominal |
|---|---|---|---|
| mean HR, mean RR, SDNN, RMSSD, single detector (NeuroKit2) | 0.254 | 0.485 (0.459) | 0.545 |
| 10 conventional HRV metrics, ensemble refined peaks | 0.473 | — (0.445) | 0.446 |
| 95-metric panel, ensemble refined peaks | 0.484 | 0.489 (0.488) | 0.332 |
| 10 conventional HRV metrics, single detector (NeuroKit2) | 0.328 | 0.399 (0.434) | 0.460 |
| best deep arm (ECGFounder 12-lead) | 0.604 [0.511, 0.696] | — | — |

Ensemble-detector arms use refined R peaks (each beat counted once); single-detector arms (NeuroKit2) use that detector's own peaks. Post hoc nested-selection arms of the extended sweep are reported in Results › What the data exclude.

## Supplementary Table S4. TRIPOD+AI reporting checklist

TRIPOD+AI (Collins et al., *BMJ* 2024;385:e078378), completed for this study. The study reports the development and internal evaluation of prediction models, a within-center cross-device evaluation of frozen models in a second cohort, and a negative result with formal bounds. Items are completed for that scope. Where an item is not applicable, or where the study falls short of the standard, this is stated rather than omitted. Section names refer to the main text.

| Item | TRIPOD+AI topic | Where addressed, and how |
|---|---|---|
| **Title and abstract** | | |
| 1 | Title identifies the study as developing/evaluating a prediction model, the target population and the outcome | Title. States the outcome (perceived stress), the population (pregnancy), the signal (maternal ECG) and that the contribution is an exclusion bound. |
| 2 | Abstract: structured summary of objectives, methods, results, conclusions | Abstract. Reports the participant-level evaluation, the bound, the cross-device nulls, the channel variance share, the cortisol null and the stability of the reference instrument across pregnancy. |
| **Introduction** | | |
| 3a | Background, rationale, references to existing models | Introduction, paragraphs 1–4, including the self-supervised ECG literature and the absence of maternal/fetal applications. |
| 3b | Objectives, including intended use and target population | Introduction, final paragraph: the three development questions and the aim of bounding the perceived-stress information in the maternal ECG. |
| **Methods** | | |
| 4a | Data sources, separately for development and evaluation | Methods › Study design and participants. FELICITy 1 (Monica AN-24, 900 Hz) for development; FELICITy 2 (Bittium Faros 360, 1000 Hz), same center, for cross-device evaluation. |
| 4b | Dates of recruitment, exposure, follow-up | Methods › Study design and participants; Results › Participant flow. |
| 5a | Eligibility criteria | Methods › Study design and participants. |
| 5b | Treatments received, if relevant | FELICITy 2 is a yoga-intervention trial; arm allocation is reported and tested (Results › Other targets; Table 3). |

| Item | TRIPOD+AI topic | Where addressed, and how |
|---|---|---|
| 6 | Outcome definition, blinding, and why it was chosen | Methods › Study design and participants (PSS-10 ≥ 19 as the registered cohort definition, not a validated clinical cutoff); Results › Other cutoffs. **Outcome stability across pregnancy is measured and reported** under three ICC forms, with the population it applies to; an orthogonal biological outcome (hair cortisol) is additionally tested. |
| 7 | Predictors: definition, timing, blinding | Methods › ECG acquisition and pre-processing; Deep-learning architecture and features. All representations are frozen before the downstream head is fitted. |
| 8 | Sample size | Methods › Statistical analysis. Fixed by parent-study enrollment; **minimum detectable effects are reported in place of a prospective power statement**. |
| 9 | Missing data | Methods › Statistical analysis › Missing data. One participant without PSS-10 and six duplicated recordings excluded; six FELICITy 2 women without ECG excluded; no outcome imputation (rare undefined grid features set to the column median, label-free); no window or recording excluded on signal quality; per-arm *n* stated. |
| 10a | Analytical methods: how predictors were handled | Methods › Statistical analysis. All fitted steps inside the training fold; nested selection where a grid was searched. |
| 10b | Model type, building procedure, internal validation | Methods › Stress classification and regression; Statistical analysis. Frozen encoders with logistic or ridge heads (scikit-learn defaults, untuned; $C = \alpha$ sweep reported); participant-level 5 × 5 cross-validation on one shared fold definition reused by all arms. |
| 10c | Handling of class imbalance or other AI-specific choices | Not applicable at the primary cutoff (51% event rate). Other cutoffs are post hoc and reported as such. |
| 10d | Measures used to assess performance | Discrimination (AUC), regression metrics ($R^2$, MAE), correlation, all with bootstrap intervals over participants; operating points with thresholds chosen in inner CV; Brier skill score raw and after recalibration inside the training fold (Supplementary Table S5). **Decision curves are not reported**, and calibration cannot be assessed in FELICITy 2 because it recruited on the outcome (Results). |
| 10e | Model updating between development and evaluation | None. Heads are fitted once on FELICITy 1 and applied unchanged to FELICITy 2 (Methods › Stress classification and regression). |
| 11 | Risk groups, if created | Not applicable. |
| 12 | Differences between development and evaluation settings | Methods › Study design and participants; Discussion › Strengths and limitations. Device, cohort, design, outcome selection, number of recording channels, sampling rate and recording conditions all differ together. |
| **AI-specific items** | | |
| 13a | Model architecture, hyperparameters, software, versions | Methods › Deep-learning architecture and features; Computational environment. |
| 13b | Initialization, training procedure, stopping | Methods. Self-supervised pretraining without labels on all FELICITy 1 segments (stated as a limitation); downstream heads with fixed regularization. |
| 13c | Fairness: performance across demographic subgroups | **Not performed.** Both cohorts are from a single tertiary center in Germany; stated as a limitation. |
| 13d | Measures to prevent data leakage | Methods › Evaluation harness and leakage controls. Participant-level fold definition asserted at load in the script that reports each result; participants per fold logged; the row-level result computed beside the participant-level one; selection nested inside the training fold; one aggregation fixed for all arms. |
| **Results** | | |
| 14 | Participant flow, with a diagram if possible | Results › Participant flow. |
| 15 | Participant characteristics | Results › FELICITy 1 cohort characteristics; Supplementary Figure S1 (segments per participant). |
| 16 | Number of participants and outcome events in each analysis | Stated in every table caption and per arm. |
| 17 | Model specification sufficient for prediction | Architecture and heads specified; **weights are not deposited** owing to third-party intellectual property. Cached feature matrices and out-of-fold predictions are available to editors and reviewers under a confidentiality agreement, so that every reported result can be verified. |

| Item | TRIPOD+AI topic | Where addressed, and how |
|---|---|---|
| 18 | Model performance, with uncertainty | Tables 1–3 and Figures 2–4, with confidence intervals; permutation *p* calibrated over the search performed; raw *p* beside every adjusted *p* (Supplementary Table S7). |
| 19 | Model updating results | Not applicable. |
| **Discussion** | | |
| 20 | Limitations | Discussion › Strengths and limitations: single center, cross-device confounding, pretraining on all participants, fetal arms not evidence of fetal physiology, unmeasured cortisol reliability, missing cortisol covariates, accumulated multiplicity. |
| 21 | Interpretation, with reference to the intended use | Discussion. A single-recording ECG screener at the clinical threshold is excluded in this population; the low stability of the reference standard limits the prediction of a woman's stable stress level, not of the state at the time of recording. |
| 22 | Implications for practice and future research | Discussion: channel rule and quality-estimator implementation to be pre-registered; repeated outcome administration; paired ECG at each administration for the within-person axis. |
| **Other information** | | |
| 23 | Supplementary information | Supplementary Figures S1–S2 and Tables S3–S7. |
| 24 | Funding | Funding. |
| 25 | Conflicts of interest | Competing interests. Discloses two provisional patent applications, including one whose claims rest in part on internal figures that do not hold under participant-level splitting. |
| 26 | Data and code availability | Data availability (individual data on request) and Code availability (code not publicly released owing to pending patent applications; code, folds and out-of-fold predictions available to editors and reviewers under a confidentiality agreement; pre-registrations on request). |
| 27 | Registration and protocol | The parent cohorts are registered. This re-analysis was not prospectively registered; **every cross-device test and every test against a target other than PSS-10 was pre-registered in a dated document before the corresponding outcome was examined**, and those documents are available on request, including for null tests. |

**Note on scope.** Three items are not fully met: fairness analysis across demographic subgroups (13c), decision-curve analysis (10d), and prospective registration of this re-analysis (27).

## Supplementary Table S5. Screening operating points with training-fold thresholds

Folds: the shared participant-level fold definition (5 seeds × 5 outer folds). Every threshold, the Platt recalibration and (fusion arms) the stacker are fitted by inner 5-fold CV on the outer-training women only, then applied to the held-out women. Operating points tested: target sensitivity 0.70, 0.75, 0.80, 0.90, 0.95 and Youden. Set A = the eleven arms carried through formal testing (Table 1 of the main text); set B = SimCLR projection-head-only arms; set C = fusion arms (participant-mean features). PPV/NPV at q% are re-weighted from the operating point's sensitivity and specificity; NPV (obs) is observed in this cohort (prevalence 51%). Ruled out = women called negative, per 150.

Cells: seed mean, or seed mean (participant-bootstrap 95% CI of the seed mean). LR CIs average finite seed values only; "inf" = a seed had specificity 0.

## Table S5.1. Models: discrimination and calibration

*FELICITy 1, n = 144 women (73 positive, observed prevalence 51%); target PSS-10 ≥ 19 (binary); unit of analysis: participant; Brier on participant-level outer-test probabilities.*

| Set | Signal | Representation / architecture | Pipeline | AUC rerun | AUC stored | Brier raw | BSS raw | Brier recal. | BSS recal. |
|---|---|---|---|---|---|---|---|---|---|
| A | maternal ECG | SimCLR ResNet-34 (1-D), multi-layer 1088-d | LR on segments, mean prob per woman; nested | 0.546 | 0.546 | 0.358 (0.310, 0.407) | -0.430 (-0.629, -0.240) | 0.252 (0.248, 0.256) | -0.009 (-0.025, 0.005) |
| A | maternal ECG | HeartWise DeepECG-SSL (wav2vec2), 768-d | LR on segments, mean prob per woman; nested | 0.573 | 0.573 | 0.318 (0.275, 0.363) | -0.272 (-0.450, -0.098) | 0.249 (0.239, 0.260) | 0.005 (-0.037, 0.044) |
| A | maternal ECG | 100-feature HRV panel (95 usable; refined ensemble R-peaks) | LR on one row per woman; nested | 0.484 | 0.484 | 0.339 (0.300, 0.382) | -0.356 (-0.525, -0.197) | 0.254 (0.252, 0.257) | -0.016 (-0.026, -0.006) |
| A | maternal ECG | hrv-omnibus panel, 32 features (refined ensemble R-peaks) | LR on one row per woman; nested | 0.478 | 0.478 | 0.286 (0.262, 0.313) | -0.144 (-0.252, -0.047) | 0.253 (0.247, 0.259) | -0.013 (-0.036, 0.011) |
| A | maternal ECG | conventional HRV, 10 features (single NeuroKit detector) | LR on one row per woman; nested | 0.328 | 0.328 | 0.274 (0.264, 0.286) | -0.097 (-0.141, -0.056) | 0.242 (0.234, 0.250) | 0.032 (0.002, 0.064) |
| A | fetal ECG | SimCLR ResNet-34 (1-D), multi-layer 1088-d | LR on segments, mean prob per woman; nested | 0.558 | 0.558 | 0.379 (0.323, 0.439) | -0.515 (-0.754, -0.295) | 0.255 (0.251, 0.259) | -0.019 (-0.036, -0.003) |
| A | abdominal ECG | SimCLR ResNet-34 (1-D), multi-layer 1088-d | LR on segments, mean prob per woman; nested | 0.519 | 0.519 | 0.337 (0.293, 0.382) | -0.346 (-0.528, -0.170) | 0.254 (0.251, 0.258) | -0.017 (-0.032, -0.005) |
| A | abdominal ECG | HeartWise DeepECG-SSL (wav2vec2), 768-d | LR on segments, mean prob per woman; nested | 0.538 | 0.538 | 0.301 (0.265, 0.339) | -0.204 (-0.355, -0.058) | 0.252 (0.248, 0.257) | -0.009 (-0.028, 0.010) |
| A | maternal ECG | ECGFounder 12-lead pathway (published weights, 507/509 tensors) | LR on participant-mean embedding; nested | 0.604 | 0.604 | 0.348 (0.295, 0.403) | -0.389 (-0.610, -0.180) | 0.248 (0.238, 0.259) | 0.008 (-0.033, 0.047) |
| A | maternal ECG | ECGFounder 1-lead pathway (published weights, 507/509 tensors) | LR on participant-mean embedding; nested | 0.600 | 0.600 | 0.348 (0.293, 0.404) | -0.391 (-0.613, -0.172) | 0.254 (0.240, 0.267) | -0.014 (-0.070, 0.039) |
| A | maternal ECG | **CONTROL: untrained encoder (weights never loaded, 0/509 tensors)** | LR on segments, mean prob per woman; nested | 0.572 | 0.572 | 0.263 (0.235, 0.290) | -0.050 (-0.161, 0.060) | 0.253 (0.244, 0.262) | -0.011 (-0.049, 0.025) |
| B | maternal ECG | SimCLR ResNet-34 (1-D), projection head only 128-d (cols 960:1088) | LR on segments, mean prob per woman; nested | 0.534 | 0.534 | 0.339 (0.296, 0.385) | -0.356 (-0.538, -0.183) | 0.254 (0.249, 0.259) | -0.015 (-0.033, 0.002) |
| B | fetal ECG | SimCLR ResNet-34 (1-D), projection head only 128-d (cols 960:1088) | LR on segments, mean prob per woman; nested | 0.553 | 0.553 | 0.353 (0.299, 0.407) | -0.411 (-0.627, -0.194) | 0.255 (0.252, 0.259) | -0.021 (-0.034, -0.009) |
| B | abdominal ECG | SimCLR ResNet-34 (1-D), projection head only 128-d (cols 960:1088) | LR on segments, mean prob per woman; nested | 0.480 | 0.480 | 0.340 (0.299, 0.381) | -0.361 (-0.522, -0.196) | 0.255 (0.251, 0.258) | -0.018 (-0.030, -0.006) |
| C | maternal ECG | HeartWise DeepECG-SSL, participant-mean | LR on participant-mean embedding; nested | 0.601 | 0.611 | 0.354 (0.299, 0.412) | -0.415 (-0.648, -0.196) | 0.252 (0.242, 0.262) | -0.006 (-0.045, 0.032) |
| C | maternal ECG | SimCLR ResNet-34 multi-layer, participant-mean | LR on participant-mean embedding; nested | 0.588 | 0.600 | 0.390 (0.337, 0.446) | -0.559 (-0.785, -0.344) | 0.252 (0.242, 0.262) | -0.006 (-0.047, 0.033) |

| Set | Signal | Representation / architecture | Pipeline | AUC rerun | AUC stored | Brier raw | BSS raw | Brier recal. | BSS recal. |
|---|---|---|---|---|---|---|---|---|---|
| C | maternal ECG | **CONTROL: HeartWise architecture, random init, participant-mean** | LR on participant-mean embedding; nested | 0.543 | 0.534 | 0.347 (0.296, 0.397) | -0.386 (-0.589, -0.183) | 0.253 (0.248, 0.260) | -0.013 (-0.037, 0.009) |
| C | maternal ECG | concat ECGFounder 12-lead + DeepECG-SSL (1792-d) | one LR on concatenated participant-mean features; nested | 0.622 | 0.625 | 0.349 (0.294, 0.406) | -0.394 (-0.625, -0.178) | 0.248 (0.235, 0.262) | 0.008 (-0.047, 0.059) |
| C | maternal ECG | late mean: ECGFounder 12-lead + DeepECG-SSL | mean of per-encoder LR probabilities; nested | 0.610 | 0.617 | 0.299 (0.256, 0.347) | -0.197 (-0.388, -0.020) | 0.246 (0.235, 0.258) | 0.016 (-0.031, 0.060) |
| C | maternal ECG | stacked: ECGFounder 12-lead + DeepECG-SSL | LR stacker on inner-CV encoder probabilities; nested | 0.509 | 0.546 | 0.253 (0.246, 0.261) | -0.013 (-0.041, 0.016) | 0.254 (0.250, 0.259) | -0.017 (-0.034, -0.001) |
| C | maternal ECG | late mean: ECGFounder 12-lead + DeepECG-SSL + SimCLR | mean of per-encoder LR probabilities; nested | 0.619 | 0.626 | 0.284 (0.245, 0.325) | -0.136 (-0.299, 0.021) | 0.246 (0.236, 0.257) | 0.017 (-0.028, 0.058) |
| C | maternal ECG | **CONTROL: late mean ECGFounder 12-lead + random-init HeartWise** | mean of per-encoder LR probabilities; nested | 0.582 | 0.577 | 0.306 (0.261, 0.351) | -0.225 (-0.400, -0.043) | 0.250 (0.243, 0.259) | -0.001 (-0.034, 0.031) |

Brier reference (training-fold prevalence for every woman) = 0.2501; BSS = 1 − Brier/reference; recal. = Platt map fitted on inner-CV scores of the training women. AUC rerun = participant-level AUC of the seed-averaged outer-test score; AUC stored = the producing script's stored value (set C: 151 women, including the three duplicated pairs).

## Table S5.2a. Models: sensitivity, specificity, LR+, LR− at target sensitivity 0.70, target sensitivity 0.75, target sensitivity 0.80

*FELICITy 1, n = 144 women (73 positive, observed prevalence 51%); target PSS-10 ≥ 19 (binary); unit of analysis: participant; seed means; CIs in the per-target tables below.*

| Set | Signal | Representation | sens @70 | spec @70 | LR+ @70 | LR− @70 | sens @75 | spec @75 | LR+ @75 | LR− @75 | sens @80 | spec @80 | LR+ @80 | LR− @80 |
|---|---|---|---|---|---|---|---|---|---|---|---|---|---|---|
| A | maternal ECG | SimCLR ResNet-34 (1-D), multi-layer 1088-d | 0.710 | 0.310 | 1.034 | 0.970 | 0.762 | 0.245 | 1.016 | 1.043 | 0.811 | 0.175 | 0.984 | 1.123 |
| A | maternal ECG | HeartWise DeepECG-SSL (wav2vec2), 768-d | 0.745 | 0.335 | 1.122 | 0.762 | 0.797 | 0.287 | 1.121 | 0.710 | 0.838 | 0.223 | 1.079 | 0.729 |
| A | maternal ECG | 100-feature HRV panel (95 usable; refined ensemble R-peaks) | 0.710 | 0.282 | 0.990 | 1.048 | 0.748 | 0.248 | 0.995 | 1.027 | 0.811 | 0.220 | 1.042 | 0.898 |
| A | maternal ECG | hrv-omnibus panel, 32 features (refined ensemble R-peaks) | 0.707 | 0.265 | 0.963 | 1.121 | 0.759 | 0.231 | 0.987 | 1.049 | 0.816 | 0.186 | 1.003 | 0.993 |
| A | maternal ECG | conventional HRV, 10 features (single NeuroKit detector) | 0.712 | 0.161 | 0.849 | 1.874 | 0.751 | 0.115 | 0.850 | 2.318 | 0.830 | 0.079 | 0.901 | 2.237 |
| A | fetal ECG | SimCLR ResNet-34 (1-D), multi-layer 1088-d | 0.685 | 0.299 | 0.981 | 1.099 | 0.748 | 0.223 | 0.968 | 1.349 | 0.811 | 0.152 | 0.960 | 1.454 |
| A | abdominal ECG | SimCLR ResNet-34 (1-D), multi-layer 1088-d | 0.699 | 0.287 | 0.983 | 1.064 | 0.767 | 0.225 | 0.994 | 1.071 | 0.808 | 0.152 | 0.955 | 1.348 |

| Set | Signal | Representation | sens @70 | spec @70 | LR+ @70 | LR− @70 | sens @75 | spec @75 | LR+ @75 | LR− @75 | sens @80 | spec @80 | LR+ @80 | LR− @80 |
|---|---|---|---|---|---|---|---|---|---|---|---|---|---|---|
| A | abdominal ECG | HeartWise DeepECG-SSL (wav2vec2), 768-d | 0.688 | 0.335 | 1.035 | 0.932 | 0.745 | 0.279 | 1.036 | 0.927 | 0.808 | 0.220 | 1.039 | 0.901 |
| A | maternal ECG | ECGFounder 12-lead pathway (published weights, 507/509 tensors) | 0.751 | 0.394 | 1.249 | 0.643 | 0.819 | 0.344 | 1.254 | 0.535 | 0.849 | 0.318 | 1.251 | 0.482 |
| A | maternal ECG | ECGFounder 1-lead pathway (published weights, 507/509 tensors) | 0.679 | 0.403 | 1.138 | 0.793 | 0.756 | 0.346 | 1.160 | 0.708 | 0.830 | 0.279 | 1.151 | 0.605 |
| A | maternal ECG | **CONTROL: untrained encoder (weights never loaded, 0/509 tensors)** | 0.762 | 0.324 | 1.129 | 0.744 | 0.805 | 0.245 | 1.068 | 0.798 | 0.836 | 0.189 | 1.030 | 0.859 |
| B | maternal ECG | SimCLR ResNet-34 (1-D), projection head only 128-d (cols 960:1088) | 0.699 | 0.315 | 1.024 | 0.971 | 0.748 | 0.268 | 1.026 | 0.970 | 0.795 | 0.189 | 0.985 | 1.204 |
| B | fetal ECG | SimCLR ResNet-34 (1-D), projection head only 128-d (cols 960:1088) | 0.696 | 0.321 | 1.030 | 0.964 | 0.745 | 0.223 | 0.960 | 1.181 | 0.795 | 0.135 | 0.922 | 1.994 |
| B | abdominal ECG | SimCLR ResNet-34 (1-D), projection head only 128-d (cols 960:1088) | 0.693 | 0.268 | 0.946 | 1.147 | 0.745 | 0.237 | 0.977 | 1.084 | 0.816 | 0.189 | 1.008 | 0.994 |
| C | maternal ECG | HeartWise DeepECG-SSL, participant-mean | 0.756 | 0.358 | 1.197 | 0.724 | 0.786 | 0.310 | 1.163 | 0.756 | 0.830 | 0.245 | 1.110 | 0.749 |
| C | maternal ECG | SimCLR ResNet-34 multi-layer, participant-mean | 0.655 | 0.397 | 1.090 | 0.876 | 0.715 | 0.338 | 1.081 | 0.844 | 0.786 | 0.301 | 1.126 | 0.709 |
| C | maternal ECG | **CONTROL: HeartWise architecture, random init, participant-mean** | 0.721 | 0.304 | 1.044 | 0.953 | 0.770 | 0.223 | 0.994 | 1.085 | 0.836 | 0.172 | 1.010 | 0.984 |
| C | maternal ECG | concat ECGFounder 12-lead + DeepECG-SSL (1792-d) | 0.696 | 0.420 | 1.204 | 0.732 | 0.767 | 0.344 | 1.172 | 0.685 | 0.816 | 0.270 | 1.121 | 0.685 |
| C | maternal ECG | late mean: ECGFounder 12-lead + DeepECG-SSL | 0.732 | 0.380 | 1.193 | 0.734 | 0.784 | 0.338 | 1.198 | 0.684 | 0.847 | 0.276 | 1.179 | 0.594 |
| C | maternal ECG | stacked: ECGFounder 12-lead + DeepECG-SSL | 0.756 | 0.296 | 1.081 | 0.840 | 0.808 | 0.242 | 1.078 | 0.856 | 0.863 | 0.186 | 1.064 | 0.781 |
| C | maternal ECG | late mean: ECGFounder 12-lead + DeepECG-SSL + SimCLR | 0.718 | 0.411 | 1.220 | 0.688 | 0.767 | 0.349 | 1.180 | 0.669 | 0.827 | 0.268 | 1.135 | 0.673 |
| C | maternal ECG | **CONTROL: late mean ECGFounder 12-lead + random-init HeartWise** | 0.734 | 0.375 | 1.183 | 0.733 | 0.784 | 0.313 | 1.148 | 0.723 | 0.830 | 0.293 | 1.178 | 0.591 |

## Table S5.2b. Models: sensitivity, specificity, LR+, LR− at target sensitivity 0.90, target sensitivity 0.95, Youden

*FELICITy 1, n = 144 women (73 positive, observed prevalence 51%); target PSS-10 ≥ 19 (binary); unit of analysis: participant; seed means; CIs in the per-target tables below.*

| Set | Signal | Representation | sens @90 | spec @90 | LR+ @90 | LR− @90 | sens @95 | spec @95 | LR+ @95 | LR− @95 | sens @Youden | spec @Youden | LR+ @Youden | LR− @Youden |
|---|---|---|---|---|---|---|---|---|---|---|---|---|---|---|
| A | maternal ECG | SimCLR ResNet-34 (1-D), multi-layer 1088-d | 0.904 | 0.110 | 1.016 | 0.861 | 0.967 | 0.062 | 1.031 | 0.460 | 0.496 | 0.515 | 1.027 | 0.978 |
| A | maternal ECG | HeartWise DeepECG-SSL (wav2vec2), 768-d | 0.915 | 0.118 | 1.038 | 0.705 | 0.975 | 0.056 | 1.034 | 0.482 | 0.411 | 0.676 | 1.308 | 0.871 |

| Set | Signal | Representation | sens @90 | spec @90 | LR+ @90 | LR− @90 | sens @95 | spec @95 | LR+ @95 | LR− @95 | sens @Youden | spec @Youden | LR+ @Youden | LR− @Youden |
|---|---|---|---|---|---|---|---|---|---|---|---|---|---|---|
| A | maternal ECG | 100-feature HRV panel (95 usable; refined ensemble R-peaks) | 0.896 | 0.138 | 1.040 | 0.755 | 0.962 | 0.073 | 1.039 | 0.593 | 0.605 | 0.363 | 0.945 | 1.064 |
| A | maternal ECG | hrv-omnibus panel, 32 features (refined ensemble R-peaks) | 0.926 | 0.082 | 1.008 | 0.872 | 0.973 | 0.034 | 1.007 | 0.843 | 0.414 | 0.561 | 0.934 | 1.044 |
| A | maternal ECG | conventional HRV, 10 features (single NeuroKit detector) | 0.929 | 0.017 | 0.945 | inf | 0.964 | 0.003 | 0.967 | n/a | 0.704 | 0.251 | 0.937 | 1.172 |
| A | fetal ECG | SimCLR ResNet-34 (1-D), multi-layer 1088-d | 0.893 | 0.062 | 0.953 | 2.206 | 0.959 | 0.017 | 0.976 | inf | 0.414 | 0.665 | 1.257 | 0.886 |
| A | abdominal ECG | SimCLR ResNet-34 (1-D), multi-layer 1088-d | 0.904 | 0.037 | 0.939 | 3.566 | 0.940 | 0.011 | 0.950 | inf | 0.373 | 0.659 | 1.123 | 0.958 |
| A | abdominal ECG | HeartWise DeepECG-SSL (wav2vec2), 768-d | 0.907 | 0.085 | 0.991 | 1.156 | 0.970 | 0.014 | 0.984 | inf | 0.438 | 0.558 | 0.989 | 1.008 |
| A | maternal ECG | ECGFounder 12-lead pathway (published weights, 507/509 tensors) | 0.926 | 0.161 | 1.105 | 0.438 | 0.970 | 0.076 | 1.051 | 0.367 | 0.677 | 0.417 | 1.165 | 0.759 |
| A | maternal ECG | ECGFounder 1-lead pathway (published weights, 507/509 tensors) | 0.915 | 0.144 | 1.069 | 0.582 | 0.956 | 0.096 | 1.058 | 0.430 | 0.636 | 0.448 | 1.163 | 0.826 |
| A | maternal ECG | **CONTROL: untrained encoder (weights never loaded, 0/509 tensors)** | 0.907 | 0.070 | 0.975 | 1.327 | 0.953 | 0.028 | 0.981 | inf | 0.373 | 0.707 | 1.326 | 0.887 |
| B | maternal ECG | SimCLR ResNet-34 (1-D), projection head only 128-d (cols 960:1088) | 0.923 | 0.065 | 0.987 | 1.232 | 0.978 | 0.037 | 1.015 | 0.713 | 0.562 | 0.470 | 1.070 | 0.933 |
| B | fetal ECG | SimCLR ResNet-34 (1-D), projection head only 128-d (cols 960:1088) | 0.915 | 0.051 | 0.965 | 2.286 | 0.964 | 0.017 | 0.981 | inf | 0.436 | 0.634 | 1.218 | 0.897 |
| B | abdominal ECG | SimCLR ResNet-34 (1-D), projection head only 128-d (cols 960:1088) | 0.907 | 0.070 | 0.976 | 1.375 | 0.962 | 0.011 | 0.973 | inf | 0.392 | 0.504 | 0.793 | 1.221 |
| C | maternal ECG | HeartWise DeepECG-SSL, participant-mean | 0.901 | 0.099 | 1.000 | 1.037 | 0.959 | 0.048 | 1.007 | 0.940 | 0.507 | 0.603 | 1.318 | 0.820 |
| C | maternal ECG | SimCLR ResNet-34 multi-layer, participant-mean | 0.901 | 0.130 | 1.036 | 0.764 | 0.953 | 0.054 | 1.008 | 0.940 | 0.479 | 0.592 | 1.239 | 0.881 |
| C | maternal ECG | **CONTROL: HeartWise architecture, random init, participant-mean** | 0.915 | 0.068 | 0.982 | 1.380 | 0.959 | 0.023 | 0.981 | 2.042 | 0.455 | 0.589 | 1.115 | 0.928 |
| C | maternal ECG | concat ECGFounder 12-lead + DeepECG-SSL (1792-d) | 0.918 | 0.146 | 1.078 | 0.599 | 0.959 | 0.090 | 1.055 | 0.480 | 0.567 | 0.532 | 1.248 | 0.822 |
| C | maternal ECG | late mean: ECGFounder 12-lead + DeepECG-SSL | 0.942 | 0.166 | 1.135 | 0.397 | 0.967 | 0.104 | 1.082 | 0.348 | 0.622 | 0.493 | 1.261 | 0.786 |
| C | maternal ECG | stacked: ECGFounder 12-lead + DeepECG-SSL | 0.940 | 0.096 | 1.040 | 0.643 | 0.967 | 0.045 | 1.014 | n/a | 0.679 | 0.338 | 1.033 | 0.975 |
| C | maternal ECG | late mean: ECGFounder 12-lead + DeepECG-SSL + SimCLR | 0.929 | 0.152 | 1.099 | 0.535 | 0.970 | 0.087 | 1.065 | 0.510 | 0.529 | 0.592 | 1.299 | 0.797 |

| Set | Signal | Representation | sens @90 | spec @90 | LR+ @90 | LR− @90 | sens @95 | spec @95 | LR+ @95 | LR− @95 | sens @Youden | spec @Youden | LR+ @Youden | LR− @Youden |
|---|---|---|---|---|---|---|---|---|---|---|---|---|---|---|
| C | maternal ECG | **CONTROL: late mean ECGFounder 12-lead + random-init HeartWise** | 0.929 | 0.144 | 1.088 | 0.549 | 0.967 | 0.085 | 1.058 | 0.487 | 0.627 | 0.448 | 1.159 | 0.822 |

## Operating point: target sensitivity 0.70

*FELICITy 1, n = 144 women (73 positive, observed prevalence 51%); target PSS-10 ≥ 19 (binary); unit of analysis: participant; operating point target sensitivity 0.70.*

| Set | Signal | Representation | sens | spec | LR+ | LR− | PPV@20 | PPV@30 | PPV@50 | PPV@70 | NPV@20 | NPV@30 | NPV@50 | NPV@70 | NPV (obs) | ruled out /150 |
|---|---|---|---|---|---|---|---|---|---|---|---|---|---|---|---|---|
| A | maternal ECG | SimCLR ResNet-34 (1-D), multi-layer 1088-d | 0.710 (0.630, 0.786) | 0.310 (0.232, 0.394) | 1.034 (0.880, 1.226) | 0.970 (0.662, 1.487) | 0.205 (0.180, 0.234) | 0.307 (0.273, 0.343) | 0.508 (0.466, 0.548) | 0.706 (0.670, 0.738) | 0.806 (0.736, 0.859) | 0.709 (0.623, 0.781) | 0.513 (0.418, 0.608) | 0.313 (0.238, 0.402) | 0.506 (0.378, 0.629) | 45.000 (36.458, 53.750) |
| A | maternal ECG | HeartWise DeepECG-SSL (wav2vec2), 768-d | 0.745 (0.668, 0.823) | 0.335 (0.244, 0.430) | 1.122 (0.952, 1.342) | 0.762 (0.496, 1.165) | 0.219 (0.192, 0.251) | 0.325 (0.289, 0.364) | 0.529 (0.487, 0.572) | 0.723 (0.688, 0.757) | 0.840 (0.776, 0.890) | 0.754 (0.670, 0.825) | 0.568 (0.467, 0.670) | 0.361 (0.274, 0.467) | 0.561 (0.430, 0.687) | 44.167 (35.000, 53.542) |
| A | maternal ECG | 100-feature HRV panel (95 usable; refined ensemble R-peaks) | 0.710 (0.629, 0.784) | 0.282 (0.200, 0.366) | 0.990 (0.841, 1.164) | 1.048 (0.721, 1.654) | 0.198 (0.173, 0.225) | 0.297 (0.264, 0.332) | 0.497 (0.456, 0.536) | 0.697 (0.661, 0.729) | 0.794 (0.712, 0.850) | 0.693 (0.593, 0.769) | 0.494 (0.389, 0.591) | 0.297 (0.217, 0.386) | 0.487 (0.363, 0.614) | 42.917 (34.375, 51.672) |
| A | maternal ECG | hrv-omnibus panel, 32 features (refined ensemble R-peaks) | 0.707 (0.619, 0.791) | 0.265 (0.185, 0.345) | 0.963 (0.813, 1.132) | 1.121 (0.745, 1.802) | 0.194 (0.168, 0.220) | 0.292 (0.258, 0.326) | 0.490 (0.447, 0.529) | 0.691 (0.653, 0.723) | 0.782 (0.693, 0.845) | 0.678 (0.571, 0.762) | 0.476 (0.367, 0.582) | 0.282 (0.201, 0.378) | 0.469 (0.338, 0.601) | 41.875 (32.917, 50.417) |
| A | maternal ECG | conventional HRV, 10 features (single NeuroKit detector) | 0.712 (0.628, 0.786) | 0.161 (0.111, 0.218) | 0.849 (0.744, 0.969) | 1.874 (1.215, 3.411) | 0.175 (0.157, 0.195) | 0.267 (0.241, 0.293) | 0.459 (0.425, 0.491) | 0.664 (0.633, 0.692) | 0.685 (0.565, 0.770) | 0.561 (0.438, 0.664) | 0.357 (0.257, 0.465) | 0.194 (0.132, 0.276) | 0.351 (0.238, 0.473) | 33.750 (27.083, 41.458) |
| A | fetal ECG | SimCLR ResNet-34 (1-D), multi-layer 1088-d | 0.685 (0.597, 0.769) | 0.299 (0.220, 0.380) | 0.981 (0.827, 1.158) | 1.099 (0.754, 1.681) | 0.197 (0.171, 0.224) | 0.295 (0.261, 0.331) | 0.494 (0.451, 0.535) | 0.695 (0.657, 0.728) | 0.787 (0.713, 0.843) | 0.684 (0.594, 0.759) | 0.484 (0.389, 0.577) | 0.288 (0.216, 0.374) | 0.477 (0.349, 0.599) | 46.042 (37.292, 55.208) |
| A | abdominal ECG | SimCLR ResNet-34 (1-D), multi-layer 1088-d | 0.699 (0.612, 0.779) | 0.287 (0.216, 0.365) | 0.983 (0.838, 1.150) | 1.064 (0.736, 1.642) | 0.197 (0.173, 0.223) | 0.296 (0.264, 0.329) | 0.495 (0.455, 0.534) | 0.696 (0.661, 0.727) | 0.790 (0.714, 0.847) | 0.688 (0.594, 0.764) | 0.487 (0.390, 0.582) | 0.290 (0.218, 0.378) | 0.480 (0.357, 0.606) | 44.167 (35.833, 52.922) |
| A | abdominal ECG | HeartWise DeepECG-SSL (wav2vec2), 768-d | 0.688 (0.603, 0.774) | 0.335 (0.253, 0.422) | 1.035 (0.866, 1.244) | 0.932 (0.641, 1.378) | 0.205 (0.178, 0.237) | 0.307 (0.270, 0.347) | 0.508 (0.463, 0.553) | 0.706 (0.667, 0.742) | 0.812 (0.746, 0.863) | 0.716 (0.634, 0.787) | 0.521 (0.428, 0.614) | 0.319 (0.244, 0.408) | 0.514 (0.389, 0.638) | 48.542 (39.167, 57.708) |
| A | maternal ECG | ECGFounder 12-lead pathway (published weights, 507/509 tensors) | 0.751 (0.678, 0.822) | 0.394 (0.302, 0.484) | 1.249 (1.048, 1.525) | 0.643 (0.431, 0.949) | 0.237 (0.207, 0.275) | 0.348 (0.308, 0.393) | 0.554 (0.509, 0.601) | 0.743 (0.707, 0.778) | 0.863 (0.811, 0.903) | 0.786 (0.715, 0.846) | 0.614 (0.522, 0.705) | 0.408 (0.322, 0.511) | 0.608 (0.487, 0.719) | 48.125 (39.167, 56.672) |

| Set | Signal | Representation | sens | spec | LR+ | LR− | PPV@20 | PPV@30 | PPV@50 | PPV@70 | NPV@20 | NPV@30 | NPV@50 | NPV@70 | NPV (obs) | ruled out /150 |
|---|---|---|---|---|---|---|---|---|---|---|---|---|---|---|---|---|
| A | maternal ECG | ECGFounder 1-lead pathway (published weights, 507/509 tensors) | 0.679 (0.600, 0.760) | 0.403 (0.308, 0.494) | 1.138 (0.930, 1.407) | 0.793 (0.561, 1.149) | 0.221 (0.188, 0.260) | 0.328 (0.284, 0.375) | 0.532 (0.481, 0.583) | 0.726 (0.683, 0.765) | 0.835 (0.778, 0.877) | 0.747 (0.672, 0.807) | 0.559 (0.468, 0.644) | 0.353 (0.275, 0.439) | 0.552 (0.432, 0.666) | 54.167 (45.000, 63.333) |
| A | maternal ECG | **CONTROL: untrained encoder (weights never loaded, 0/509 tensors)** | 0.762 (0.681, 0.838) | 0.324 (0.235, 0.414) | 1.129 (0.965, 1.348) | 0.744 (0.475, 1.152) | 0.220 (0.194, 0.252) | 0.326 (0.292, 0.366) | 0.529 (0.489, 0.573) | 0.724 (0.691, 0.758) | 0.844 (0.780, 0.895) | 0.761 (0.677, 0.834) | 0.580 (0.477, 0.687) | 0.375 (0.284, 0.491) | 0.573 (0.438, 0.703) | 42.083 (33.333, 51.458) |
| B | maternal ECG | SimCLR ResNet-34 (1-D), projection head only 128-d (cols 960:1088) | 0.699 (0.613, 0.781) | 0.315 (0.231, 0.403) | 1.024 (0.867, 1.224) | 0.971 (0.661, 1.475) | 0.204 (0.178, 0.234) | 0.304 (0.270, 0.343) | 0.505 (0.462, 0.548) | 0.704 (0.667, 0.738) | 0.806 (0.736, 0.860) | 0.709 (0.621, 0.783) | 0.513 (0.416, 0.610) | 0.313 (0.236, 0.408) | 0.506 (0.383, 0.639) | 46.250 (37.708, 55.833) |
| B | fetal ECG | SimCLR ResNet-34 (1-D), projection head only 128-d (cols 960:1088) | 0.696 (0.611, 0.778) | 0.321 (0.247, 0.400) | 1.030 (0.871, 1.234) | 0.964 (0.661, 1.435) | 0.205 (0.179, 0.235) | 0.306 (0.271, 0.345) | 0.507 (0.465, 0.551) | 0.706 (0.669, 0.740) | 0.806 (0.737, 0.859) | 0.708 (0.622, 0.781) | 0.511 (0.417, 0.606) | 0.309 (0.236, 0.399) | 0.504 (0.379, 0.627) | 46.875 (38.542, 55.833) |
| B | abdominal ECG | SimCLR ResNet-34 (1-D), projection head only 128-d (cols 960:1088) | 0.693 (0.612, 0.774) | 0.268 (0.186, 0.354) | 0.946 (0.808, 1.119) | 1.147 (0.771, 1.790) | 0.191 (0.168, 0.219) | 0.288 (0.257, 0.324) | 0.486 (0.446, 0.527) | 0.688 (0.653, 0.722) | 0.778 (0.693, 0.839) | 0.671 (0.571, 0.754) | 0.467 (0.366, 0.569) | 0.273 (0.200, 0.363) | 0.460 (0.334, 0.591) | 43.125 (34.167, 52.083) |
| C | maternal ECG | HeartWise DeepECG-SSL, participant-mean | 0.756 (0.679, 0.832) | 0.358 (0.269, 0.452) | 1.197 (1.002, 1.456) | 0.724 (0.464, 1.146) | 0.229 (0.199, 0.265) | 0.337 (0.299, 0.381) | 0.542 (0.497, 0.588) | 0.733 (0.696, 0.767) | 0.849 (0.784, 0.897) | 0.768 (0.684, 0.837) | 0.592 (0.489, 0.693) | 0.389 (0.297, 0.500) | 0.585 (0.454, 0.709) | 45.000 (36.250, 53.958) |
| C | maternal ECG | SimCLR ResNet-34 multi-layer, participant-mean | 0.655 (0.573, 0.737) | 0.397 (0.306, 0.487) | 1.090 (0.906, 1.311) | 0.876 (0.640, 1.240) | 0.214 (0.184, 0.246) | 0.318 (0.279, 0.359) | 0.520 (0.473, 0.565) | 0.716 (0.676, 0.752) | 0.821 (0.766, 0.863) | 0.729 (0.658, 0.787) | 0.537 (0.456, 0.616) | 0.333 (0.267, 0.411) | 0.530 (0.410, 0.643) | 55.625 (46.250, 65.208) |
| C | maternal ECG | **CONTROL: HeartWise architecture, random init, participant-mean** | 0.721 (0.636, 0.803) | 0.304 (0.223, 0.388) | 1.044 (0.881, 1.244) | 0.953 (0.632, 1.492) | 0.207 (0.180, 0.236) | 0.309 (0.274, 0.346) | 0.510 (0.467, 0.552) | 0.708 (0.671, 0.742) | 0.809 (0.734, 0.865) | 0.712 (0.620, 0.789) | 0.517 (0.414, 0.619) | 0.316 (0.235, 0.414) | 0.510 (0.379, 0.639) | 43.750 (34.792, 52.708) |
| C | maternal ECG | concat ECGFounder 12-lead + DeepECG-SSL (1792-d) | 0.696 (0.616, 0.775) | 0.420 (0.332, 0.513) | 1.204 (0.992, 1.482) | 0.732 (0.514, 1.044) | 0.231 (0.198, 0.270) | 0.339 (0.297, 0.387) | 0.545 (0.496, 0.594) | 0.736 (0.696, 0.773) | 0.846 (0.795, 0.887) | 0.763 (0.695, 0.821) | 0.582 (0.497, 0.667) | 0.376 (0.300, 0.467) | 0.575 (0.457, 0.685) | 54.167 (45.208, 63.125) |
| C | maternal ECG | late mean: ECGFounder 12-lead + DeepECG-SSL | 0.732 (0.659, 0.806) | 0.380 (0.289, 0.473) | 1.193 (0.999, 1.451) | 0.734 (0.500, 1.095) | 0.229 (0.199, 0.265) | 0.337 (0.299, 0.381) | 0.542 (0.498, 0.588) | 0.733 (0.697, 0.768) | 0.847 (0.790, 0.890) | 0.764 (0.690, 0.826) | 0.585 (0.494, 0.674) | 0.381 (0.300, 0.477) | 0.578 (0.455, 0.693) | 48.542 (39.792, 57.500) |
| C | maternal ECG | stacked: ECGFounder 12-lead + DeepECG-SSL | 0.756 (0.703, 0.810) | 0.296 (0.233, 0.360) | 1.081 (0.962, 1.232) | 0.840 (0.618, 1.224) | 0.212 (0.194, 0.234) | 0.316 (0.291, 0.343) | 0.518 (0.489, 0.548) | 0.715 (0.690, 0.738) | 0.827 (0.769, 0.868) | 0.737 (0.663, 0.795) | 0.547 (0.463, 0.627) | 0.342 (0.273, 0.426) | 0.540 (0.424, 0.652) | 40.417 (34.167, 46.667) |

| Set | Signal | Representation | sens | spec | LR+ | LR− | PPV@20 | PPV@30 | PPV@50 | PPV@70 | NPV@20 | NPV@30 | NPV@50 | NPV@70 | NPV (obs) | ruled out /150 |
|---|---|---|---|---|---|---|---|---|---|---|---|---|---|---|---|---|
| C | maternal ECG | late mean: ECGFounder 12-lead + DeepECG-SSL + SimCLR | 0.718 (0.635, 0.803) | 0.411 (0.319, 0.503) | 1.220 (1.014, 1.503) | 0.688 (0.465, 1.000) | 0.233 (0.202, 0.273) | 0.343 (0.302, 0.391) | 0.549 (0.502, 0.599) | 0.739 (0.701, 0.777) | 0.854 (0.802, 0.897) | 0.774 (0.704, 0.836) | 0.596 (0.507, 0.689) | 0.390 (0.309, 0.493) | 0.590 (0.464, 0.705) | 51.875 (42.083, 61.255) |
| C | maternal ECG | **CONTROL: late mean ECGFounder 12-lead + random-init HeartWise** | 0.734 (0.652, 0.816) | 0.375 (0.283, 0.467) | 1.183 (0.981, 1.447) | 0.733 (0.479, 1.134) | 0.228 (0.196, 0.264) | 0.335 (0.295, 0.380) | 0.540 (0.493, 0.588) | 0.732 (0.693, 0.768) | 0.847 (0.784, 0.894) | 0.765 (0.682, 0.833) | 0.586 (0.484, 0.685) | 0.381 (0.290, 0.489) | 0.579 (0.452, 0.700) | 47.917 (38.958, 56.875) |

## Operating point: target sensitivity 0.75

*FELICITy 1, n = 144 women (73 positive, observed prevalence 51%); target PSS-10 ≥ 19 (binary); unit of analysis: participant; operating point target sensitivity 0.75.*

| Set | Signal | Representation | sens | spec | LR+ | LR− | PPV@20 | PPV@30 | PPV@50 | PPV@70 | NPV@20 | NPV@30 | NPV@50 | NPV@70 | NPV (obs) | ruled out /150 |
|---|---|---|---|---|---|---|---|---|---|---|---|---|---|---|---|---|
| A | maternal ECG | SimCLR ResNet-34 (1-D), multi-layer 1088-d | 0.762 (0.687, 0.830) | 0.245 (0.172, 0.325) | 1.016 (0.882, 1.175) | 1.043 (0.660, 1.773) | 0.202 (0.180, 0.227) | 0.303 (0.274, 0.334) | 0.503 (0.467, 0.539) | 0.702 (0.671, 0.731) | 0.797 (0.709, 0.860) | 0.698 (0.593, 0.784) | 0.502 (0.395, 0.613) | 0.306 (0.223, 0.410) | 0.495 (0.357, 0.629) | 36.250 (28.542, 44.167) |
| A | maternal ECG | HeartWise DeepECG-SSL (wav2vec2), 768-d | 0.797 (0.730, 0.865) | 0.287 (0.202, 0.377) | 1.121 (0.972, 1.317) | 0.710 (0.439, 1.146) | 0.219 (0.195, 0.247) | 0.324 (0.294, 0.360) | 0.528 (0.492, 0.567) | 0.723 (0.693, 0.753) | 0.849 (0.780, 0.902) | 0.767 (0.675, 0.843) | 0.585 (0.474, 0.700) | 0.377 (0.280, 0.504) | 0.579 (0.440, 0.713) | 36.667 (28.750, 44.797) |
| A | maternal ECG | 100-feature HRV panel (95 usable; refined ensemble R-peaks) | 0.748 (0.675, 0.816) | 0.248 (0.169, 0.332) | 0.995 (0.860, 1.155) | 1.027 (0.670, 1.689) | 0.199 (0.177, 0.224) | 0.299 (0.269, 0.331) | 0.498 (0.461, 0.535) | 0.698 (0.666, 0.728) | 0.797 (0.708, 0.857) | 0.697 (0.588, 0.779) | 0.499 (0.383, 0.605) | 0.301 (0.213, 0.400) | 0.492 (0.361, 0.627) | 37.500 (29.583, 45.833) |
| A | maternal ECG | hrv-omnibus panel, 32 features (refined ensemble R-peaks) | 0.759 (0.678, 0.836) | 0.231 (0.156, 0.308) | 0.987 (0.850, 1.141) | 1.049 (0.661, 1.780) | 0.198 (0.175, 0.222) | 0.297 (0.267, 0.328) | 0.496 (0.459, 0.532) | 0.697 (0.664, 0.726) | 0.793 (0.698, 0.860) | 0.692 (0.576, 0.783) | 0.492 (0.372, 0.610) | 0.295 (0.204, 0.406) | 0.485 (0.340, 0.626) | 35.417 (27.500, 43.130) |
| A | maternal ECG | conventional HRV, 10 features (single NeuroKit detector) | 0.751 (0.666, 0.823) | 0.115 (0.071, 0.166) | 0.850 (0.747, 0.954) | 2.318 (1.424, 5.592) | 0.175 (0.157, 0.192) | 0.267 (0.242, 0.290) | 0.459 (0.427, 0.488) | 0.664 (0.634, 0.689) | 0.641 (0.482, 0.747) | 0.513 (0.363, 0.637) | 0.315 (0.203, 0.437) | 0.167 (0.101, 0.254) | 0.310 (0.191, 0.441) | 27.500 (20.833, 35.208) |
| A | fetal ECG | SimCLR ResNet-34 (1-D), multi-layer 1088-d | 0.748 (0.662, 0.819) | 0.223 (0.159, 0.288) | 0.968 (0.835, 1.111) | 1.349 (0.835, 2.809) | 0.194 (0.172, 0.217) | 0.292 (0.263, 0.321) | 0.490 (0.453, 0.524) | 0.691 (0.659, 0.719) | 0.760 (0.654, 0.833) | 0.655 (0.541, 0.748) | 0.460 (0.352, 0.574) | 0.275 (0.194, 0.378) | 0.454 (0.317, 0.589) | 35.625 (28.125, 43.750) |
| A | abdominal ECG | SimCLR ResNet-34 (1-D), multi-layer 1088-d | 0.767 (0.684, 0.843) | 0.225 (0.164, 0.294) | 0.994 (0.862, 1.132) | 1.071 (0.680, 1.801) | 0.199 (0.177, 0.220) | 0.298 (0.270, 0.326) | 0.498 (0.463, 0.530) | 0.698 (0.667, 0.724) | 0.790 (0.699, 0.856) | 0.687 (0.580, 0.777) | 0.487 (0.377, 0.604) | 0.290 (0.208, 0.400) | 0.480 (0.350, 0.618) | 34.375 (26.667, 42.292) |
| A | abdominal ECG | HeartWise DeepECG-SSL (wav2vec2), 768-d | 0.745 (0.666, 0.823) | 0.279 (0.205, 0.357) | 1.036 (0.893, 1.215) | 0.927 (0.596, 1.475) | 0.206 (0.182, 0.233) | 0.307 (0.276, 0.342) | 0.508 (0.470, 0.547) | 0.707 (0.674, 0.738) | 0.813 (0.737, 0.872) | 0.718 (0.622, 0.799) | 0.524 (0.420, 0.634) | 0.322 (0.240, 0.434) | 0.517 (0.381, 0.655) | 40.000 (31.661, 48.333) |

| Set | Signal | Representation | sens | spec | LR+ | LR− | PPV@20 | PPV@30 | PPV@50 | PPV@70 | NPV@20 | NPV@30 | NPV@50 | NPV@70 | NPV (obs) | ruled out /150 |
|---|---|---|---|---|---|---|---|---|---|---|---|---|---|---|---|---|
| A | maternal ECG | ECGFounder 12-lead pathway (published weights, 507/509 tensors) | 0.819 (0.753, 0.881) | 0.344 (0.254, 0.431) | 1.254 (1.072, 1.485) | 0.535 (0.333, 0.853) | 0.238 (0.211, 0.270) | 0.349 (0.314, 0.388) | 0.555 (0.516, 0.596) | 0.744 (0.713, 0.774) | 0.883 (0.827, 0.924) | 0.815 (0.738, 0.877) | 0.656 (0.550, 0.757) | 0.452 (0.348, 0.577) | 0.649 (0.521, 0.765) | 39.167 (30.625, 47.708) |
| A | maternal ECG | ECGFounder 1-lead pathway (published weights, 507/509 tensors) | 0.756 (0.679, 0.830) | 0.346 (0.259, 0.436) | 1.160 (0.980, 1.386) | 0.708 (0.463, 1.092) | 0.225 (0.197, 0.257) | 0.332 (0.295, 0.372) | 0.537 (0.494, 0.580) | 0.730 (0.695, 0.763) | 0.850 (0.787, 0.897) | 0.768 (0.684, 0.836) | 0.587 (0.484, 0.687) | 0.379 (0.289, 0.488) | 0.580 (0.451, 0.702) | 44.167 (35.625, 52.917) |
| A | maternal ECG | **CONTROL: untrained encoder (weights never loaded, 0/509 tensors)** | 0.805 (0.727, 0.879) | 0.245 (0.167, 0.323) | 1.068 (0.937, 1.230) | 0.798 (0.470, 1.322) | 0.211 (0.190, 0.235) | 0.314 (0.286, 0.345) | 0.516 (0.483, 0.551) | 0.713 (0.685, 0.741) | 0.835 (0.755, 0.896) | 0.747 (0.646, 0.835) | 0.561 (0.443, 0.689) | 0.357 (0.258, 0.494) | 0.555 (0.408, 0.705) | 32.917 (24.995, 41.458) |
| B | maternal ECG | SimCLR ResNet-34 (1-D), projection head only 128-d (cols 960:1088) | 0.748 (0.665, 0.824) | 0.268 (0.188, 0.348) | 1.026 (0.884, 1.194) | 0.970 (0.642, 1.582) | 0.204 (0.181, 0.229) | 0.305 (0.274, 0.338) | 0.506 (0.468, 0.543) | 0.704 (0.672, 0.734) | 0.806 (0.723, 0.863) | 0.709 (0.606, 0.787) | 0.513 (0.404, 0.617) | 0.313 (0.229, 0.413) | 0.506 (0.376, 0.642) | 38.958 (30.833, 47.917) |
| B | fetal ECG | SimCLR ResNet-34 (1-D), projection head only 128-d (cols 960:1088) | 0.745 (0.665, 0.822) | 0.223 (0.156, 0.294) | 0.960 (0.837, 1.108) | 1.181 (0.761, 1.986) | 0.194 (0.173, 0.217) | 0.291 (0.264, 0.321) | 0.490 (0.455, 0.525) | 0.691 (0.661, 0.720) | 0.773 (0.678, 0.841) | 0.666 (0.556, 0.757) | 0.462 (0.351, 0.574) | 0.270 (0.192, 0.370) | 0.455 (0.326, 0.592) | 35.833 (28.125, 44.167) |
| B | abdominal ECG | SimCLR ResNet-34 (1-D), projection head only 128-d (cols 960:1088) | 0.745 (0.667, 0.824) | 0.237 (0.158, 0.320) | 0.977 (0.842, 1.138) | 1.084 (0.681, 1.792) | 0.196 (0.174, 0.221) | 0.295 (0.265, 0.327) | 0.494 (0.457, 0.531) | 0.695 (0.662, 0.726) | 0.787 (0.695, 0.856) | 0.684 (0.572, 0.776) | 0.482 (0.366, 0.600) | 0.286 (0.200, 0.394) | 0.475 (0.337, 0.617) | 36.875 (28.333, 45.417) |
| C | maternal ECG | HeartWise DeepECG-SSL, participant-mean | 0.786 (0.708, 0.859) | 0.310 (0.227, 0.400) | 1.163 (0.992, 1.388) | 0.756 (0.463, 1.276) | 0.224 (0.197, 0.255) | 0.330 (0.296, 0.369) | 0.534 (0.493, 0.575) | 0.727 (0.693, 0.759) | 0.844 (0.770, 0.898) | 0.762 (0.667, 0.839) | 0.585 (0.473, 0.697) | 0.385 (0.287, 0.504) | 0.579 (0.441, 0.710) | 39.167 (30.625, 47.708) |
| C | maternal ECG | SimCLR ResNet-34 multi-layer, participant-mean | 0.715 (0.641, 0.789) | 0.338 (0.252, 0.425) | 1.081 (0.924, 1.281) | 0.844 (0.589, 1.243) | 0.213 (0.187, 0.242) | 0.316 (0.283, 0.354) | 0.519 (0.479, 0.561) | 0.716 (0.682, 0.748) | 0.826 (0.765, 0.872) | 0.735 (0.657, 0.800) | 0.544 (0.452, 0.633) | 0.339 (0.263, 0.429) | 0.537 (0.408, 0.657) | 46.667 (37.917, 55.422) |
| C | maternal ECG | **CONTROL: HeartWise architecture, random init, participant-mean** | 0.770 (0.696, 0.844) | 0.223 (0.154, 0.300) | 0.994 (0.869, 1.147) | 1.085 (0.669, 1.899) | 0.199 (0.178, 0.222) | 0.299 (0.271, 0.329) | 0.498 (0.464, 0.533) | 0.698 (0.668, 0.726) | 0.789 (0.690, 0.859) | 0.687 (0.571, 0.782) | 0.487 (0.368, 0.609) | 0.292 (0.203, 0.407) | 0.480 (0.338, 0.627) | 33.958 (26.042, 42.083) |
| C | maternal ECG | concat ECGFounder 12-lead + DeepECG-SSL (1792-d) | 0.767 (0.694, 0.841) | 0.344 (0.258, 0.434) | 1.172 (0.995, 1.407) | 0.685 (0.441, 1.051) | 0.226 (0.199, 0.260) | 0.334 (0.298, 0.375) | 0.539 (0.498, 0.583) | 0.731 (0.698, 0.765) | 0.855 (0.794, 0.902) | 0.775 (0.694, 0.843) | 0.598 (0.496, 0.701) | 0.391 (0.299, 0.505) | 0.591 (0.459, 0.713) | 43.125 (34.792, 51.667) |

| Set | Signal | Representation | sens | spec | LR+ | LR− | PPV@20 | PPV@30 | PPV@50 | PPV@70 | NPV@20 | NPV@30 | NPV@50 | NPV@70 | NPV (obs) | ruled out /150 |
|---|---|---|---|---|---|---|---|---|---|---|---|---|---|---|---|---|
| C | maternal ECG | late mean: ECGFounder 12-lead + DeepECG-SSL | 0.784 (0.717, 0.847) | 0.338 (0.254, 0.425) | 1.198 (1.029, 1.424) | 0.684 (0.451, 1.060) | 0.230 (0.204, 0.261) | 0.338 (0.305, 0.377) | 0.543 (0.505, 0.583) | 0.734 (0.704, 0.765) | 0.856 (0.798, 0.900) | 0.778 (0.701, 0.841) | 0.605 (0.509, 0.700) | 0.402 (0.316, 0.510) | 0.599 (0.471, 0.717) | 41.458 (33.328, 49.792) |
| C | maternal ECG | stacked: ECGFounder 12-lead + DeepECG-SSL | 0.808 (0.762, 0.855) | 0.242 (0.186, 0.300) | 1.078 (0.978, 1.206) | 0.856 (0.593, 1.519) | 0.212 (0.196, 0.231) | 0.315 (0.295, 0.339) | 0.518 (0.493, 0.543) | 0.714 (0.694, 0.734) | 0.826 (0.750, 0.872) | 0.735 (0.643, 0.802) | 0.547 (0.450, 0.640) | 0.344 (0.267, 0.441) | 0.540 (0.417, 0.661) | 32.500 (26.875, 37.917) |
| C | maternal ECG | late mean: ECGFounder 12-lead + DeepECG-SSL + SimCLR | 0.767 (0.693, 0.841) | 0.349 (0.257, 0.447) | 1.180 (0.998, 1.417) | 0.669 (0.436, 1.026) | 0.228 (0.199, 0.261) | 0.336 (0.299, 0.377) | 0.541 (0.499, 0.585) | 0.733 (0.699, 0.767) | 0.857 (0.798, 0.902) | 0.779 (0.699, 0.844) | 0.603 (0.502, 0.702) | 0.396 (0.305, 0.509) | 0.596 (0.463, 0.719) | 43.542 (34.167, 52.505) |
| C | maternal ECG | **CONTROL: late mean ECGFounder 12-lead + random-init HeartWise** | 0.784 (0.709, 0.858) | 0.313 (0.225, 0.402) | 1.148 (0.978, 1.364) | 0.723 (0.446, 1.178) | 0.222 (0.196, 0.253) | 0.329 (0.294, 0.367) | 0.533 (0.492, 0.574) | 0.726 (0.692, 0.758) | 0.849 (0.778, 0.901) | 0.768 (0.676, 0.842) | 0.591 (0.479, 0.699) | 0.388 (0.290, 0.507) | 0.585 (0.449, 0.713) | 39.583 (31.250, 48.542) |

## Operating point: target sensitivity 0.80

*FELICITy 1, n = 144 women (73 positive, observed prevalence 51%); target PSS-10 ≥ 19 (binary); unit of analysis: participant; operating point target sensitivity 0.80.*

| Set | Signal | Representation | sens | spec | LR+ | LR− | PPV@20 | PPV@30 | PPV@50 | PPV@70 | NPV@20 | NPV@30 | NPV@50 | NPV@70 | NPV (obs) | ruled out /150 |
|---|---|---|---|---|---|---|---|---|---|---|---|---|---|---|---|---|
| A | maternal ECG | SimCLR ResNet-34 (1-D), multi-layer 1088-d | 0.811 (0.745, 0.872) | 0.175 (0.108, 0.248) | 0.984 (0.878, 1.107) | 1.123 (0.674, 2.162) | 0.197 (0.180, 0.217) | 0.296 (0.273, 0.321) | 0.496 (0.467, 0.525) | 0.696 (0.671, 0.720) | 0.783 (0.666, 0.859) | 0.680 (0.545, 0.782) | 0.479 (0.347, 0.610) | 0.285 (0.189, 0.410) | 0.472 (0.320, 0.622) | 27.292 (20.208, 34.583) |
| A | maternal ECG | HeartWise DeepECG-SSL (wav2vec2), 768-d | 0.838 (0.779, 0.893) | 0.223 (0.146, 0.306) | 1.079 (0.957, 1.229) | 0.729 (0.443, 1.298) | 0.212 (0.193, 0.235) | 0.316 (0.291, 0.345) | 0.519 (0.489, 0.551) | 0.716 (0.690, 0.741) | 0.846 (0.759, 0.901) | 0.763 (0.649, 0.842) | 0.581 (0.445, 0.701) | 0.373 (0.259, 0.507) | 0.574 (0.421, 0.715) | 28.750 (21.667, 36.667) |
| A | maternal ECG | 100-feature HRV panel (95 usable; refined ensemble R-peaks) | 0.811 (0.746, 0.869) | 0.220 (0.147, 0.297) | 1.042 (0.924, 1.187) | 0.898 (0.559, 1.564) | 0.206 (0.187, 0.228) | 0.308 (0.283, 0.336) | 0.509 (0.478, 0.541) | 0.707 (0.680, 0.732) | 0.823 (0.738, 0.881) | 0.736 (0.631, 0.816) | 0.560 (0.445, 0.667) | 0.371 (0.269, 0.488) | 0.554 (0.420, 0.683) | 30.625 (23.745, 38.125) |
| A | maternal ECG | hrv-omnibus panel, 32 features (refined ensemble R-peaks) | 0.816 (0.744, 0.882) | 0.186 (0.118, 0.255) | 1.003 (0.884, 1.133) | 0.993 (0.593, 1.884) | 0.200 (0.181, 0.221) | 0.301 (0.275, 0.327) | 0.501 (0.469, 0.531) | 0.700 (0.673, 0.725) | 0.802 (0.689, 0.872) | 0.703 (0.566, 0.800) | 0.506 (0.363, 0.635) | 0.306 (0.200, 0.432) | 0.499 (0.340, 0.652) | 27.708 (20.625, 35.000) |
| A | maternal ECG | conventional HRV, 10 features (single NeuroKit detector) | 0.830 (0.757, 0.891) | 0.079 (0.043, 0.121) | 0.901 (0.814, 0.986) | 2.237 (1.231, 5.697) | 0.184 (0.169, 0.198) | 0.279 (0.258, 0.297) | 0.474 (0.448, 0.496) | 0.678 (0.655, 0.697) | 0.644 (0.437, 0.769) | 0.515 (0.326, 0.664) | 0.314 (0.178, 0.464) | 0.165 (0.086, 0.279) | 0.308 (0.167, 0.476) | 18.750 (13.328, 24.797) |

| Set | Signal | Representation | sens | spec | LR+ | LR− | PPV@20 | PPV@30 | PPV@50 | PPV@70 | NPV@20 | NPV@30 | NPV@50 | NPV@70 | NPV (obs) | ruled out /150 |
|---|---|---|---|---|---|---|---|---|---|---|---|---|---|---|---|---|
| A | fetal ECG | SimCLR ResNet-34 (1-D), multi-layer 1088-d | 0.811 (0.734, 0.873) | 0.152 (0.102, 0.209) | 0.960 (0.862, 1.071) | 1.454 (0.833, 3.565) | 0.193 (0.177, 0.211) | 0.291 (0.269, 0.313) | 0.489 (0.461, 0.515) | 0.690 (0.666, 0.712) | 0.748 (0.628, 0.833) | 0.641 (0.514, 0.749) | 0.446 (0.328, 0.577) | 0.265 (0.178, 0.389) | 0.439 (0.298, 0.585) | 25.625 (19.375, 32.708) |
| A | abdominal ECG | SimCLR ResNet-34 (1-D), multi-layer 1088-d | 0.808 (0.732, 0.878) | 0.152 (0.101, 0.208) | 0.955 (0.853, 1.064) | 1.348 (0.799, 2.655) | 0.193 (0.176, 0.210) | 0.290 (0.267, 0.313) | 0.488 (0.460, 0.515) | 0.690 (0.665, 0.712) | 0.751 (0.628, 0.836) | 0.640 (0.503, 0.751) | 0.436 (0.313, 0.571) | 0.252 (0.167, 0.370) | 0.429 (0.289, 0.586) | 25.833 (18.953, 33.125) |
| A | abdominal ECG | HeartWise DeepECG-SSL (wav2vec2), 768-d | 0.808 (0.729, 0.879) | 0.220 (0.151, 0.295) | 1.039 (0.911, 1.190) | 0.901 (0.529, 1.595) | 0.206 (0.185, 0.229) | 0.308 (0.281, 0.337) | 0.509 (0.476, 0.543) | 0.707 (0.679, 0.734) | 0.818 (0.724, 0.885) | 0.725 (0.608, 0.819) | 0.534 (0.406, 0.662) | 0.333 (0.233, 0.463) | 0.527 (0.378, 0.678) | 30.833 (23.542, 38.958) |
| A | maternal ECG | ECGFounder 12-lead pathway (published weights, 507/509 tensors) | 0.849 (0.787, 0.906) | 0.318 (0.231, 0.406) | 1.251 (1.084, 1.463) | 0.482 (0.284, 0.795) | 0.238 (0.213, 0.267) | 0.349 (0.317, 0.384) | 0.555 (0.519, 0.592) | 0.744 (0.716, 0.772) | 0.893 (0.837, 0.934) | 0.830 (0.751, 0.892) | 0.679 (0.568, 0.783) | 0.478 (0.364, 0.614) | 0.673 (0.542, 0.794) | 35.000 (26.875, 43.125) |
| A | maternal ECG | ECGFounder 1-lead pathway (published weights, 507/509 tensors) | 0.830 (0.770, 0.887) | 0.279 (0.195, 0.365) | 1.151 (1.006, 1.333) | 0.605 (0.374, 0.990) | 0.223 (0.201, 0.250) | 0.330 (0.301, 0.363) | 0.535 (0.501, 0.570) | 0.729 (0.700, 0.756) | 0.869 (0.804, 0.915) | 0.796 (0.707, 0.864) | 0.627 (0.511, 0.736) | 0.422 (0.314, 0.552) | 0.621 (0.475, 0.743) | 33.542 (26.036, 41.255) |
| A | maternal ECG | **CONTROL: untrained encoder (weights never loaded, 0/509 tensors)** | 0.836 (0.759, 0.903) | 0.189 (0.119, 0.268) | 1.030 (0.910, 1.175) | 0.859 (0.450, 1.659) | 0.205 (0.185, 0.227) | 0.306 (0.280, 0.335) | 0.507 (0.476, 0.540) | 0.706 (0.679, 0.732) | 0.824 (0.712, 0.900) | 0.732 (0.594, 0.842) | 0.541 (0.388, 0.700) | 0.337 (0.215, 0.510) | 0.534 (0.362, 0.709) | 26.458 (18.958, 34.375) |
| B | maternal ECG | SimCLR ResNet-34 (1-D), projection head only 128-d (cols 960:1088) | 0.795 (0.716, 0.863) | 0.189 (0.128, 0.254) | 0.985 (0.875, 1.110) | 1.204 (0.743, 2.289) | 0.197 (0.179, 0.217) | 0.296 (0.272, 0.322) | 0.495 (0.465, 0.524) | 0.696 (0.670, 0.720) | 0.774 (0.665, 0.848) | 0.669 (0.545, 0.767) | 0.471 (0.352, 0.592) | 0.281 (0.197, 0.393) | 0.465 (0.325, 0.605) | 29.583 (22.500, 37.292) |
| B | fetal ECG | SimCLR ResNet-34 (1-D), projection head only 128-d (cols 960:1088) | 0.795 (0.721, 0.865) | 0.135 (0.083, 0.192) | 0.922 (0.825, 1.035) | 1.994 (1.054, 4.925) | 0.187 (0.171, 0.205) | 0.283 (0.261, 0.307) | 0.479 (0.451, 0.508) | 0.682 (0.657, 0.706) | 0.693 (0.535, 0.799) | 0.577 (0.424, 0.703) | 0.381 (0.257, 0.518) | 0.215 (0.135, 0.326) | 0.375 (0.237, 0.524) | 25.625 (18.750, 32.500) |
| B | abdominal ECG | SimCLR ResNet-34 (1-D), projection head only 128-d (cols 960:1088) | 0.816 (0.748, 0.884) | 0.189 (0.119, 0.263) | 1.008 (0.895, 1.143) | 0.994 (0.578, 1.804) | 0.201 (0.183, 0.222) | 0.301 (0.277, 0.329) | 0.502 (0.472, 0.533) | 0.701 (0.676, 0.727) | 0.802 (0.695, 0.874) | 0.703 (0.575, 0.804) | 0.505 (0.372, 0.640) | 0.305 (0.204, 0.437) | 0.498 (0.347, 0.653) | 27.917 (20.208, 35.417) |
| C | maternal ECG | HeartWise DeepECG-SSL, participant-mean | 0.830 (0.756, 0.897) | 0.245 (0.169, 0.325) | 1.110 (0.968, 1.287) | 0.749 (0.416, 1.376) | 0.217 (0.194, 0.242) | 0.322 (0.292, 0.353) | 0.525 (0.490, 0.559) | 0.720 (0.691, 0.747) | 0.844 (0.753, 0.907) | 0.762 (0.645, 0.852) | 0.583 (0.449, 0.716) | 0.380 (0.265, 0.527) | 0.576 (0.425, 0.729) | 31.042 (23.333, 39.167) |
| C | maternal ECG | SimCLR ResNet-34 multi-layer, participant-mean | 0.786 (0.720, 0.849) | 0.301 (0.218, 0.387) | 1.126 (0.982, 1.310) | 0.709 (0.468, 1.093) | 0.220 (0.197, 0.246) | 0.325 (0.296, 0.359) | 0.529 (0.495, 0.566) | 0.724 (0.695, 0.752) | 0.850 (0.786, 0.896) | 0.768 (0.684, 0.835) | 0.587 (0.485, 0.688) | 0.380 (0.290, 0.489) | 0.580 (0.445, 0.704) | 38.542 (30.417, 46.875) |

| Set | Signal | Representation | sens | spec | LR+ | LR− | PPV@20 | PPV@30 | PPV@50 | PPV@70 | NPV@20 | NPV@30 | NPV@50 | NPV@70 | NPV (obs) | ruled out /150 |
|---|---|---|---|---|---|---|---|---|---|---|---|---|---|---|---|---|
| C | maternal ECG | **CONTROL: HeartWise architecture, random init, participant-mean** | 0.836 (0.765, 0.903) | 0.172 (0.110, 0.238) | 1.010 (0.904, 1.136) | 0.984 (0.545, 1.968) | 0.202 (0.184, 0.221) | 0.302 (0.279, 0.327) | 0.502 (0.475, 0.531) | 0.702 (0.678, 0.725) | 0.803 (0.686, 0.881) | 0.704 (0.565, 0.814) | 0.506 (0.367, 0.656) | 0.306 (0.205, 0.454) | 0.499 (0.338, 0.667) | 25.208 (18.542, 32.500) |
| C | maternal ECG | concat ECGFounder 12-lead + DeepECG-SSL (1792-d) | 0.816 (0.751, 0.881) | 0.270 (0.189, 0.358) | 1.121 (0.981, 1.307) | 0.685 (0.420, 1.123) | 0.219 (0.197, 0.246) | 0.324 (0.295, 0.358) | 0.528 (0.494, 0.565) | 0.723 (0.694, 0.752) | 0.855 (0.785, 0.906) | 0.776 (0.683, 0.851) | 0.600 (0.485, 0.714) | 0.395 (0.293, 0.525) | 0.594 (0.443, 0.729) | 33.958 (26.250, 41.880) |
| C | maternal ECG | late mean: ECGFounder 12-lead + DeepECG-SSL | 0.847 (0.789, 0.900) | 0.276 (0.197, 0.363) | 1.179 (1.039, 1.366) | 0.594 (0.355, 1.010) | 0.227 (0.206, 0.254) | 0.335 (0.307, 0.368) | 0.540 (0.508, 0.575) | 0.732 (0.706, 0.759) | 0.872 (0.807, 0.920) | 0.801 (0.713, 0.871) | 0.638 (0.526, 0.748) | 0.437 (0.331, 0.572) | 0.632 (0.495, 0.760) | 32.083 (24.578, 40.208) |
| C | maternal ECG | stacked: ECGFounder 12-lead + DeepECG-SSL | 0.863 (0.818, 0.904) | 0.186 (0.135, 0.238) | 1.064 (0.984, 1.165) | 0.781 (0.490, 1.499) | 0.210 (0.197, 0.225) | 0.313 (0.296, 0.332) | 0.515 (0.495, 0.537) | 0.712 (0.696, 0.729) | 0.839 (0.753, 0.893) | 0.754 (0.649, 0.832) | 0.572 (0.458, 0.688) | 0.368 (0.275, 0.499) | 0.565 (0.424, 0.701) | 24.167 (19.167, 29.167) |
| C | maternal ECG | late mean: ECGFounder 12-lead + DeepECG-SSL + SimCLR | 0.827 (0.764, 0.889) | 0.268 (0.184, 0.359) | 1.135 (0.990, 1.319) | 0.673 (0.399, 1.178) | 0.221 (0.198, 0.248) | 0.327 (0.297, 0.361) | 0.531 (0.497, 0.568) | 0.725 (0.697, 0.754) | 0.857 (0.780, 0.910) | 0.778 (0.678, 0.856) | 0.604 (0.480, 0.721) | 0.398 (0.290, 0.535) | 0.597 (0.450, 0.733) | 32.917 (24.792, 41.250) |
| C | maternal ECG | **CONTROL: late mean ECGFounder 12-lead + random-init HeartWise** | 0.830 (0.766, 0.891) | 0.293 (0.209, 0.381) | 1.178 (1.024, 1.376) | 0.591 (0.362, 0.980) | 0.227 (0.204, 0.255) | 0.335 (0.304, 0.370) | 0.540 (0.505, 0.577) | 0.733 (0.704, 0.761) | 0.872 (0.806, 0.918) | 0.800 (0.710, 0.868) | 0.633 (0.518, 0.740) | 0.428 (0.320, 0.554) | 0.627 (0.489, 0.753) | 34.583 (26.667, 42.708) |

## Operating point: target sensitivity 0.90

*FELICITy 1, n = 144 women (73 positive, observed prevalence 51%); target PSS-10 ≥ 19 (binary); unit of analysis: participant; operating point target sensitivity 0.90.*

| Set | Signal | Representation | sens | spec | LR+ | LR− | PPV@20 | PPV@30 | PPV@50 | PPV@70 | NPV@20 | NPV@30 | NPV@50 | NPV@70 | NPV (obs) | ruled out /150 |
|---|---|---|---|---|---|---|---|---|---|---|---|---|---|---|---|---|
| A | maternal ECG | SimCLR ResNet-34 (1-D), multi-layer 1088-d | 0.904 (0.855, 0.947) | 0.110 (0.058, 0.176) | 1.016 (0.936, 1.111) | 0.861 (0.429, 2.109) | 0.202 (0.190, 0.217) | 0.303 (0.286, 0.322) | 0.504 (0.483, 0.526) | 0.703 (0.685, 0.721) | 0.825 (0.672, 0.907) | 0.735 (0.552, 0.852) | 0.547 (0.357, 0.721) | 0.347 (0.199, 0.550) | 0.541 (0.337, 0.728) | 15.417 (10.208, 21.667) |
| A | maternal ECG | HeartWise DeepECG-SSL (wav2vec2), 768-d | 0.915 (0.866, 0.955) | 0.118 (0.061, 0.183) | 1.038 (0.957, 1.132) | 0.705 (0.329, 1.737) | 0.206 (0.193, 0.220) | 0.308 (0.291, 0.326) | 0.509 (0.489, 0.530) | 0.708 (0.690, 0.725) | 0.853 (0.715, 0.926) | 0.773 (0.601, 0.881) | 0.600 (0.401, 0.770) | 0.398 (0.229, 0.616) | 0.593 (0.382, 0.777) | 15.208 (9.792, 21.042) |
| A | maternal ECG | 100-feature HRV panel (95 usable; refined ensemble R-peaks) | 0.896 (0.849, 0.937) | 0.138 (0.076, 0.203) | 1.040 (0.955, 1.141) | 0.755 (0.399, 1.700) | 0.206 (0.193, 0.222) | 0.308 (0.290, 0.328) | 0.509 (0.488, 0.533) | 0.708 (0.689, 0.727) | 0.845 (0.723, 0.911) | 0.763 (0.615, 0.858) | 0.590 (0.431, 0.730) | 0.396 (0.258, 0.558) | 0.584 (0.411, 0.735) | 18.125 (12.292, 23.958) |

| Set | Signal | Representation | sens | spec | LR+ | LR− | PPV@20 | PPV@30 | PPV@50 | PPV@70 | NPV@20 | NPV@30 | NPV@50 | NPV@70 | NPV (obs) | ruled out /150 |
|---|---|---|---|---|---|---|---|---|---|---|---|---|---|---|---|---|
| A | maternal ECG | hrv-omnibus panel, 32 features (refined ensemble R-peaks) | 0.926 (0.881, 0.966) | 0.082 (0.038, 0.131) | 1.008 (0.943, 1.085) | 0.872 (0.363, 2.602) | 0.201 (0.191, 0.213) | 0.302 (0.288, 0.317) | 0.502 (0.485, 0.520) | 0.702 (0.687, 0.717) | 0.823 (0.610, 0.919) | 0.732 (0.495, 0.871) | 0.544 (0.316, 0.751) | 0.343 (0.172, 0.592) | 0.537 (0.308, 0.760) | 11.667 (6.875, 16.458) |
| A | maternal ECG | conventional HRV, 10 features (single NeuroKit detector) | 0.929 (0.875, 0.974) | 0.017 (0.003, 0.033) | 0.945 (0.887, 0.993) | inf (0.293, 6.967) | 0.191 (0.181, 0.199) | 0.288 (0.275, 0.298) | 0.486 (0.469, 0.498) | 0.688 (0.673, 0.698) | 0.428 (0.146, 0.588) | 0.348 (0.113, 0.520) | 0.240 (0.065, 0.408) | 0.156 (0.033, 0.327) | 0.237 (0.065, 0.412) | 6.667 (3.125, 10.833) |
| A | fetal ECG | SimCLR ResNet-34 (1-D), multi-layer 1088-d | 0.893 (0.836, 0.946) | 0.062 (0.031, 0.097) | 0.953 (0.884, 1.023) | 2.206 (0.904, 5.198) | 0.192 (0.181, 0.203) | 0.290 (0.275, 0.304) | 0.488 (0.469, 0.505) | 0.690 (0.673, 0.704) | 0.665 (0.404, 0.813) | 0.545 (0.325, 0.722) | 0.352 (0.192, 0.548) | 0.196 (0.098, 0.365) | 0.346 (0.176, 0.550) | 12.708 (8.125, 17.708) |
| A | abdominal ECG | SimCLR ResNet-34 (1-D), multi-layer 1088-d | 0.904 (0.842, 0.958) | 0.037 (0.009, 0.071) | 0.939 (0.869, 1.001) | 3.566 (1.031, 7.377) | 0.190 (0.178, 0.200) | 0.287 (0.271, 0.300) | 0.484 (0.464, 0.500) | 0.686 (0.669, 0.700) | 0.593 (0.221, 0.778) | 0.472 (0.165, 0.684) | 0.290 (0.087, 0.511) | 0.154 (0.043, 0.345) | 0.285 (0.083, 0.511) | 10.000 (5.417, 15.208) |
| A | abdominal ECG | HeartWise DeepECG-SSL (wav2vec2), 768-d | 0.907 (0.853, 0.950) | 0.085 (0.048, 0.127) | 0.991 (0.924, 1.064) | 1.156 (0.556, 3.224) | 0.199 (0.188, 0.210) | 0.298 (0.283, 0.313) | 0.498 (0.480, 0.515) | 0.698 (0.683, 0.713) | 0.780 (0.595, 0.882) | 0.677 (0.485, 0.817) | 0.478 (0.304, 0.669) | 0.285 (0.164, 0.483) | 0.471 (0.279, 0.677) | 13.333 (9.161, 18.333) |
| A | maternal ECG | ECGFounder 12-lead pathway (published weights, 507/509 tensors) | 0.926 (0.883, 0.963) | 0.161 (0.101, 0.225) | 1.105 (1.019, 1.215) | 0.438 (0.196, 0.912) | 0.216 (0.203, 0.233) | 0.321 (0.304, 0.342) | 0.525 (0.505, 0.548) | 0.720 (0.704, 0.739) | 0.902 (0.819, 0.954) | 0.845 (0.728, 0.924) | 0.704 (0.543, 0.845) | 0.513 (0.348, 0.718) | 0.699 (0.524, 0.853) | 17.500 (12.083, 23.333) |
| A | maternal ECG | ECGFounder 1-lead pathway (published weights, 507/509 tensors) | 0.915 (0.873, 0.951) | 0.144 (0.083, 0.208) | 1.069 (0.987, 1.169) | 0.582 (0.298, 1.260) | 0.211 (0.198, 0.226) | 0.314 (0.297, 0.334) | 0.516 (0.496, 0.539) | 0.714 (0.697, 0.731) | 0.874 (0.775, 0.932) | 0.803 (0.673, 0.889) | 0.641 (0.477, 0.783) | 0.439 (0.291, 0.627) | 0.634 (0.449, 0.793) | 17.083 (11.875, 22.714) |
| A | maternal ECG | **CONTROL: untrained encoder (weights never loaded, 0/509 tensors)** | 0.907 (0.850, 0.955) | 0.070 (0.027, 0.123) | 0.975 (0.904, 1.054) | 1.327 (0.552, 4.485) | 0.196 (0.184, 0.209) | 0.295 (0.279, 0.311) | 0.494 (0.474, 0.513) | 0.695 (0.678, 0.711) | 0.755 (0.466, 0.883) | 0.646 (0.360, 0.819) | 0.445 (0.213, 0.670) | 0.262 (0.109, 0.494) | 0.439 (0.193, 0.676) | 12.292 (7.292, 17.917) |
| B | maternal ECG | SimCLR ResNet-34 (1-D), projection head only 128-d (cols 960:1088) | 0.923 (0.873, 0.964) | 0.065 (0.032, 0.104) | 0.987 (0.925, 1.053) | 1.232 (0.506, 3.527) | 0.198 (0.188, 0.208) | 0.297 (0.284, 0.311) | 0.497 (0.480, 0.513) | 0.697 (0.683, 0.711) | 0.766 (0.495, 0.889) | 0.657 (0.398, 0.826) | 0.453 (0.240, 0.682) | 0.264 (0.126, 0.505) | 0.446 (0.229, 0.692) | 10.625 (6.458, 15.417) |
| B | fetal ECG | SimCLR ResNet-34 (1-D), projection head only 128-d (cols 960:1088) | 0.915 (0.865, 0.960) | 0.051 (0.021, 0.085) | 0.965 (0.904, 1.026) | 2.286 (0.865, 5.582) | 0.194 (0.184, 0.204) | 0.292 (0.279, 0.305) | 0.491 (0.475, 0.506) | 0.692 (0.678, 0.705) | 0.666 (0.342, 0.814) | 0.550 (0.283, 0.730) | 0.362 (0.181, 0.567) | 0.206 (0.097, 0.403) | 0.356 (0.167, 0.580) | 10.208 (5.833, 14.792) |
| B | abdominal ECG | SimCLR ResNet-34 (1-D), projection head only 128-d (cols 960:1088) | 0.907 (0.851, 0.958) | 0.070 (0.038, 0.109) | 0.976 (0.910, 1.050) | 1.375 (0.604, 3.773) | 0.196 (0.185, 0.208) | 0.295 (0.280, 0.310) | 0.494 (0.476, 0.512) | 0.695 (0.679, 0.710) | 0.747 (0.536, 0.874) | 0.634 (0.420, 0.805) | 0.429 (0.251, 0.649) | 0.245 (0.131, 0.457) | 0.422 (0.238, 0.648) | 12.292 (7.083, 17.292) |

| Set | Signal | Representation | sens | spec | LR+ | LR− | PPV@20 | PPV@30 | PPV@50 | PPV@70 | NPV@20 | NPV@30 | NPV@50 | NPV@70 | NPV (obs) | ruled out /150 |
|---|---|---|---|---|---|---|---|---|---|---|---|---|---|---|---|---|
| C | maternal ECG | HeartWise DeepECG-SSL, participant-mean | 0.901 (0.843, 0.951) | 0.099 (0.051, 0.155) | 1.000 (0.918, 1.090) | 1.037 (0.433, 2.867) | 0.200 (0.186, 0.214) | 0.300 (0.282, 0.318) | 0.500 (0.478, 0.521) | 0.700 (0.681, 0.718) | 0.796 (0.559, 0.905) | 0.697 (0.457, 0.849) | 0.501 (0.292, 0.718) | 0.305 (0.160, 0.539) | 0.494 (0.274, 0.726) | 14.792 (9.583, 20.417) |
| C | maternal ECG | SimCLR ResNet-34 multi-layer, participant-mean | 0.901 (0.848, 0.944) | 0.130 (0.077, 0.183) | 1.036 (0.954, 1.126) | 0.764 (0.398, 1.671) | 0.206 (0.192, 0.220) | 0.307 (0.290, 0.325) | 0.509 (0.488, 0.529) | 0.707 (0.690, 0.724) | 0.841 (0.722, 0.911) | 0.757 (0.611, 0.857) | 0.574 (0.413, 0.725) | 0.370 (0.239, 0.545) | 0.568 (0.381, 0.741) | 17.083 (11.875, 22.708) |
| C | maternal ECG | **CONTROL: HeartWise architecture, random init, participant-mean** | 0.915 (0.860, 0.961) | 0.068 (0.031, 0.113) | 0.982 (0.914, 1.054) | 1.380 (0.542, 3.912) | 0.197 (0.186, 0.208) | 0.296 (0.281, 0.311) | 0.495 (0.477, 0.513) | 0.696 (0.681, 0.711) | 0.747 (0.450, 0.885) | 0.635 (0.358, 0.820) | 0.430 (0.220, 0.675) | 0.247 (0.114, 0.491) | 0.424 (0.197, 0.678) | 11.458 (6.875, 16.875) |
| C | maternal ECG | concat ECGFounder 12-lead + DeepECG-SSL (1792-d) | 0.918 (0.872, 0.961) | 0.146 (0.089, 0.209) | 1.078 (0.990, 1.180) | 0.599 (0.279, 1.362) | 0.212 (0.198, 0.228) | 0.316 (0.298, 0.336) | 0.518 (0.497, 0.541) | 0.715 (0.698, 0.733) | 0.872 (0.761, 0.937) | 0.801 (0.660, 0.897) | 0.638 (0.467, 0.798) | 0.438 (0.285, 0.649) | 0.632 (0.450, 0.807) | 17.083 (11.875, 22.708) |
| C | maternal ECG | late mean: ECGFounder 12-lead + DeepECG-SSL | 0.942 (0.904, 0.974) | 0.166 (0.105, 0.233) | 1.135 (1.045, 1.250) | 0.397 (0.162, 0.940) | 0.221 (0.207, 0.238) | 0.327 (0.309, 0.348) | 0.531 (0.510, 0.554) | 0.725 (0.708, 0.744) | 0.912 (0.829, 0.962) | 0.861 (0.749, 0.937) | 0.734 (0.583, 0.869) | 0.555 (0.394, 0.756) | 0.729 (0.571, 0.874) | 16.667 (11.250, 22.708) |
| C | maternal ECG | stacked: ECGFounder 12-lead + DeepECG-SSL | 0.940 (0.912, 0.965) | 0.096 (0.062, 0.132) | 1.040 (0.992, 1.095) | 0.643 (0.325, 1.646) | 0.206 (0.199, 0.215) | 0.308 (0.298, 0.319) | 0.510 (0.498, 0.522) | 0.708 (0.698, 0.718) | 0.863 (0.668, 0.927) | 0.787 (0.598, 0.884) | 0.617 (0.439, 0.778) | 0.412 (0.274, 0.638) | 0.610 (0.411, 0.788) | 11.667 (8.333, 15.000) |
| C | maternal ECG | late mean: ECGFounder 12-lead + DeepECG-SSL + SimCLR | 0.929 (0.887, 0.965) | 0.152 (0.092, 0.221) | 1.099 (1.012, 1.205) | 0.535 (0.251, 1.302) | 0.215 (0.202, 0.231) | 0.320 (0.302, 0.340) | 0.523 (0.502, 0.545) | 0.719 (0.702, 0.736) | 0.885 (0.783, 0.942) | 0.821 (0.693, 0.907) | 0.674 (0.516, 0.814) | 0.485 (0.334, 0.675) | 0.668 (0.497, 0.821) | 16.667 (11.250, 22.708) |
| C | maternal ECG | **CONTROL: late mean ECGFounder 12-lead + random-init HeartWise** | 0.929 (0.889, 0.964) | 0.144 (0.085, 0.210) | 1.088 (1.001, 1.195) | 0.549 (0.251, 1.344) | 0.214 (0.200, 0.229) | 0.318 (0.300, 0.338) | 0.520 (0.499, 0.543) | 0.717 (0.699, 0.735) | 0.882 (0.771, 0.942) | 0.816 (0.675, 0.907) | 0.665 (0.496, 0.817) | 0.475 (0.315, 0.683) | 0.660 (0.473, 0.823) | 16.042 (10.833, 21.875) |

## Operating point: target sensitivity 0.95

*FELICITy 1, n = 144 women (73 positive, observed prevalence 51%); target PSS-10 ≥ 19 (binary); unit of analysis: participant; operating point target sensitivity 0.95.*

| Set | Signal | Representation | sens | spec | LR+ | LR− | PPV@20 | PPV@30 | PPV@50 | PPV@70 | NPV@20 | NPV@30 | NPV@50 | NPV@70 | NPV (obs) | ruled out /150 |
|---|---|---|---|---|---|---|---|---|---|---|---|---|---|---|---|---|
| A | maternal ECG | SimCLR ResNet-34 (1-D), multi-layer 1088-d | 0.967 (0.938, 0.989) | 0.062 (0.026, 0.110) | 1.031 (0.983, 1.093) | 0.460 (0.133, 1.874) | 0.205 (0.197, 0.215) | 0.306 (0.296, 0.319) | 0.508 (0.496, 0.522) | 0.706 (0.696, 0.718) | 0.902 (0.685, 0.968) | 0.847 (0.588, 0.948) | 0.723 (0.414, 0.894) | 0.572 (0.250, 0.809) | 0.719 (0.400, 0.893) | 7.083 (3.750, 11.042) |

| Set | Signal | Representation | sens | spec | LR+ | LR− | PPV@20 | PPV@30 | PPV@50 | PPV@70 | NPV@20 | NPV@30 | NPV@50 | NPV@70 | NPV (obs) | ruled out /150 |
|---|---|---|---|---|---|---|---|---|---|---|---|---|---|---|---|---|
| A | maternal ECG | HeartWise DeepECG-SSL (wav2vec2), 768-d | 0.975 (0.941, 0.997) | 0.056 (0.018, 0.108) | 1.034 (0.981, 1.100) | 0.482 (0.053, 2.137) | 0.205 (0.197, 0.216) | 0.307 (0.296, 0.320) | 0.508 (0.495, 0.524) | 0.707 (0.696, 0.719) | 0.895 (0.458, 0.985) | 0.835 (0.393, 0.976) | 0.692 (0.288, 0.951) | 0.507 (0.177, 0.908) | 0.687 (0.279, 0.950) | 6.042 (2.708, 10.417) |
| A | maternal ECG | 100-feature HRV panel (95 usable; refined ensemble R-peaks) | 0.962 (0.930, 0.986) | 0.073 (0.030, 0.125) | 1.039 (0.981, 1.106) | 0.593 (0.189, 2.408) | 0.206 (0.197, 0.216) | 0.308 (0.296, 0.321) | 0.509 (0.495, 0.525) | 0.708 (0.696, 0.720) | 0.874 (0.575, 0.956) | 0.805 (0.499, 0.929) | 0.648 (0.358, 0.859) | 0.456 (0.221, 0.745) | 0.642 (0.349, 0.860) | 8.333 (4.578, 12.917) |
| A | maternal ECG | hrv-omnibus panel, 32 features (refined ensemble R-peaks) | 0.973 (0.951, 0.991) | 0.034 (0.009, 0.062) | 1.007 (0.974, 1.043) | 0.843 (0.217, 2.571) | 0.201 (0.196, 0.207) | 0.301 (0.294, 0.309) | 0.502 (0.493, 0.511) | 0.701 (0.694, 0.709) | 0.832 (0.368, 0.943) | 0.746 (0.314, 0.913) | 0.570 (0.203, 0.839) | 0.375 (0.116, 0.754) | 0.563 (0.205, 0.842) | 4.583 (2.292, 7.089) |
| A | maternal ECG | conventional HRV, 10 features (single NeuroKit detector) | 0.964 (0.931, 0.989) | 0.003 (0.000, 0.009) | 0.967 (0.933, 0.993) | n/a (0.999, 10.571) | 0.195 (0.189, 0.199) | 0.293 (0.286, 0.299) | 0.492 (0.482, 0.498) | 0.693 (0.685, 0.698) | n/a (0.000, 0.346) | n/a (0.000, 0.286) | n/a (0.000, 0.182) | n/a (0.000, 0.104) | n/a (0.000, 0.167) | 2.917 (1.042, 5.417) |
| A | fetal ECG | SimCLR ResNet-34 (1-D), multi-layer 1088-d | 0.959 (0.922, 0.987) | 0.017 (0.006, 0.031) | 0.976 (0.936, 1.008) | inf (0.160, 4.984) | 0.196 (0.190, 0.201) | 0.295 (0.286, 0.302) | 0.494 (0.483, 0.502) | 0.695 (0.686, 0.702) | 0.533 (0.181, 0.855) | 0.449 (0.154, 0.803) | 0.309 (0.099, 0.735) | 0.187 (0.051, 0.667) | 0.305 (0.094, 0.733) | 4.375 (1.875, 7.292) |
| A | abdominal ECG | SimCLR ResNet-34 (1-D), multi-layer 1088-d | 0.940 (0.889, 0.982) | 0.011 (0.000, 0.026) | 0.950 (0.897, 0.995) | inf (0.432, 7.478) | 0.192 (0.183, 0.199) | 0.289 (0.278, 0.299) | 0.487 (0.473, 0.499) | 0.689 (0.676, 0.699) | 0.394 (0.000, 0.623) | 0.292 (0.000, 0.546) | 0.161 (0.000, 0.420) | 0.079 (0.000, 0.327) | 0.157 (0.000, 0.423) | 5.417 (1.875, 9.375) |
| A | abdominal ECG | HeartWise DeepECG-SSL (wav2vec2), 768-d | 0.970 (0.942, 0.991) | 0.014 (0.003, 0.028) | 0.984 (0.955, 1.009) | inf (0.143, 4.118) | 0.197 (0.193, 0.201) | 0.297 (0.290, 0.302) | 0.496 (0.488, 0.502) | 0.696 (0.690, 0.702) | 0.581 (0.139, 0.859) | 0.489 (0.113, 0.806) | 0.329 (0.069, 0.712) | 0.188 (0.035, 0.637) | 0.324 (0.067, 0.714) | 3.333 (1.667, 5.422) |
| A | maternal ECG | ECGFounder 12-lead pathway (published weights, 507/509 tensors) | 0.970 (0.942, 0.990) | 0.076 (0.038, 0.117) | 1.051 (1.002, 1.108) | 0.367 (0.119, 1.294) | 0.208 (0.200, 0.217) | 0.310 (0.300, 0.322) | 0.512 (0.500, 0.525) | 0.710 (0.700, 0.721) | 0.919 (0.765, 0.972) | 0.871 (0.692, 0.955) | 0.756 (0.539, 0.912) | 0.602 (0.367, 0.843) | 0.751 (0.522, 0.917) | 7.917 (4.792, 11.667) |
| A | maternal ECG | ECGFounder 1-lead pathway (published weights, 507/509 tensors) | 0.956 (0.920, 0.984) | 0.096 (0.053, 0.145) | 1.058 (0.997, 1.129) | 0.430 (0.126, 1.269) | 0.209 (0.199, 0.220) | 0.312 (0.299, 0.326) | 0.514 (0.499, 0.530) | 0.712 (0.699, 0.725) | 0.904 (0.746, 0.971) | 0.848 (0.655, 0.953) | 0.709 (0.483, 0.904) | 0.519 (0.303, 0.829) | 0.704 (0.469, 0.905) | 10.417 (6.667, 15.000) |
| A | maternal ECG | **CONTROL: untrained encoder (weights never loaded, 0/509 tensors)** | 0.953 (0.917, 0.983) | 0.028 (0.008, 0.051) | 0.981 (0.939, 1.021) | inf (0.431, 4.053) | 0.197 (0.190, 0.203) | 0.296 (0.287, 0.304) | 0.495 (0.484, 0.505) | 0.696 (0.687, 0.704) | 0.564 (0.194, 0.823) | 0.468 (0.151, 0.740) | 0.306 (0.089, 0.608) | 0.170 (0.045, 0.475) | 0.301 (0.083, 0.617) | 5.625 (2.917, 8.958) |
| B | maternal ECG | SimCLR ResNet-34 (1-D), projection head only 128-d (cols 960:1088) | 0.978 (0.951, 0.997) | 0.037 (0.013, 0.066) | 1.015 (0.977, 1.054) | 0.713 (0.000, 2.364) | 0.202 (0.196, 0.209) | 0.303 (0.295, 0.311) | 0.504 (0.494, 0.513) | 0.703 (0.695, 0.711) | 0.865 (0.530, 0.984) | 0.803 (0.497, 0.975) | 0.684 (0.442, 0.950) | 0.570 (0.362, 0.912) | 0.680 (0.434, 0.950) | 4.375 (1.875, 7.292) |
| B | fetal ECG | SimCLR ResNet-34 (1-D), projection head only 128-d (cols 960:1088) | 0.964 (0.936, 0.988) | 0.017 (0.003, 0.034) | 0.981 (0.951, 1.012) | inf (0.434, 4.667) | 0.197 (0.192, 0.202) | 0.296 (0.289, 0.302) | 0.495 (0.487, 0.503) | 0.696 (0.689, 0.702) | 0.547 (0.146, 0.842) | 0.451 (0.117, 0.770) | 0.292 (0.069, 0.627) | 0.162 (0.034, 0.531) | 0.287 (0.067, 0.634) | 3.958 (1.875, 6.250) |

| Set | Signal | Representation | sens | spec | LR+ | LR− | PPV@20 | PPV@30 | PPV@50 | PPV@70 | NPV@20 | NPV@30 | NPV@50 | NPV@70 | NPV (obs) | ruled out /150 |
|---|---|---|---|---|---|---|---|---|---|---|---|---|---|---|---|---|
| B | abdominal ECG | SimCLR ResNet-34 (1-D), projection head only 128-d (cols 960:1088) | 0.962 (0.922, 0.991) | 0.011 (0.003, 0.024) | 0.973 (0.933, 1.004) | inf (0.000, 4.617) | 0.196 (0.189, 0.201) | 0.294 (0.286, 0.301) | 0.493 (0.482, 0.501) | 0.694 (0.685, 0.701) | 0.395 (0.094, 0.600) | 0.329 (0.069, 0.563) | 0.227 (0.037, 0.500) | 0.139 (0.017, 0.447) | 0.223 (0.033, 0.500) | 3.750 (1.250, 6.875) |
| C | maternal ECG | HeartWise DeepECG-SSL, participant-mean | 0.959 (0.923, 0.989) | 0.048 (0.013, 0.089) | 1.007 (0.956, 1.064) | 0.940 (0.211, 3.044) | 0.201 (0.193, 0.210) | 0.302 (0.291, 0.313) | 0.502 (0.489, 0.515) | 0.701 (0.690, 0.713) | 0.818 (0.344, 0.952) | 0.729 (0.291, 0.922) | 0.547 (0.195, 0.851) | 0.350 (0.106, 0.744) | 0.540 (0.192, 0.847) | 6.667 (3.125, 10.417) |
| C | maternal ECG | SimCLR ResNet-34 multi-layer, participant-mean | 0.953 (0.924, 0.977) | 0.054 (0.026, 0.085) | 1.008 (0.966, 1.052) | 0.940 (0.405, 2.792) | 0.201 (0.194, 0.208) | 0.302 (0.293, 0.311) | 0.502 (0.491, 0.512) | 0.701 (0.692, 0.710) | 0.820 (0.559, 0.910) | 0.732 (0.480, 0.861) | 0.556 (0.349, 0.761) | 0.366 (0.205, 0.641) | 0.550 (0.324, 0.767) | 7.500 (4.583, 10.625) |
| C | maternal ECG | **CONTROL: HeartWise architecture, random init, participant-mean** | 0.959 (0.923, 0.988) | 0.023 (0.003, 0.050) | 0.981 (0.940, 1.022) | 2.042 (0.331, 5.170) | 0.197 (0.190, 0.204) | 0.296 (0.287, 0.305) | 0.495 (0.484, 0.505) | 0.696 (0.686, 0.704) | 0.697 (0.145, 0.893) | 0.586 (0.113, 0.843) | 0.397 (0.064, 0.752) | 0.230 (0.032, 0.679) | 0.390 (0.062, 0.753) | 4.792 (2.083, 8.125) |
| C | maternal ECG | concat ECGFounder 12-lead + DeepECG-SSL (1792-d) | 0.959 (0.924, 0.988) | 0.090 (0.046, 0.139) | 1.055 (0.992, 1.127) | 0.480 (0.133, 1.484) | 0.209 (0.199, 0.220) | 0.311 (0.298, 0.325) | 0.513 (0.498, 0.529) | 0.711 (0.698, 0.724) | 0.893 (0.702, 0.969) | 0.831 (0.618, 0.948) | 0.680 (0.448, 0.892) | 0.480 (0.276, 0.796) | 0.674 (0.433, 0.894) | 9.792 (5.833, 14.167) |
| C | maternal ECG | late mean: ECGFounder 12-lead + DeepECG-SSL | 0.967 (0.938, 0.990) | 0.104 (0.058, 0.158) | 1.082 (1.021, 1.162) | 0.348 (0.103, 1.041) | 0.213 (0.203, 0.225) | 0.317 (0.304, 0.332) | 0.519 (0.505, 0.536) | 0.716 (0.704, 0.729) | 0.921 (0.724, 0.975) | 0.872 (0.677, 0.959) | 0.749 (0.543, 0.912) | 0.567 (0.367, 0.837) | 0.744 (0.528, 0.918) | 10.208 (6.042, 14.792) |
| C | maternal ECG | stacked: ECGFounder 12-lead + DeepECG-SSL | 0.967 (0.947, 0.984) | 0.045 (0.026, 0.064) | 1.014 (0.985, 1.046) | n/a (0.274, 1.997) | 0.202 (0.198, 0.207) | 0.303 (0.297, 0.309) | 0.503 (0.496, 0.511) | 0.703 (0.697, 0.709) | n/a (0.459, 0.913) | n/a (0.415, 0.862) | n/a (0.298, 0.744) | n/a (0.180, 0.607) | n/a (0.286, 0.758) | 5.833 (3.958, 7.917) |
| C | maternal ECG | late mean: ECGFounder 12-lead + DeepECG-SSL + SimCLR | 0.970 (0.935, 0.994) | 0.087 (0.047, 0.133) | 1.065 (1.008, 1.130) | 0.510 (0.116, 1.740) | 0.210 (0.201, 0.220) | 0.313 (0.302, 0.326) | 0.515 (0.502, 0.530) | 0.713 (0.701, 0.724) | 0.895 (0.690, 0.972) | 0.839 (0.627, 0.954) | 0.715 (0.490, 0.908) | 0.555 (0.327, 0.845) | 0.710 (0.474, 0.913) | 8.750 (5.000, 13.125) |
| C | maternal ECG | **CONTROL: late mean ECGFounder 12-lead + random-init HeartWise** | 0.967 (0.939, 0.988) | 0.085 (0.039, 0.138) | 1.058 (0.999, 1.132) | 0.487 (0.130, 1.459) | 0.209 (0.200, 0.221) | 0.312 (0.300, 0.327) | 0.514 (0.500, 0.531) | 0.711 (0.699, 0.725) | 0.895 (0.677, 0.968) | 0.836 (0.609, 0.947) | 0.699 (0.470, 0.893) | 0.519 (0.321, 0.810) | 0.693 (0.450, 0.893) | 8.750 (4.792, 13.333) |

## Operating point: Youden

*FELICITy 1, n = 144 women (73 positive, observed prevalence 51%); target PSS-10 ≥ 19 (binary); unit of analysis: participant; operating point Youden.*

| Set | Signal | Representation | sens | spec | LR+ | LR− | PPV@20 | PPV@30 | PPV@50 | PPV@70 | NPV@20 | NPV@30 | NPV@50 | NPV@70 | NPV (obs) | ruled out /150 |
|---|---|---|---|---|---|---|---|---|---|---|---|---|---|---|---|---|

| Set | Signal | Representation | sens | spec | LR+ | LR− | PPV@20 | PPV@30 | PPV@50 | PPV@70 | NPV@20 | NPV@30 | NPV@50 | NPV@70 | NPV (obs) | ruled out /150 |
|---|---|---|---|---|---|---|---|---|---|---|---|---|---|---|---|---|
| A | maternal ECG | SimCLR ResNet-34 (1-D), multi-layer 1088-d | 0.496 (0.426, 0.561) | 0.515 (0.442, 0.589) | 1.027 (0.840, 1.281) | 0.978 (0.811, 1.222) | 0.204 (0.173, 0.241) | 0.306 (0.264, 0.352) | 0.506 (0.454, 0.558) | 0.705 (0.659, 0.746) | 0.804 (0.767, 0.832) | 0.705 (0.658, 0.743) | 0.506 (0.453, 0.554) | 0.305 (0.263, 0.348) | 0.499 (0.401, 0.593) | 76.458 (69.167, 83.750) |
| A | maternal ECG | HeartWise DeepECG-SSL (wav2vec2), 768-d | 0.411 (0.336, 0.489) | 0.676 (0.610, 0.740) | 1.308 (0.999, 1.833) | 0.871 (0.736, 1.027) | 0.246 (0.199, 0.308) | 0.358 (0.298, 0.430) | 0.564 (0.494, 0.634) | 0.751 (0.694, 0.799) | 0.821 (0.796, 0.845) | 0.728 (0.695, 0.761) | 0.535 (0.494, 0.578) | 0.330 (0.295, 0.370) | 0.528 (0.436, 0.621) | 94.792 (86.875, 102.292) |
| A | maternal ECG | 100-feature HRV panel (95 usable; refined ensemble R-peaks) | 0.605 (0.538, 0.672) | 0.363 (0.297, 0.429) | 0.945 (0.801, 1.117) | 1.064 (0.820, 1.472) | 0.191 (0.166, 0.218) | 0.287 (0.254, 0.323) | 0.484 (0.441, 0.526) | 0.686 (0.645, 0.720) | 0.791 (0.735, 0.833) | 0.690 (0.619, 0.745) | 0.490 (0.414, 0.563) | 0.293 (0.234, 0.364) | 0.483 (0.377, 0.593) | 56.875 (49.792, 64.167) |
| A | maternal ECG | hrv-omnibus panel, 32 features (refined ensemble R-peaks) | 0.414 (0.356, 0.469) | 0.561 (0.494, 0.621) | 0.934 (0.765, 1.174) | 1.044 (0.904, 1.233) | 0.189 (0.160, 0.225) | 0.286 (0.245, 0.332) | 0.483 (0.429, 0.536) | 0.685 (0.635, 0.728) | 0.793 (0.765, 0.816) | 0.691 (0.656, 0.722) | 0.489 (0.450, 0.528) | 0.291 (0.260, 0.324) | 0.482 (0.390, 0.572) | 86.042 (79.792, 92.083) |
| A | maternal ECG | conventional HRV, 10 features (single NeuroKit detector) | 0.704 (0.654, 0.753) | 0.251 (0.211, 0.292) | 0.937 (0.859, 1.035) | 1.172 (0.937, 1.591) | 0.190 (0.176, 0.205) | 0.286 (0.268, 0.306) | 0.483 (0.460, 0.507) | 0.685 (0.664, 0.705) | 0.775 (0.725, 0.813) | 0.668 (0.609, 0.718) | 0.464 (0.404, 0.526) | 0.272 (0.228, 0.329) | 0.458 (0.358, 0.562) | 41.042 (36.453, 45.833) |
| A | fetal ECG | SimCLR ResNet-34 (1-D), multi-layer 1088-d | 0.414 (0.327, 0.499) | 0.665 (0.589, 0.737) | 1.257 (0.911, 1.767) | 0.886 (0.734, 1.079) | 0.238 (0.184, 0.303) | 0.349 (0.279, 0.425) | 0.555 (0.474, 0.631) | 0.744 (0.676, 0.798) | 0.819 (0.788, 0.845) | 0.725 (0.685, 0.761) | 0.531 (0.483, 0.578) | 0.327 (0.286, 0.371) | 0.524 (0.429, 0.617) | 93.750 (85.417, 102.292) |
| A | abdominal ECG | SimCLR ResNet-34 (1-D), multi-layer 1088-d | 0.373 (0.289, 0.458) | 0.659 (0.588, 0.733) | 1.123 (0.821, 1.585) | 0.958 (0.806, 1.150) | 0.219 (0.170, 0.280) | 0.324 (0.259, 0.398) | 0.527 (0.446, 0.603) | 0.721 (0.652, 0.779) | 0.807 (0.778, 0.832) | 0.709 (0.672, 0.744) | 0.511 (0.468, 0.555) | 0.310 (0.274, 0.349) | 0.504 (0.414, 0.598) | 96.458 (88.120, 104.792) |
| A | abdominal ECG | HeartWise DeepECG-SSL (wav2vec2), 768-d | 0.438 (0.361, 0.520) | 0.558 (0.478, 0.643) | 0.989 (0.772, 1.318) | 1.008 (0.825, 1.247) | 0.198 (0.161, 0.246) | 0.297 (0.248, 0.358) | 0.496 (0.433, 0.565) | 0.696 (0.639, 0.751) | 0.799 (0.763, 0.830) | 0.699 (0.654, 0.740) | 0.499 (0.448, 0.551) | 0.299 (0.258, 0.345) | 0.492 (0.396, 0.587) | 83.958 (75.417, 92.500) |
| A | maternal ECG | ECGFounder 12-lead pathway (published weights, 507/509 tensors) | 0.677 (0.613, 0.738) | 0.417 (0.331, 0.501) | 1.165 (0.969, 1.431) | 0.759 (0.558, 1.074) | 0.225 (0.194, 0.262) | 0.332 (0.292, 0.378) | 0.537 (0.489, 0.585) | 0.730 (0.690, 0.767) | 0.841 (0.791, 0.879) | 0.756 (0.689, 0.810) | 0.572 (0.490, 0.649) | 0.366 (0.295, 0.447) | 0.565 (0.449, 0.672) | 55.417 (47.292, 63.339) |
| A | maternal ECG | ECGFounder 1-lead pathway (published weights, 507/509 tensors) | 0.636 (0.557, 0.713) | 0.448 (0.363, 0.532) | 1.163 (0.952, 1.455) | 0.826 (0.621, 1.128) | 0.225 (0.191, 0.265) | 0.331 (0.288, 0.381) | 0.536 (0.484, 0.588) | 0.728 (0.686, 0.768) | 0.830 (0.783, 0.867) | 0.740 (0.679, 0.793) | 0.552 (0.479, 0.625) | 0.348 (0.285, 0.420) | 0.545 (0.434, 0.650) | 60.833 (52.292, 69.375) |
| A | maternal ECG | **CONTROL: untrained encoder (weights never loaded, 0/509 tensors)** | 0.373 (0.289, 0.458) | 0.707 (0.633, 0.778) | 1.326 (0.924, 1.989) | 0.887 (0.745, 1.060) | 0.247 (0.185, 0.326) | 0.358 (0.279, 0.448) | 0.563 (0.470, 0.650) | 0.749 (0.671, 0.809) | 0.819 (0.791, 0.844) | 0.725 (0.689, 0.759) | 0.531 (0.487, 0.575) | 0.328 (0.290, 0.368) | 0.525 (0.426, 0.617) | 100.000 (91.667, 108.750) |
| B | maternal ECG | SimCLR ResNet-34 (1-D), projection head only 128-d (cols 960:1088) | 0.562 (0.484, 0.634) | 0.470 (0.387, 0.551) | 1.070 (0.866, 1.366) | 0.933 (0.728, 1.223) | 0.211 (0.177, 0.253) | 0.314 (0.269, 0.366) | 0.516 (0.461, 0.573) | 0.713 (0.665, 0.756) | 0.812 (0.768, 0.847) | 0.716 (0.660, 0.764) | 0.520 (0.457, 0.583) | 0.318 (0.266, 0.378) | 0.513 (0.413, 0.615) | 68.125 (60.411, 76.667) |

| Set | Signal | Representation | sens | spec | LR+ | LR− | PPV@20 | PPV@30 | PPV@50 | PPV@70 | NPV@20 | NPV@30 | NPV@50 | NPV@70 | NPV (obs) | ruled out /150 |
|---|---|---|---|---|---|---|---|---|---|---|---|---|---|---|---|---|
| B | fetal ECG | SimCLR ResNet-34 (1-D), projection head only 128-d (cols 960:1088) | 0.436 (0.351, 0.522) | 0.634 (0.554, 0.712) | 1.218 (0.906, 1.699) | 0.897 (0.733, 1.098) | 0.233 (0.184, 0.295) | 0.342 (0.278, 0.417) | 0.547 (0.471, 0.623) | 0.737 (0.673, 0.792) | 0.817 (0.785, 0.845) | 0.723 (0.681, 0.762) | 0.528 (0.479, 0.579) | 0.324 (0.283, 0.372) | 0.521 (0.423, 0.616) | 89.792 (81.458, 98.542) |
| B | abdominal ECG | SimCLR ResNet-34 (1-D), projection head only 128-d (cols 960:1088) | 0.392 (0.328, 0.466) | 0.504 (0.429, 0.583) | 0.793 (0.632, 1.010) | 1.221 (1.016, 1.508) | 0.165 (0.136, 0.201) | 0.253 (0.212, 0.301) | 0.442 (0.383, 0.500) | 0.648 (0.590, 0.698) | 0.767 (0.728, 0.798) | 0.657 (0.611, 0.698) | 0.451 (0.404, 0.498) | 0.261 (0.226, 0.299) | 0.444 (0.356, 0.538) | 83.542 (75.411, 91.042) |
| C | maternal ECG | HeartWise DeepECG-SSL, participant-mean | 0.507 (0.436, 0.585) | 0.603 (0.518, 0.689) | 1.318 (1.010, 1.836) | 0.820 (0.656, 1.029) | 0.247 (0.201, 0.307) | 0.359 (0.300, 0.430) | 0.565 (0.498, 0.634) | 0.751 (0.697, 0.800) | 0.830 (0.797, 0.860) | 0.741 (0.696, 0.782) | 0.551 (0.496, 0.607) | 0.345 (0.298, 0.400) | 0.544 (0.444, 0.640) | 82.083 (73.542, 90.422) |
| C | maternal ECG | SimCLR ResNet-34 multi-layer, participant-mean | 0.479 (0.397, 0.562) | 0.592 (0.514, 0.663) | 1.239 (0.934, 1.730) | 0.881 (0.719, 1.105) | 0.235 (0.189, 0.292) | 0.344 (0.284, 0.410) | 0.549 (0.479, 0.613) | 0.738 (0.681, 0.784) | 0.820 (0.785, 0.848) | 0.726 (0.681, 0.766) | 0.533 (0.479, 0.585) | 0.328 (0.284, 0.378) | 0.526 (0.428, 0.621) | 83.333 (74.167, 92.083) |
| C | maternal ECG | **CONTROL: HeartWise architecture, random init, participant-mean** | 0.455 (0.381, 0.532) | 0.589 (0.508, 0.672) | 1.115 (0.852, 1.500) | 0.928 (0.765, 1.153) | 0.218 (0.175, 0.270) | 0.323 (0.266, 0.388) | 0.526 (0.458, 0.594) | 0.721 (0.662, 0.772) | 0.812 (0.777, 0.840) | 0.716 (0.670, 0.754) | 0.519 (0.467, 0.568) | 0.317 (0.274, 0.362) | 0.512 (0.412, 0.605) | 85.000 (76.458, 93.125) |
| C | maternal ECG | concat ECGFounder 12-lead + DeepECG-SSL (1792-d) | 0.567 (0.490, 0.650) | 0.532 (0.451, 0.613) | 1.248 (0.983, 1.599) | 0.822 (0.636, 1.078) | 0.237 (0.197, 0.282) | 0.347 (0.295, 0.400) | 0.552 (0.493, 0.606) | 0.741 (0.693, 0.781) | 0.830 (0.789, 0.863) | 0.740 (0.686, 0.787) | 0.550 (0.484, 0.613) | 0.344 (0.289, 0.406) | 0.543 (0.434, 0.644) | 72.292 (63.333, 80.625) |
| C | maternal ECG | late mean: ECGFounder 12-lead + DeepECG-SSL | 0.622 (0.544, 0.703) | 0.493 (0.406, 0.577) | 1.261 (1.023, 1.592) | 0.786 (0.595, 1.047) | 0.238 (0.202, 0.282) | 0.348 (0.302, 0.401) | 0.554 (0.501, 0.608) | 0.742 (0.700, 0.782) | 0.836 (0.794, 0.871) | 0.749 (0.692, 0.799) | 0.563 (0.495, 0.633) | 0.357 (0.298, 0.427) | 0.556 (0.445, 0.662) | 65.208 (56.042, 73.958) |
| C | maternal ECG | stacked: ECGFounder 12-lead + DeepECG-SSL | 0.679 (0.622, 0.737) | 0.338 (0.280, 0.397) | 1.033 (0.911, 1.187) | 0.975 (0.766, 1.320) | 0.205 (0.185, 0.228) | 0.306 (0.280, 0.335) | 0.507 (0.475, 0.539) | 0.705 (0.677, 0.731) | 0.805 (0.758, 0.841) | 0.708 (0.649, 0.756) | 0.512 (0.448, 0.575) | 0.312 (0.261, 0.372) | 0.505 (0.399, 0.607) | 49.375 (43.328, 55.417) |
| C | maternal ECG | late mean: ECGFounder 12-lead + DeepECG-SSL + SimCLR | 0.529 (0.442, 0.617) | 0.592 (0.503, 0.675) | 1.299 (1.007, 1.714) | 0.797 (0.635, 1.015) | 0.244 (0.200, 0.298) | 0.356 (0.299, 0.419) | 0.563 (0.498, 0.626) | 0.750 (0.696, 0.796) | 0.834 (0.799, 0.864) | 0.746 (0.699, 0.788) | 0.558 (0.500, 0.615) | 0.352 (0.302, 0.408) | 0.551 (0.449, 0.645) | 79.583 (70.411, 88.542) |
| C | maternal ECG | **CONTROL: late mean ECGFounder 12-lead + random-init HeartWise** | 0.627 (0.554, 0.699) | 0.448 (0.365, 0.527) | 1.159 (0.953, 1.456) | 0.822 (0.606, 1.131) | 0.224 (0.191, 0.265) | 0.330 (0.287, 0.381) | 0.533 (0.483, 0.586) | 0.726 (0.684, 0.766) | 0.832 (0.784, 0.870) | 0.745 (0.682, 0.799) | 0.560 (0.485, 0.637) | 0.358 (0.291, 0.441) | 0.554 (0.437, 0.660) | 61.458 (53.125, 69.589) |

## Label-permutation null: ECGFounder 12-lead pathway (published weights, 507/509 tensors), maternal ECG

*FELICITy 1, n = 144 women (73 positive, observed prevalence 51%); target PSS-10 ≥ 19 (binary); unit of analysis: participant; 1000 permutations of the participant labels, whole nested pipeline rerun each time.*

Best arm = highest stored AUC of the eleven set-A arms, fixed before this analysis. Bonferroni over 11 arms = min(1, 11·p).

| statistic | observed | null median | null 2.5–97.5% | p (one-sided) | Bonferroni ×11 |
|---|---|---|---|---|---|
| LR− at 70% sensitivity | 0.643 | 1.030 | 0.659–1.816 | 0.0220 | 0.242 |
| LR+ at 70% sensitivity | 1.249 | 0.996 | 0.832–1.272 | 0.0410 | 0.451 |
| Specificity at 70% sensitivity | 0.394 | 0.299 | 0.175–0.445 | 0.1019 | 1.000 |
| LR− at 70% sensitivity, pooled | 0.632 | 1.018 | 0.655–1.780 | 0.0200 | 0.220 |
| LR− at 75% sensitivity | 0.535 | 1.045 | 0.630–1.986 | 0.0030 | 0.033 |
| LR+ at 75% sensitivity | 1.254 | 0.992 | 0.848–1.233 | 0.0180 | 0.198 |
| Specificity at 75% sensitivity | 0.344 | 0.248 | 0.135–0.392 | 0.0969 | 1.000 |
| LR− at 75% sensitivity, pooled | 0.526 | 1.028 | 0.623–1.906 | 0.0030 | 0.033 |
| LR− at 80% sensitivity | 0.482 | 1.062 | 0.594–2.100 | 0.0030 | 0.033 |
| LR+ at 80% sensitivity | 1.251 | 0.994 | 0.871–1.199 | 0.0120 | 0.132 |
| Specificity at 80% sensitivity | 0.318 | 0.197 | 0.104–0.324 | 0.0310 | 0.341 |
| LR− at 80% sensitivity, pooled | 0.473 | 1.031 | 0.587–2.024 | 0.0020 | 0.022 |
| LR− at 90% sensitivity | 0.438 | 1.131 | 0.511–3.826 | 0.0090 | 0.099 |
| LR+ at 90% sensitivity | 1.105 | 0.995 | 0.917–1.115 | 0.0400 | 0.440 |
| Specificity at 90% sensitivity | 0.161 | 0.101 | 0.042–0.194 | 0.1049 | 1.000 |
| LR− at 90% sensitivity, pooled | 0.461 | 1.045 | 0.508–2.676 | 0.0170 | 0.187 |
| LR− at 95% sensitivity | 0.367 | 1.264 | 0.425–n/a | 0.0161 | 0.177 |
| LR+ at 95% sensitivity | 1.051 | 0.996 | 0.950–1.074 | 0.0789 | 0.868 |
| Specificity at 95% sensitivity | 0.076 | 0.054 | 0.014–0.115 | 0.2208 | 1.000 |
| LR− at 95% sensitivity, pooled | 0.396 | 1.081 | 0.405–3.894 | 0.0230 | 0.253 |
| LR− at the Youden point | 0.759 | 1.033 | 0.749–1.440 | 0.0310 | 0.341 |
| LR+ at the Youden point | 1.165 | 0.987 | 0.784–1.483 | 0.1578 | 1.000 |
| Specificity at the Youden point | 0.417 | 0.490 | 0.239–0.744 | 0.7373 | 1.000 |
| LR− at the Youden point, pooled | 0.775 | 1.025 | 0.751–1.311 | 0.0380 | 0.418 |
| Brier score | 0.348 | 0.418 | 0.342–0.490 | 0.0380 | 0.418 |
| Brier skill score | -0.389 | -0.646 | -0.936–-0.345 | 0.0509 | 0.560 |
| Brier score, recalibrated | 0.248 | 0.258 | 0.248–0.265 | 0.0300 | 0.330 |
| Brier skill score, recalibrated | 0.008 | -0.017 | -0.040–0.026 | 0.0789 | 0.868 |
| auc (mean of per-seed AUCs) | 0.599 | 0.492 | 0.382–0.603 | 0.0330 | 0.363 |

Fallback diagnostic on the best arm (threshold cross-fitted on stored OOF predictions instead of inner CV): sens70: sens 0.701, spec 0.425, LR− 0.702; sens75: sens 0.745, spec 0.420, LR− 0.607; sens80: sens 0.786, spec 0.361, LR− 0.594; sens90: sens 0.904, spec 0.225, LR− 0.428; sens95: sens 0.951, spec 0.208, LR− 0.237; youden: sens 0.800, spec 0.293, LR− 0.686; Brier 0.3058, BSS -0.223.

## Supplementary Table S6. Complementarity of representations and nested ensembles

Scores = participant-level outer-test probabilities, averaged over the 5 seeds. Calls = per-seed threshold calls at each target sensitivity (thresholds fitted inside the training fold). Arm abbreviations: SimCLR-m = maternal SimCLR, multi-layer; HeartWise-m = maternal DeepECG-SSL; HRV100-m = maternal 95-metric HRV panel (ensemble R peaks); omnibus-m = maternal hrv-omnibus HRV panel (ensemble R peaks); convHRV-m = maternal HRV, single detector (NeuroKit2); SimCLR-f = fetal SimCLR, multi-layer; SimCLR-a = abdominal SimCLR, multi-layer; HeartWise-a = abdominal DeepECG-SSL; ECGF12-m = maternal ECGFounder, 12-lead input; ECGF1-m = maternal ECGFounder, single-lead input; untrained-m = maternal ECGFounder architecture with no pretrained weights loaded, the untrained-encoder control.

### Table S6.1. Spearman correlation of arm scores

*FELICITy 1, n = 144 women (73 positive); target PSS-10 ≥ 19; unit of analysis: participant; Spearman rho of seed-averaged outer-test scores.*

| | SimCLR-m | HeartWise-m | HRV100-m | omnibus-m | convHRV-m | SimCLR-f | SimCLR-a | HeartWise-a | ECGF12-m | ECGF1-m | untrained-m |
|---|---|---|---|---|---|---|---|---|---|---|---|
| SimCLR-m | 1 | +0.54 | -0.08 | -0.23 | -0.15 | -0.04 | +0.42 | +0.31 | +0.39 | +0.39 | +0.49 |
| HeartWise-m | +0.54 | 1 | -0.05 | -0.15 | -0.06 | -0.11 | +0.15 | +0.44 | +0.43 | +0.51 | +0.55 |
| HRV100-m | -0.08 | -0.05 | 1 | +0.44 | +0.09 | -0.03 | -0.05 | -0.12 | +0.08 | -0.03 | -0.10 |
| omnibus-m | -0.23 | -0.15 | +0.44 | 1 | +0.15 | -0.07 | -0.17 | -0.13 | -0.07 | -0.10 | -0.15 |
| convHRV-m | -0.15 | -0.06 | +0.09 | +0.15 | 1 | -0.01 | -0.04 | +0.00 | -0.06 | +0.04 | -0.08 |
| SimCLR-f | -0.04 | -0.11 | -0.03 | -0.07 | -0.01 | 1 | +0.03 | -0.04 | +0.02 | -0.16 | -0.01 |
| SimCLR-a | +0.42 | +0.15 | -0.05 | -0.17 | -0.04 | +0.03 | 1 | +0.30 | +0.13 | +0.14 | +0.14 |
| HeartWise-a | +0.31 | +0.44 | -0.12 | -0.13 | +0.00 | -0.04 | +0.30 | 1 | +0.25 | +0.22 | +0.41 |
| ECGF12-m | +0.39 | +0.43 | +0.08 | -0.07 | -0.06 | +0.02 | +0.13 | +0.25 | 1 | +0.53 | +0.44 |
| ECGF1-m | +0.39 | +0.51 | -0.03 | -0.10 | +0.04 | -0.16 | +0.14 | +0.22 | +0.53 | 1 | +0.40 |
| untrained-m | +0.49 | +0.55 | -0.10 | -0.15 | -0.08 | -0.01 | +0.14 | +0.41 | +0.44 | +0.40 | 1 |

### Table S6.2. Cohen kappa of calls at target sensitivity 0.80

*FELICITy 1, n = 144 women (73 positive); target PSS-10 ≥ 19; unit of analysis: participant; kappa of the two arms' per-seed calls, seed mean.*

| | SimCLR-m | HeartWise-m | HRV100-m | omnibus-m | convHRV-m | SimCLR-f | SimCLR-a | HeartWise-a | ECGF12-m | ECGF1-m | untrained-m |
|---|---|---|---|---|---|---|---|---|---|---|---|
| SimCLR-m | 1 | +0.25 | -0.01 | -0.12 | -0.08 | -0.02 | +0.19 | +0.20 | +0.21 | +0.27 | +0.29 |
| HeartWise-m | +0.25 | 1 | -0.01 | -0.13 | -0.09 | +0.00 | +0.08 | +0.29 | +0.21 | +0.27 | +0.28 |
| HRV100-m | -0.01 | -0.01 | 1 | +0.32 | +0.12 | -0.02 | +0.01 | -0.02 | -0.01 | +0.02 | -0.01 |
| omnibus-m | -0.12 | -0.13 | +0.32 | 1 | +0.19 | -0.01 | -0.14 | -0.07 | -0.08 | -0.11 | -0.16 |
| convHRV-m | -0.08 | -0.09 | +0.12 | +0.19 | 1 | +0.09 | -0.01 | +0.04 | -0.06 | -0.00 | -0.10 |
| SimCLR-f | -0.02 | +0.00 | -0.02 | -0.01 | +0.09 | 1 | +0.00 | -0.09 | +0.02 | -0.04 | -0.04 |

| | SimCLR-m | HeartWise-m | HRV100-m | omnibus-m | convHRV-m | SimCLR-f | SimCLR-a | HeartWise-a | ECGF12-m | ECGF1-m | untrained-m |
|---|---|---|---|---|---|---|---|---|---|---|---|
| SimCLR-a | +0.19 | +0.08 | +0.01 | -0.14 | -0.01 | +0.00 | 1 | +0.19 | +0.08 | +0.01 | +0.07 |
| HeartWise-a | +0.20 | +0.29 | -0.02 | -0.07 | +0.04 | -0.09 | +0.19 | 1 | +0.11 | +0.17 | +0.22 |
| ECGF12-m | +0.21 | +0.21 | -0.01 | -0.08 | -0.06 | +0.02 | +0.08 | +0.11 | 1 | +0.41 | +0.30 |
| ECGF1-m | +0.27 | +0.27 | +0.02 | -0.11 | -0.00 | -0.04 | +0.01 | +0.17 | +0.41 | 1 | +0.38 |
| untrained-m | +0.29 | +0.28 | -0.01 | -0.16 | -0.10 | -0.04 | +0.07 | +0.22 | +0.30 | +0.38 | 1 |

## Table S6.3. Agreement and error overlap by family

*FELICITy 1, n = 144 women (73 positive); target PSS-10 ≥ 19; unit of analysis: participant; pair means; error overlap = |both wrong| / |either wrong| at the target, vs. expected if errors were independent within each class.*

| pairs | n pairs | Spearman mean [min, max] | kappa @70 | kappa @75 | kappa @80 | kappa @90 | kappa @95 | kappa @Youden | overlap @80 obs / expected (class-wise) / expected (marginal) | overlap @95 obs / exp. (class-wise) |
|---|---|---|---|---|---|---|---|---|---|---|
| within DL encoders | 21 | +0.23 [-0.16, +0.54] | +0.13 | +0.14 | +0.13 | +0.12 | +0.10 | +0.11 | 0.56 / 0.50 / 0.31 | 0.86 / 0.85 |
| within HRV panels | 3 | +0.23 [+0.09, +0.44] | +0.21 | +0.21 | +0.21 | +0.18 | +0.10 | +0.02 | 0.64 / 0.56 / 0.34 | 0.88 / 0.87 |
| DL vs HRV | 21 | -0.07 [-0.23, +0.08] | -0.05 | -0.05 | -0.04 | -0.02 | -0.01 | -0.02 | 0.52 / 0.53 / 0.32 | 0.86 / 0.86 |
| untrained control vs DL | 7 | +0.35 [-0.01, +0.55] | +0.22 | +0.22 | +0.22 | +0.17 | +0.11 | +0.14 | 0.61 / 0.52 / 0.31 | 0.87 / 0.86 |
| same encoder, mECG vs fECG vs aECG (SimCLR m/f/a, HeartWise m/a) | 4 | +0.21 [-0.04, +0.44] | +0.11 | +0.09 | +0.12 | +0.10 | +0.06 | +0.09 | 0.59 / 0.54 / 0.33 | 0.88 / 0.87 |
| SimCLR m/f/a | 3 | +0.14 [-0.04, +0.42] | +0.06 | +0.04 | +0.06 | +0.05 | +0.03 | +0.08 | 0.58 / 0.56 / 0.34 | 0.87 / 0.87 |
| HeartWise m/a | 1 | +0.44 [+0.44, +0.44] | +0.27 | +0.25 | +0.29 | +0.24 | +0.14 | +0.11 | 0.62 / 0.50 / 0.31 | 0.90 / 0.88 |
| different encoders, same signal (mECG DL) | 6 | +0.46 [+0.39, +0.54] | +0.28 | +0.30 | +0.27 | +0.25 | +0.25 | +0.19 | 0.58 / 0.47 / 0.29 | 0.85 / 0.81 |

## Table S6.4. Error overlap per pair at target sensitivity 0.80

*FELICITy 1, n = 144 women (73 positive); target PSS-10 ≥ 19; unit of analysis: participant; |both wrong| / |either wrong|, seed mean.*

| pair | Spearman | kappa | overlap obs | expected (class-wise indep.) | obs − expected |
|---|---|---|---|---|---|
| SimCLR-m–HeartWise-m | +0.54 | +0.25 | 0.62 | 0.52 | +0.10 |
| SimCLR-m–HRV100-m | -0.08 | -0.01 | 0.52 | 0.52 | -0.00 |
| SimCLR-m–omnibus-m | -0.23 | -0.12 | 0.50 | 0.54 | -0.04 |
| SimCLR-m–convHRV-m | -0.15 | -0.08 | 0.58 | 0.60 | -0.03 |

| pair | Spearman | kappa | overlap obs | expected (class-wise indep.) | obs − expected |
|---|---|---|---|---|---|
| SimCLR-m–SimCLR-f | -0.04 | -0.02 | 0.55 | 0.56 | -0.01 |
| SimCLR-m–SimCLR-a | +0.42 | +0.19 | 0.63 | 0.55 | +0.07 |
| SimCLR-m–HeartWise-a | +0.31 | +0.20 | 0.59 | 0.52 | +0.08 |
| SimCLR-m–ECGF12-m | +0.39 | +0.21 | 0.56 | 0.47 | +0.09 |
| SimCLR-m–ECGF1-m | +0.39 | +0.27 | 0.60 | 0.49 | +0.11 |
| SimCLR-m–untrained-m | +0.49 | +0.29 | 0.65 | 0.54 | +0.11 |
| HeartWise-m–HRV100-m | -0.05 | -0.01 | 0.50 | 0.50 | -0.00 |
| HeartWise-m–omnibus-m | -0.15 | -0.13 | 0.47 | 0.52 | -0.04 |
| HeartWise-m–convHRV-m | -0.06 | -0.09 | 0.55 | 0.58 | -0.03 |
| HeartWise-m–SimCLR-f | -0.11 | +0.00 | 0.54 | 0.54 | +0.00 |
| HeartWise-m–SimCLR-a | +0.15 | +0.08 | 0.56 | 0.53 | +0.03 |
| HeartWise-m–HeartWise-a | +0.44 | +0.29 | 0.62 | 0.50 | +0.12 |
| HeartWise-m–ECGF12-m | +0.43 | +0.21 | 0.54 | 0.45 | +0.08 |
| HeartWise-m–ECGF1-m | +0.51 | +0.27 | 0.58 | 0.47 | +0.11 |
| HeartWise-m–untrained-m | +0.55 | +0.28 | 0.63 | 0.52 | +0.11 |
| HRV100-m–omnibus-m | +0.44 | +0.32 | 0.64 | 0.52 | +0.13 |
| HRV100-m–convHRV-m | +0.09 | +0.12 | 0.62 | 0.57 | +0.04 |
| HRV100-m–SimCLR-f | -0.03 | -0.02 | 0.53 | 0.53 | -0.01 |
| HRV100-m–SimCLR-a | -0.05 | +0.01 | 0.53 | 0.53 | +0.00 |
| HRV100-m–HeartWise-a | -0.12 | -0.02 | 0.49 | 0.49 | -0.00 |
| HRV100-m–ECGF12-m | +0.08 | -0.01 | 0.44 | 0.45 | -0.01 |
| HRV100-m–ECGF1-m | -0.03 | +0.02 | 0.48 | 0.47 | +0.01 |
| HRV100-m–untrained-m | -0.10 | -0.01 | 0.52 | 0.52 | -0.00 |
| omnibus-m–convHRV-m | +0.15 | +0.19 | 0.66 | 0.59 | +0.06 |
| omnibus-m–SimCLR-f | -0.07 | -0.01 | 0.55 | 0.55 | -0.00 |
| omnibus-m–SimCLR-a | -0.17 | -0.14 | 0.50 | 0.55 | -0.05 |
| omnibus-m–HeartWise-a | -0.13 | -0.07 | 0.49 | 0.51 | -0.03 |
| omnibus-m–ECGF12-m | -0.07 | -0.08 | 0.44 | 0.46 | -0.03 |
| omnibus-m–ECGF1-m | -0.10 | -0.11 | 0.45 | 0.49 | -0.04 |
| omnibus-m–untrained-m | -0.15 | -0.16 | 0.48 | 0.54 | -0.05 |
| convHRV-m–SimCLR-f | -0.01 | +0.09 | 0.64 | 0.62 | +0.03 |
| convHRV-m–SimCLR-a | -0.04 | -0.01 | 0.61 | 0.61 | -0.00 |
| convHRV-m–HeartWise-a | +0.00 | +0.04 | 0.58 | 0.57 | +0.01 |
| convHRV-m–ECGF12-m | -0.06 | -0.06 | 0.50 | 0.51 | -0.01 |
| convHRV-m–ECGF1-m | +0.04 | -0.00 | 0.54 | 0.54 | +0.01 |
| convHRV-m–untrained-m | -0.08 | -0.10 | 0.57 | 0.60 | -0.03 |
| SimCLR-f–SimCLR-a | +0.03 | +0.00 | 0.57 | 0.57 | -0.00 |
| SimCLR-f–HeartWise-a | -0.04 | -0.09 | 0.50 | 0.53 | -0.03 |
| SimCLR-f–ECGF12-m | +0.02 | +0.02 | 0.49 | 0.48 | +0.01 |
| SimCLR-f–ECGF1-m | -0.16 | -0.04 | 0.48 | 0.50 | -0.01 |
| SimCLR-f–untrained-m | -0.01 | -0.04 | 0.54 | 0.55 | -0.01 |
| SimCLR-a–HeartWise-a | +0.30 | +0.19 | 0.60 | 0.53 | +0.07 |
| SimCLR-a–ECGF12-m | +0.13 | +0.08 | 0.51 | 0.48 | +0.04 |
| SimCLR-a–ECGF1-m | +0.14 | +0.01 | 0.50 | 0.50 | +0.01 |
| SimCLR-a–untrained-m | +0.14 | +0.07 | 0.58 | 0.55 | +0.03 |
| HeartWise-a–ECGF12-m | +0.25 | +0.11 | 0.49 | 0.45 | +0.04 |

| pair | Spearman | kappa | overlap obs | expected (class-wise indep.) | obs − expected |
|---|---|---|---|---|---|
| HeartWise-a–ECGF1-m | +0.22 | +0.17 | 0.53 | 0.47 | +0.07 |
| HeartWise-a–untrained-m | +0.41 | +0.22 | 0.60 | 0.51 | +0.09 |
| ECGF12-m–ECGF1-m | +0.53 | +0.41 | 0.61 | 0.43 | +0.18 |
| ECGF12-m–untrained-m | +0.44 | +0.30 | 0.59 | 0.47 | +0.13 |
| ECGF1-m–untrained-m | +0.40 | +0.38 | 0.65 | 0.49 | +0.16 |

## Table S6.5. Nested ensembles

*FELICITy 1, n = 144 women (73 positive); target PSS-10 ≥ 19; unit of analysis: participant; AUC of the seed-averaged ensemble score with participant-bootstrap 95% CI; differences use a paired bootstrap; operating point = target sensitivity 0.80 (all targets in the JSON).*

| group | method | AUC (95% CI) | best member (AUC) | Δ vs best member (95% CI) | Δ vs best single arm ECGF12-m (95% CI) | Brier raw | BSS raw | BSS recal. | sens@80 | spec@80 | LR+@80 | LR−@80 | ruled out /150 |
|---|---|---|---|---|---|---|---|---|---|---|---|---|---|
| i: 3 DL encoders, mECG | mean | 0.583 (0.490, 0.676) | ECGF12-m (0.604) | -0.021 (-0.086, +0.040) | -0.021 (-0.086, +0.040) | 0.287 | -0.148 | +0.015 | 0.841 | 0.273 | 1.159 | 0.592 | 32.3 |
| i: 3 DL encoders, mECG | stack | 0.555 (0.457, 0.648) | ECGF12-m (0.604) | -0.049 (-0.117, +0.020) | -0.049 (-0.117, +0.020) | 0.252 | -0.007 | +0.003 | 0.841 | 0.166 | 1.013 | 1.017 | 24.4 |
| ii: all trained mECG arms | mean | 0.585 (0.494, 0.676) | ECGF12-m (0.604) | -0.019 (-0.084, +0.052) | -0.019 (-0.084, +0.052) | 0.255 | -0.022 | -0.007 | 0.827 | 0.234 | 1.082 | 0.760 | 30.4 |
| ii: all trained mECG arms | stack | 0.592 (0.499, 0.686) | ECGF12-m (0.604) | -0.012 (-0.102, +0.077) | -0.012 (-0.102, +0.077) | 0.259 | -0.034 | +0.004 | 0.816 | 0.223 | 1.056 | 0.934 | 30.4 |
| iii: 3 mECG DL encoders + 2 ensemble-R-peak HRV panels | mean | 0.582 (0.491, 0.674) | ECGF12-m (0.604) | -0.021 (-0.090, +0.046) | -0.021 (-0.090, +0.046) | 0.261 | -0.042 | -0.010 | 0.838 | 0.248 | 1.117 | 0.668 | 30.6 |
| iii: 3 mECG DL encoders + 2 ensemble-R-peak HRV panels | stack | 0.537 (0.439, 0.631) | ECGF12-m (0.604) | -0.066 (-0.137, +0.008) | -0.066 (-0.137, +0.008) | 0.259 | -0.036 | -0.001 | 0.830 | 0.206 | 1.051 | 0.885 | 28.1 |
| iv-a: SimCLR across mECG+fECG+aECG | mean | 0.563 (0.477, 0.657) | SimCLR-f (0.558) | +0.005 (-0.082, +0.098) | -0.041 (-0.151, +0.071) | 0.281 | -0.125 | -0.010 | 0.797 | 0.256 | 1.071 | 0.778 | 34.4 |
| iv-a: SimCLR across mECG+fECG+aECG | stack | 0.437 (0.339, 0.539) | SimCLR-f (0.558) | -0.121 (-0.220, -0.018) | -0.167 (-0.277, -0.050) | 0.261 | -0.045 | -0.005 | 0.816 | 0.115 | 0.923 | 1.567 | 22.5 |
| iv-b: HeartWise across mECG+aECG | mean | 0.565 (0.472, 0.654) | HeartWise-m (0.573) | -0.008 (-0.058, +0.038) | -0.039 (-0.143, +0.059) | 0.283 | -0.131 | -0.001 | 0.811 | 0.248 | 1.080 | 0.775 | 32.7 |

| group | method | AUC (95% CI) | best member (AUC) | Δ vs best member (95% CI) | Δ vs best single arm ECGF12-m (95% CI) | Brier raw | BSS raw | BSS recal. | sens@80 | spec@80 | LR+@80 | LR−@80 | ruled out /150 |
|---|---|---|---|---|---|---|---|---|---|---|---|---|---|
| iv-b: HeartWise across mECG+aECG | stack | 0.541 (0.450, 0.638) | HeartWise-m (0.573) | -0.032 (-0.071, +0.007) | -0.062 (-0.169, +0.045) | 0.252 | -0.007 | -0.006 | 0.800 | 0.194 | 0.995 | 1.059 | 29.6 |
| v: all 10 trained arms (control excluded) | mean | 0.589 (0.496, 0.678) | ECGF12-m (0.604) | -0.014 (-0.085, +0.061) | -0.014 (-0.085, +0.061) | 0.249 | +0.006 | -0.000 | 0.836 | 0.296 | 1.187 | 0.551 | 34.4 |
| v: all 10 trained arms (control excluded) | stack | 0.550 (0.454, 0.648) | ECGF12-m (0.604) | -0.054 (-0.144, +0.038) | -0.054 (-0.144, +0.038) | 0.272 | -0.088 | -0.008 | 0.795 | 0.223 | 1.027 | 0.984 | 32.1 |

Single-arm AUCs (same scores): SimCLR-m 0.546; HeartWise-m 0.573; HRV100-m 0.484; omnibus-m 0.478; convHRV-m 0.328; SimCLR-f 0.558; SimCLR-a 0.519; HeartWise-a 0.538; ECGF12-m 0.604; ECGF1-m 0.600; untrained-m 0.572.

## Table S6.6. Label-permutation null for the best ensemble (ii: all trained mECG arms | stack)

*FELICITy 1, n = 144 women (73 positive); target PSS-10 ≥ 19; unit of analysis: participant; 500 permutations; SURROGATE: segment-level members refitted with participant-mean heads; everything else identical; labels permuted, whole nested stack (members, inner CV, stacker, thresholds, Platt) refitted.*

Observed AUC with the published heads 0.592; with the surrogate heads 0.574. The ensemble was chosen as the best of 12 after seeing results, so the p below is for a selected ensemble; Bonferroni over the 12 ensembles is shown.

| statistic | observed (surrogate) | null median | null 2.5–97.5% | p (one-sided) | Bonferroni ×12 |
|---|---|---|---|---|---|
| AUC of the seed-averaged score | 0.574 | 0.472 | 0.309–0.628 | 0.1058 | 1.000 |
| auc | 0.538 | 0.479 | 0.393–0.593 | 0.1637 | 1.000 |
| LR− at 80% sensitivity | 0.871 | 1.053 | 0.627–1.867 | 0.2635 | 1.000 |
| Specificity at 80% sensitivity | 0.217 | 0.208 | 0.121–0.337 | 0.4212 | 1.000 |
| Brier score | 0.265 | 0.278 | 0.254–0.298 | 0.1517 | 1.000 |
| Brier skill score | -0.060 | -0.095 | -0.169–-0.004 | 0.2176 | 1.000 |
| Brier score, recalibrated | 0.252 | 0.260 | 0.250–0.267 | 0.0479 | 0.575 |
| Brier skill score, recalibrated | -0.007 | -0.024 | -0.054–0.015 | 0.1417 | 1.000 |
| LR− at 70% sensitivity | 0.977 | 1.032 | 0.679–1.604 | 0.4192 | 1.000 |
| LR− at 90% sensitivity | 0.749 | 1.114 | 0.500–n/a | 0.1637 | 1.000 |

## Table S6.7. What the scores track

*FELICITy 1, n = 144 women (73 positive); target PSS-10 ≥ 19; unit of analysis: participant; Spearman rho (raw p) of the seed-averaged score with each covariate.*

| arm | Maternal age, years (n=144) | BMI (n=144) | GA at inclusion, weeks (n=144) | Mean RR, ms (n=144) | SDNN, ms (n=144) |
|---|---|---|---|---|---|
| SimCLR-m | -0.01 (0.878) | +0.00 (0.979) | -0.03 (0.706) | -0.04 (0.674) | -0.02 (0.779) |
| HeartWise-m | -0.02 (0.812) | +0.08 (0.330) | +0.04 (0.605) | +0.03 (0.688) | +0.06 (0.474) |
| HRV100-m | +0.04 (0.632) | +0.19 (0.020) | +0.03 (0.714) | -0.02 (0.842) | +0.00 (0.963) |
| omnibus-m | +0.11 (0.185) | +0.14 (0.093) | -0.00 (0.986) | -0.07 (0.373) | -0.15 (0.079) |
| convHRV-m | -0.12 (0.152) | -0.09 (0.272) | -0.16 (0.062) | -0.12 (0.139) | -0.01 (0.925) |
| SimCLR-f | -0.04 (0.608) | -0.12 (0.161) | -0.19 (0.019) | +0.19 (0.020) | +0.06 (0.469) |
| SimCLR-a | -0.01 (0.902) | +0.05 (0.589) | -0.02 (0.822) | -0.08 (0.355) | -0.08 (0.364) |
| HeartWise-a | +0.09 (0.296) | +0.19 (0.023) | -0.04 (0.651) | -0.18 (0.031) | -0.14 (0.083) |
| ECGF12-m | -0.07 (0.383) | +0.21 (0.012) | -0.04 (0.640) | -0.08 (0.357) | -0.08 (0.362) |
| ECGF1-m | -0.16 (0.050) | +0.11 (0.173) | -0.01 (0.929) | -0.10 (0.229) | +0.02 (0.857) |
| untrained-m | +0.03 (0.733) | +0.07 (0.398) | -0.07 (0.414) | -0.01 (0.910) | -0.08 (0.338) |
| label PSS-10 ≥ 19 | -0.08 (0.316) | +0.22 (0.007) | -0.08 (0.362) | -0.03 (0.746) | -0.00 (0.960) |

Variables: mean RR interval and SDNN from the 95-metric HRV panel on ensemble R peaks; heart rate = 60000/mean RR, so $\rho(HR) = -\rho(RR)$; body-mass index, maternal age and gestational age at inclusion from the cohort covariate sheet, with gestational age at inclusion taken as that at the recording visit.

## Supplementary Table S7. Results with raw $p < 0.05$, with multiplicity-adjusted values

Because the cohorts are small, we report every result with raw $p < 0.05$, with the multiplicity-adjusted value beside it. Excluded as not evidence of a stress signal: non-superiority tests (they support the bound), cross-validated scores significantly *below* chance (expected under a true null), and covariate tracking (BMI, RR, SDNN). Fetal and coupling rows lead with the primary fetal set (n = 113).

*All results with raw $p < 0.05$, FELICITy 1 at n = 144 (n = 113 for the primary fetal set; n = 144 in brackets). Predicted variable: in FELICITy 1 (F1), the binary label PSS-10 ≥ 19, one out-of-fold prediction per participant (the cutoff-sweep row predicts PSS-10 below each of 20 cutoffs); in FELICITy 2 (F2), the continuous PSS-10 score of the administration nearest in date to the recording, or of the same visit where stated (visit-specific labels), one prediction per recording, with the bootstrap clustered on women.*

| Cohort, n | Analysis | Status | Effect | Raw p | Adjusted p |
|---|---|---|---|---|---|
| F1, 113 (144) | Fetal sample-entropy decay, univariate | post hoc | AUC 0.632 (0.589) | 0.003 (0.007) | Holm (32) 0.096 (0.22) |
| F1, 113 (144) | Fetal 95-metric panel, nested single feature | post hoc | AUC 0.644 (0.617) | 0.011 (0.016) | Holm 0.30 (0.91) |
| F1, 113 (144) | Fetal entropy, nested single-summary selection | post hoc | AUC 0.632 (0.587) | 0.012 (0.043) | Holm (6) 0.072 (0.26) |
| F1, 113 (144) | Maternal→fetal TE decay, univariate | post hoc | AUC 0.532 (0.564) | 0.072 (0.020) | Holm (32) 1.00 (0.62) |
| F1, 113 (144) | Conditioned maternal–fetal RR entropy grid, nested | post hoc | AUC 0.542 (0.617) | 0.17 (0.017) | Holm 1.00 (0.95) |
| F1, 113 (144) | Transfer-entropy (TE) feature panel, nested | post hoc | AUC 0.523 (0.589) | 0.20 (0.042) | Holm 1.00 (1.00) |
| F1, 144 | Screening, maternal ECGFounder, LR− at 80% sensitivity | pre-specified range | LR− 0.48 | 0.003 | Bonferroni 0.033 |

| Cohort, n | Analysis | Status | Effect | Raw p | Adjusted p |
|---|---|---|---|---|---|
| F1, 144 | Low-PSS cutoff sweep, four arms (each calibrated over 20 cutoffs) | post hoc | max AUC 0.66–0.71 | 0.004–0.034 | Bonferroni (36): none < 0.00139 |
| F2, 22 women | Maternal DeepECG-SSL, baseline visit only, two channel rules | pre-specified sensitivity | r +0.44 to +0.50 | 0.0004–0.0006 | Holm (8 rows) 0.0032–0.0042 |
| F2, 44 rec / 22 women | Maternal DeepECG-SSL, visit-specific labels, F1 spectral scaling | pre-specified sensitivity | r +0.26 | 0.0015 | — |
| F2, 44 / 22 | Abdominal, omnibus channel (fixed detector), visit-specific labels | variant | r +0.28 | 0.0003 | — |
| F2, 44 / 22 | ECGFounder, mean of three channels, visit-specific labels | post hoc | r +0.31 | 0.011 | max-statistic over 20 channel-rule × endpoint combinations 0.58 |
| F2, 44 / 22 | ECGFounder, channel 1 of 3, visit-specific labels | post hoc | r +0.29 | 0.032 | max-statistic as above 0.58 |
| F2, 22 women | ECGFounder, within-woman paired change (score vs PSS-10), nearest-date / visit-specific labels | post hoc | r +0.462 / +0.472 | 0.016 / 0.014 (permutation) | max-statistic over 10 channel-rule × label combinations 0.103 |
| F2, 44 / 22 | SimCLR maternal, visit-specific labels | pre-specified sensitivity (coherence check) | r +0.198 | 0.025 | Holm (8 SimCLR endpoints) 0.199 |
| F2, 44 / 22 | **Pre-registered primary** (abdominal) / **secondary** (maternal) | pre-registered | r −0.03 / +0.19 | 0.54 / 0.058 | secondary Bonferroni (2) 0.12 |

Only two results survive their adjustment: the screening LR− and the F2 maternal baseline-only subset (22 women, at the ~15% power stated in the pre-registration, where a positive estimate is expected to be inflated in size). The fetal entropy and HRV arms are consistent in direction across several post hoc analyses and strengthen on the primary fetal set, but none survives correction.

## Supplementary Note 1. Window-level versus person-level evaluation of a published self-supervised ECG network

**Rationale.** An earlier analysis of the FELICITy cohort from our group reported an AUROC of 0.982 for classifying stressed versus control pregnancies from maternal ECG (Sarkar et al., 2021), using a self-supervised network pretrained on four public affect datasets (Sarkar & Etemad, 2022). Both studies stated that they cross-validated over 10-s windows, so windows from one person could appear in training and test folds; this subject-dependent protocol and the subject-independent protocol used in the present study answer different questions. Because the present study reports chance-level discrimination between women in the same cohort, we re-evaluated the published network under both protocols, with an analysis plan fixed before any feature was extracted, to show how the two figures relate.

**Methods.** We used the authors' public pretrained network (SSL-ECGv2, commit 4c16923) with the encoder frozen, extracting the 128-dimensional pooled features per 10-s window, and the authors' downstream classifier (two dense layers of 512 units; Adam, learning rate 0.001, batch 128; 100 epochs for FELICITy and 250 for the affect datasets, as in the original code and paper; no validation-based epoch selection). Only the split differed between comparisons. *Window-level (W)*: for FELICITy, the authors' own split function (five folds within each person); for WESAD and DREAMER, 10-fold cross-validation over shuffled windows . *Person-level (P)*: for FELICITy, the participant folds used throughout this paper; for WESAD and DREAMER, leave-one-subject-out . Absence of shared persons was asserted in every person-level fold. Data: FELICITy 1, the 144 women of this study (41,277 windows; the 2021 analysis used 107 women with complete covariates, a subset that cannot be reconstructed from the stored windows, and removed low-quality windows, which we did not); WESAD chest ECG, 15 subjects, baseline/stress/amusement (3,301 windows); DREAMER, 23 subjects, arousal and valence binarized at the mean rating (8,372 windows). The pretrained weights were confirmed loaded (12/12 convolutional tensors; feature similarity to a randomly initialized copy, linear CKA 0.21), and the same network with random weights was run as a control. Primary endpoints were the W − P difference (Δ) in window-level AUROC for FELICITy and in window-level accuracy for WESAD, with 95% confidence intervals and two-sided $p$ from 2,000 bootstrap resamples of persons, Holm-adjusted over the two.

**Results.** Under window-level splitting the published figures were reproduced: FELICITy AUROC 0.996 (published 0.982) and WESAD accuracy 0.962 (published 0.969). Under person-level splitting they fell to chance: FELICITy AUROC 0.455 and WESAD accuracy 0.493 (Supplementary Table S8). The differences were Δ = 0.541 (95% CI 0.468–0.614) and Δ = 0.469 (0.373–0.562), each $p < 0.0005$ raw and Holm-adjusted (no bootstrap resample reached zero). At the participant level, FELICITy AUROC was 1.000 under W and 0.450 under P, the latter consistent with the main result of this paper. The randomly initialized network reached FELICITy AUROC 0.956 under W, above the pre-specified reproduction threshold (0.952), and 0.490 under P. DREAMER did not reach its reproduction thresholds under W (arousal 0.809, valence 0.797) and fell to 0.501 and 0.530 under P.

**Interpretation.** The published results are reproducible under the protocol their authors described. The two protocols differ in what they can show, and the difference depends on whether the label varies within a person. In WESAD and DREAMER it does (conditions and clip ratings change within each subject), so a subject-dependent split still tests discrimination between conditions within a person, a legitimate personalized question; the subject-independent figures answer the separate question of generalization to unseen people. In FELICITy the label is fixed per woman. Once a woman's windows appear in both training and test folds, a classifier can score well by recognizing the woman without using stress-related information, as the untrained network shows (AUROC 0.956 under the window-level split). In this cohort the window-level figure therefore cannot separate stress from identity; this follows from the task design, not from an error in the method. With each woman confined to one fold, discrimination was at chance, consistent with the main result of this paper. This note concerns only how the earlier and present figures relate. Results from one network under a different evaluation protocol do not establish that no generalizable physiological stress signal exists in the maternal ECG; whether such a signal can be detected in people not seen in training remains open and needs further research, ideally in larger cohorts with repeated stress measures.

**Supplementary Table S8.** Published self-supervised ECG network (SSL-ECGv2) evaluated under window-level, subject-dependent (W) and person-level, subject-independent (P) protocols. Unit of resampling: person.

| Cohort (persons; windows) | Target | Metric (per window) | Published | Pretrained W | Pretrained P | Δ (95% CI) | Random-init W | Random-init P |
|---|---|---|---|---|---|---|---|---|
| FELICITy 1 (144; 41,277) | stress group | AUROC | 0.982 | 0.996 | 0.455 | 0.541 (0.468–0.614) | 0.956 | 0.490 |
| WESAD (15; 3,301) | 3-class affect | accuracy | 0.969 | 0.962 | 0.493 | 0.469 (0.373–0.562) | 0.860 | 0.537 |
| DREAMER (23; 8,372) | arousal | accuracy | 0.859 | 0.809 | 0.501 | 0.308 (0.247–0.368) | 0.656 | 0.493 |
| DREAMER (23; 8,372) | valence | accuracy | 0.850 | 0.797 | 0.530 | 0.267 (0.228–0.309) | 0.651 | 0.556 |

*Primary endpoints (FELICITy, WESAD): raw and Holm $p < 0.0005$. DREAMER rows descriptive; published thresholds not reached under W.*

**Disclosure.** M.G.F. is a co-author of Sarkar et al. (2021). The authors of both earlier studies were informed of these results before submission.